\documentclass[twocolumn,twocolappendix]{aastex631}

\usepackage{amsmath}
\usepackage{amssymb}
\usepackage{tabularx}
\usepackage{hyperref}

\newcommand{\gcrtb}{ASKAP J075024$-$205945}
\newcommand{\gcrta}{ASKAP J074913$-$155457}
\newcommand{\gcrtc}{ASKAP J160646$-$513843}
\newcommand{\gcrtd}{ASKAP J163248$-$420307}
\newcommand{\gcrte}{ASKAP J172523$-$303720}
\newcommand{\gcrtf}{ASKAP J183418$-$092720}
\newcommand{\lptgcrt}{GCRT J1745$-$3009}
\shorttitle{VAST sample of GRTs}
\shortauthors{Anumarlapudi et al.}
\graphicspath{{./}{figures/}}

\begin{document}

\title{A sample of short-lived Galactic radio transients from ASKAP VAST}

\correspondingauthor{Akash Anumarlapudi}
\email{akasha@unc.edu}

\author[0000-0002-8935-9882]{Akash Anumarlapudi}
\affiliation{Department of Physics and Astronomy, University of North Carolina at Chapel Hill, 120 E. Cameron Ave, Chapel Hill, NC, 27599, USA}
\affiliation{Department of Physics, University of Wisconsin-Milwaukee, P.O. Box 413, Milwaukee, WI 53201, USA}

\author[0000-0001-6295-2881]{David L. Kaplan}
\affiliation{Department of Physics, University of Wisconsin-Milwaukee, P.O. Box 413, Milwaukee, WI 53201, USA}

\author[0000-0002-5119-4808]{Natasha Hurley Walker}
\affiliation{International Centre for Radio Astronomy Research, Curtin University, Kent St, Bentley WA 6102, Australia}

\author[0000-0002-4941-5333]{Stella~Koch~Ocker}
\affiliation{Cahill Center for Astronomy and Astrophysics, California Institute of Technology, Pasadena, CA 91125, USA}
\affiliation{Observatories of the Carnegie Institution for Science, Pasadena, CA 91101, USA}

\author[0000-0003-4727-4327]{Daniel Kelson}
\affiliation{Observatories of the Carnegie Institution for Science, Pasadena, CA 91101, USA}

\author[0000-0003-0699-7019]{Dougal Dobie}
\affiliation{Sydney Institute for Astronomy, School of Physics, University of Sydney, NSW 2006, Australia}
\affiliation{ARC Centre of Excellence for Gravitational Wave Discovery (OzGrav), Australia}

\author[0000-0002-4405-3273]{Laura Driessen}
\affiliation{Sydney Institute for Astronomy, School of Physics, University of Sydney, NSW 2006, Australia}

\author[0000-0002-2686-438X]{Tara Murphy}
\affiliation{Sydney Institute for Astronomy, School of Physics, University of Sydney, NSW 2006, Australia}
\affiliation{ARC Centre of Excellence for Gravitational Wave Discovery (OzGrav), Australia}

\author[0000-0003-1575-5249]{Joshua Pritchard}
\affiliation{Australia Telescope National Facility, CSIRO Space and Astronomy, PO Box 76, Epping, NSW 1710, Australia}

\begin{abstract}

Galactic radio transients (GRTs) are mysterious short-lived ($\sim$days to months) radio transients that are quiet at all other wavelengths. Until now, roughly half a dozen such sources have been reported, predominantly towards the Galactic center. However, no unifying properties have been identified among these, leaving their nature, emission mechanism, and even classification poorly understood. Due to the lack of periodic and uniform radio observations over wide areas of the Galactic plane until now, the sample size of such transients remained limited. Here, we use radio observations from the Australian SKA Pathfinder's Variables and Slow Transients survey to discover six new radio transients along the Galactic plane that resemble GRTs. Detailed investigation of archival data suggests that these sources may be divided into two classes: sources that exhibit sporadic, pulse-like (minutes) radio emission, and sources that exhibit long-term (weeks) flaring-type radio emission. For the short-time variable sources, we draw similarities between optically bright long-period radio transients and our sample to propose wide-orbit ($\sim$days) white dwarf binaries as underlying sources. For sources that show long-term outbursts, we draw comparisons between dust-obscured outbursts from WD binaries and our sample. These results could imply that the ongoing wide-field radio surveys are uncovering radio emission from sub-populations of WD binaries that were previously unexplored. 

\end{abstract}


\section{Introduction} \label{sec:intro}

The timescales over which Galactic transients at radio wavelengths show variability vary from nanoseconds to a few months \citep{Hankins2003,chomiuk2021}. However, sources that vary on intermediate timescales (minutes--hours--days) are far less explored compared to shorter timescales ($\sim$seconds and below). This is mainly because sources that vary on intermediate and long timescales are predominantly discovered in the image domain \citep{radiosky_dawes2026}, and discovering a large sample of such sources requires i) good cadence to sample the evolution and ii) a large field of view for untargeted searches. Historically, due to the narrow field of view of radio interferometers, this meant that many large-scale surveys were restricted to single-epochs \citep{nvss,first,sumss,tgss} or sparse samplings \citep{vlass}. Hence, examples of such sources were often restricted to serendipitous single-source discoveries \citep[e.g.,][]{zhao_transient_1992,hyman_low-frequency_2002,hyman_powerful_2005,hyman_gcrt_2009} rather than populations, leaving their origin unclear. In recent times, with the availability of wide-field radio interferometers, new classes of sources with minutes--day timescale radio emission are emerging\footnote{It is unclear if these are new source classes or new radio emission phenomena in known source classes or a combination of both.}, two of which are the long-period radio transients \citep[LPTs;][]{hurley-walker_radio_2022,hurley-walker_long-period_2023,caleb_emission-state-switching_2024,dong_discovery_2024,Dobie2024,de_ruiter_white_2024,hurley-walker_29-hour_2024,wang2025,lee2025,askap1448,bloot2025} and the so-called ``Galactic Center radio transients'' \citep[GCRTs;][]{davies_transient_1976,zhao_transient_1992,hyman_low-frequency_2002,hyman_powerful_2005,hyman_new_2006,wang_discovery_2021}. 

LPTs are periodic radio sources, with periods varying from a few minutes to a few hours \citep[e.g., ][]{hurley-walker_radio_2022,hurley-walker_long-period_2023,lee2025,askap1448}. Some LPTs are observed to be transient (visible only for a few days/months; \citealp{hurley-walker_radio_2022,caleb_emission-state-switching_2024,lee2025}), while others were found to be persistent \citep{hurley-walker_long-period_2023,wang2025,askap1448}. While LPTs with optical counterparts are traced to magnetic WD binaries \citep{bloot2025,de_ruiter_white_2024,hurley-walker_29-hour_2024}, the nature of LPTs that lack multi-wavelength counterparts is still debated \citep{caleb_emission-state-switching_2024,wang2025}. It is also possible that the current population of LPTs stems from multiple classes of sources.

GCRTs, on the other hand, are radio transients of unknown origin that were mostly found towards the Galactic center. The lack of distinguishing radio signatures (like sharp pulsations in pulsars), coupled with the absence of counterparts at other wavelengths \citep{kaplan_search_2008,wang_discovery_2021}, meant that their nature has remained elusive. Until now, fewer than half a dozen such sources have been reported. Only one of these sources \lptgcrt\ was found to be periodic \citep[LPT-like;][]{hyman_powerful_2005}, with broad bursts ($\approx$12\,minute), and a long (77-minute) period. Given its short-lived nature, this source has not been observed to be active for the past two decades. The remaining sources, however, did not show a similar bursting nature (periodic or otherwise) and were more consistent with being persistent on short timescales (minutes) but transient on longer timescales (days--months). Many of the GCRTs were found to have steep spectra (power law index $\alpha<-1$, where the flux density at frequency $\nu$ is given by $S_{\nu}\propto\nu^{\alpha}$), hinting that their emission is not thermal in nature.
Hence, proposals explaining the radio emission often involved magnetospheric processes powering the radio emission \citep{kulkarni_astronomy_2005,zhu_gcrt_2006,turolla_is_2005}. 

Unlike LPTs, which at least have dispersion measure-based distance estimates, there have been no distance estimates to any of the GCRTs, and hence it is unclear if GCRTs are indeed located at/near the GC or just happen to lie along the line of sight. Since there is a preponderance of radio observations towards the GC (vs the entire Galactic plane), it is possible that the current sample is a result of this search bias. If so, the Galactic center reference can be misleading, and hence we refrain from using ``GCRT'' (despite the initial naming) from here on and refer to these sources as Galactic radio transients (GRTs). However, we caution that this generic naming is neither reflective of the nature of the underlying sources nor any emission characteristic. In addition, while \lptgcrt\ closely resembles transient LPTs, such an explanation is ruled out in other GRTs \citep{zhao_transient_1992,hyman_gcrt_2009,wang_discovery_2021}. Hence, it is possible that sources other than \lptgcrt\ form a representative sample of ``GRTs'', although it is unclear currently what constitutes/defines such a sample, given the very limited information. 

Identifying more such events itself poses a tricky question: which are the observed radio properties that can be used to define or classify a ``GRT''? Due to the lack of optical/infrared counterparts or a definitive radio signature, it is unclear whether GRTs form a separate (sub-)population of radio bright sources or a previously unexplored radio phenomenon in known sources. This lack of a precise classification will have implications for any ``unbiased'' search for these objects. 
Uniform observations of the entire Galactic plane (and perhaps even at higher latitudes) are needed to discover more GRTs. However, multi-epoch radio surveys that target large areas of the Galactic plane have not been available until very recently, with surveys like the Variable and Slow transients survey \citep[VAST;][]{vast,vastpilot}, and the Galactic Plane Monitor from the 
Murchison Widefield Array \citep[GPM; see Methods of][for a brief description]{hurley-walker_long-period_2023}
leading the effort. VAST is a survey science project at the Australian SKA Pathfinder \citep[ASKAP;][]{askap} that scans the Galactic sky every two weeks and the extragalactic sky every two months, targeting sources that are variable over a few seconds to year(s). In an individual 12-minute exposure, VAST achieves a rms sensitivity of 250\,$\mu$Jy at 887.5\,MHz. The main VAST survey started in December 2022, providing roughly 35 snapshots of the Galactic plane in 1.5\,yrs,  
which is extremely useful in discovering GRTs where the emission varies on similar timescales. In this article, we use 1.5\,years of data from the VAST survey to search for these objects. 

This article is organized as follows: we provide a brief summary of our data acquisition and reduction techniques in  \S\ref{sec:vast}. In \S\ref{sec:sample} we detail our sample selection methods, followed by our methods in \S\ref{sec:methods} and a brief discussion of all the individual sources in \S\ref{sec:res}. We discuss the population properties of these sources in  \S\ref{sec:discussion}, including the expected rate of such events, before concluding in  \S\ref{sec:conclusion}.

\section{Sample Acquisition} \label{sec:data}

\subsection{Data Collection} \label{sec:vast}

VAST data were processed using standard software, \textsc{ASKAPSoft} \citep{askapsoft}, to generate images and noise maps. A custom source finder for ASKAP data, \textsc{selavy} \citep{selavy}, was used to perform source extraction on images, generating a catalog of sources (5-$\sigma$ significance; where $\sigma$ is the local background noise) for every observing epoch. Observations comprising 35\,epochs taken between December 2022 and August 2024 were selected to look for GRTs. Sources from individual epochs were collated and paired across the epochs by \textsc{vasttools} \citep{vasttools} to form a master catalog of unique sources.

\subsection{Sample Selection} \label{sec:sample}

The combined catalog over 35 epochs of VAST Galactic observations comprises $\sim$ 660,000 sources. The first step was to select unresolved point sources. By default, the source finding software for ASKAP data, \textsc{selavy}, labels a source to have \textit{siblings} (equivalently \texttt{n\_siblings}$>0$) when a source is fit with multiple Gaussian components --- we restricted sources to have \texttt{n\_siblings}=0. In addition, we demanded that the \textit{compactness} of the source --- the ratio of integrated to peak flux density be less than 1.5 to select for unresolved sources. 
Applying these two cuts resulted in a total of 290,000 sources.

The next step was to look for bright sources. We required that the maximum signal-to-noise (SNR) ratio of the source be at least 8. Doing so reduced the number of candidates to 90,000. We then cross-matched these sources with existing databases/catalogs, including  \textit{Simbad} \citep{simbad}, 
known active galactic nuclei (AGN) catalogs \citep{wiseagn,milliquas,sdssqso}, pulsar catalog \citep{psrcat}, and the X-ray binary catalog \citep{blackcat}, to remove VAST sources that are within 2.5\arcsec\ of a known source. The choice of this radius was motivated by the current positional uncertainties from ASKAP data --- up to 2.5\arcsec\ including systematic errors. After removing known sources, we ended up with a sample of $\sim$ 76,000 sources.

Most GRTs have been observed to be active for a short time \citep[$\sim$months;][]{hyman_powerful_2005,hyman_faint_2007,wang_discovery_2021,wangz2022}. We therefore required a source to be detected in less than 80\% of the observations to select against persistent sources. In addition, we removed sources that were detected in the Rapid ASKAP Continuum survey \citep[RACS; observed in March 2019]{racscatalogpaper} to account for unidentified persistent radio sources. We caution that this might remove sources of interest that have multiple active episodes, two of which might have randomly coincided with the VAST and RACS observations, but the number of such sources will be very low. Applying these cuts, the number of candidates was reduced to 3600.

The final step of filtering involved quantifying the variability in the light curve. We used the variability metrics $\eta$ (reduced $\chi^2$) and $V$ (fractional variability) from \citet{LoTraP} that quantify the deviation of a light curve from a steady source model, given by
\begin{align}
    \eta &= \frac{1}{N-1}\sum_{i=1}^N \left(\frac{f_{\nu,i} - \bar{f_{\nu}}}{\sigma_{\nu,i}}\right)^2 \nonumber \text{and,} \\
    V &= \frac{\sigma_{f_{\nu}}}{\bar{f_{\nu}}} \nonumber 
\end{align}
where $f_{\nu,i}$, and $\sigma_{\nu,i}$ are the $i^{th}$ flux density and error measurements, $\bar{f_{\nu}}$ is the error-weighted mean flux density of $N$ measurements, and $\sigma_{f_{\nu}}$ is the sample standard deviation of these measurements.

We removed sources with $\eta<10$ and $V<0.25$\footnote{The specific values of $\eta$ and $V$ are motivated by the balance between retaining variable sources and achieving a manageable sample for further examination.}. After applying these cuts, we ended up with 550 variable sources.

\begin{deluxetable}{lr}[!b]
    \caption{Summary of the filtering steps used to select the final sample of GRTs}\label{tab:sel}
    \tablehead{
    \colhead{Filtering step} & \colhead{Sources remaining}
    }
    \startdata
    Master catalog & $\sim$ 660,000 (100\%) \\
    Removal of extended sources & 290,000 (44\%)\\
    Removal of faint sources & 90,000 (14\%) \\
    Removal of cataloged sources & 76,000 (12\%) \\
    Selecting variable sources & 550 (0.08\%) \\
    Manual inspection & 6 (10\,ppm)\\
    \enddata
\end{deluxetable}


We performed forced photometry at the location of each candidate in the individual VAST images using the synthesized beam information. We then manually inspected all the candidates to look for low-level persistent/scintillating sources. \textsc{Selavy}, by default, identifies $5 \sigma$ confident sources ($\sigma$ is the local noise), and hence we looked for weaker detections ($>3 \sigma$) and excluded sources that had multiple weaker detections. 
After manual inspection of all the 550 sources, we ended up with a sample of 6 candidates 
A summary of this selection is provided in Table~\ref{tab:sel}.

\begin{figure*}[!htb]
\gridline{\fig{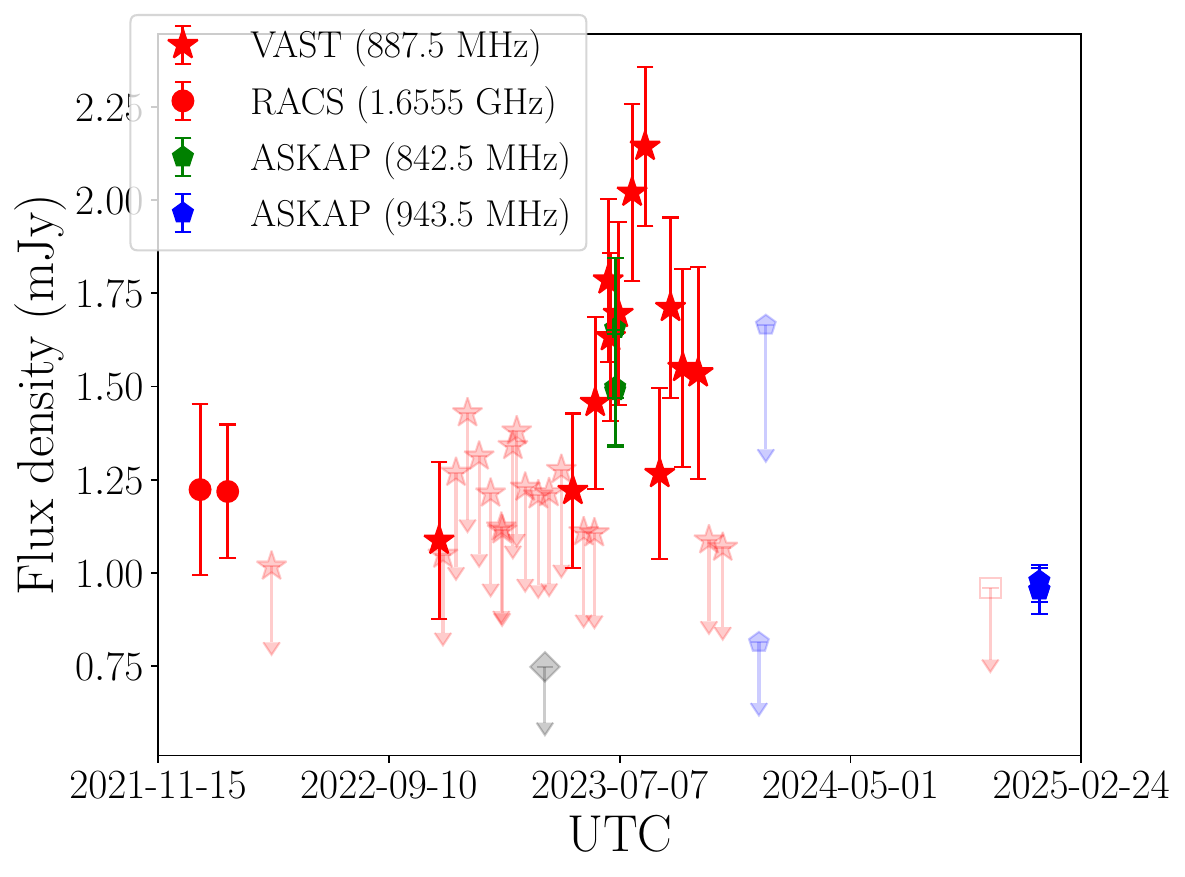}{0.32\textwidth}{(a) \gcrta}
          \fig{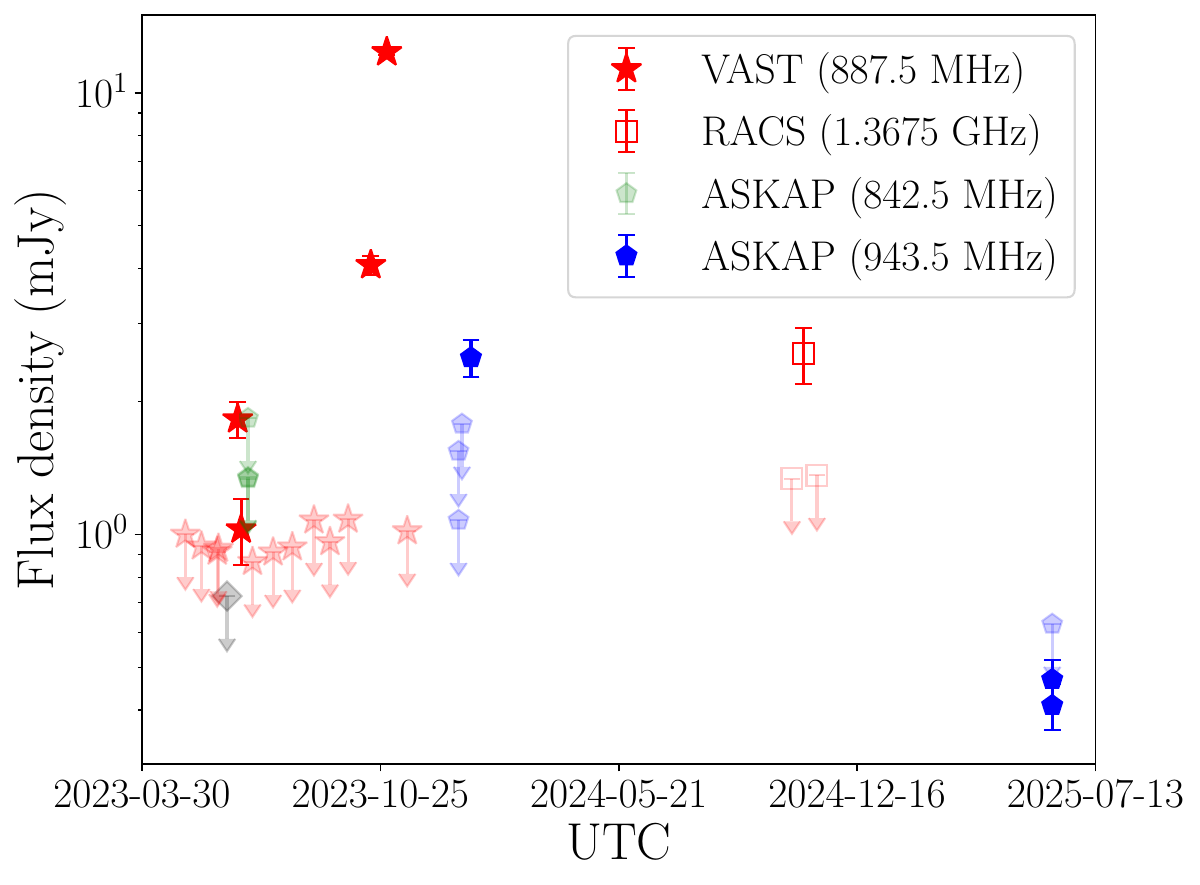}{0.32\textwidth}{(b) \gcrtb}
          \fig{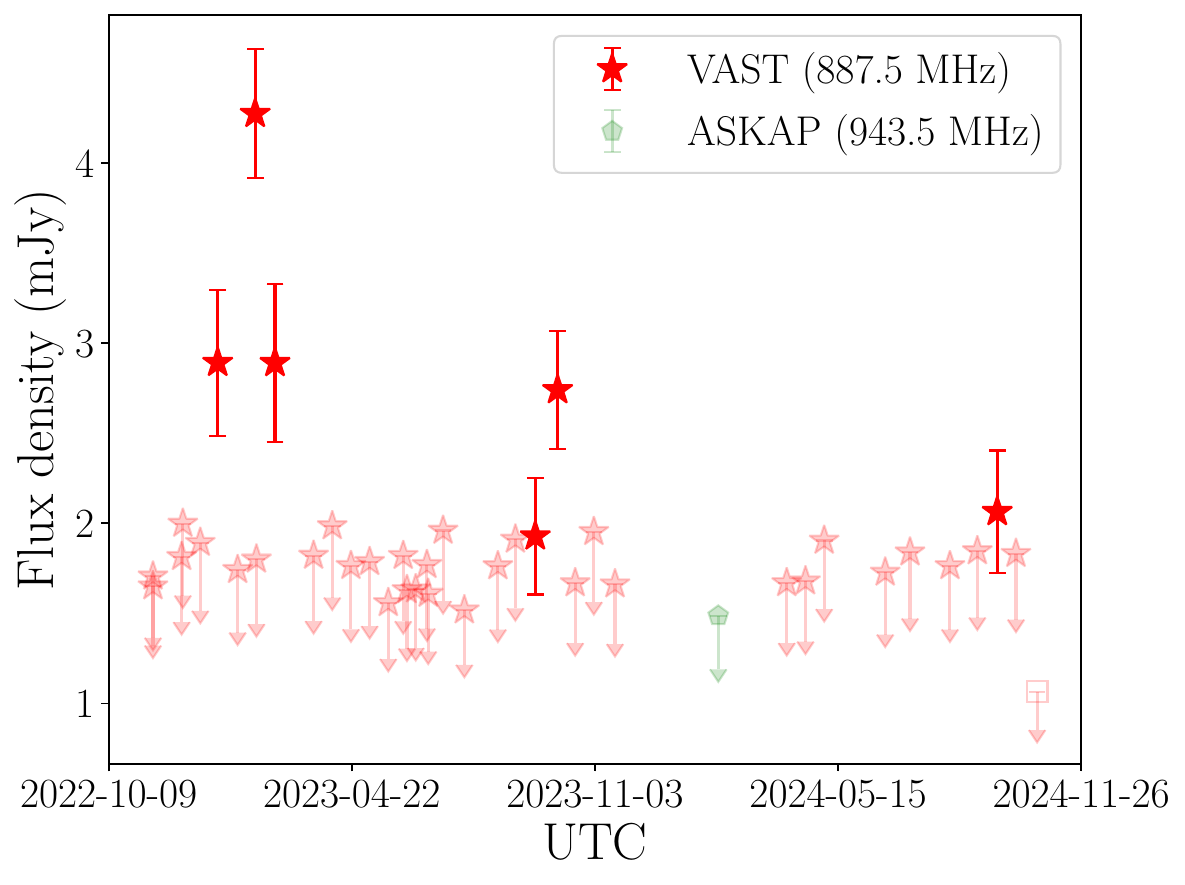}{0.32\textwidth}{(c) \gcrtc}}
\gridline{\fig{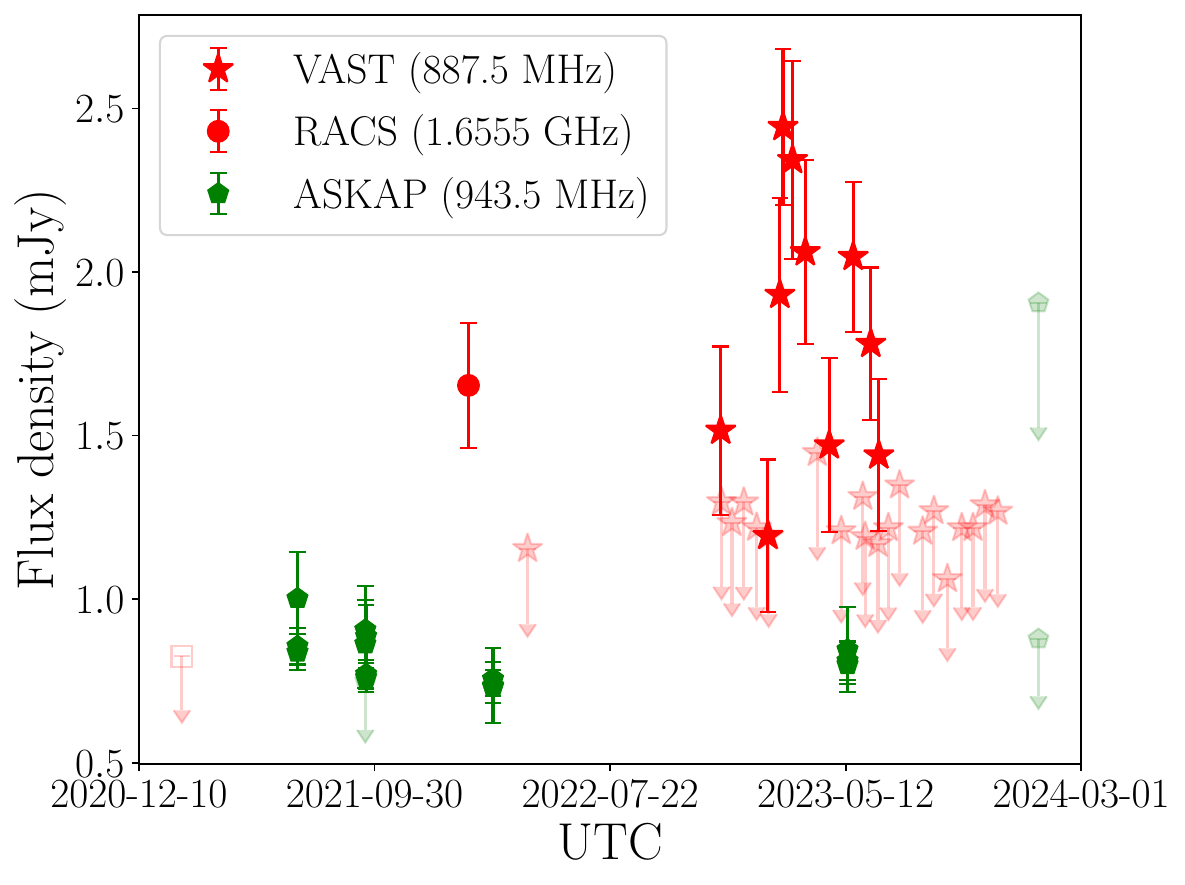}{0.32\textwidth}{(d) \gcrtd}
          \fig{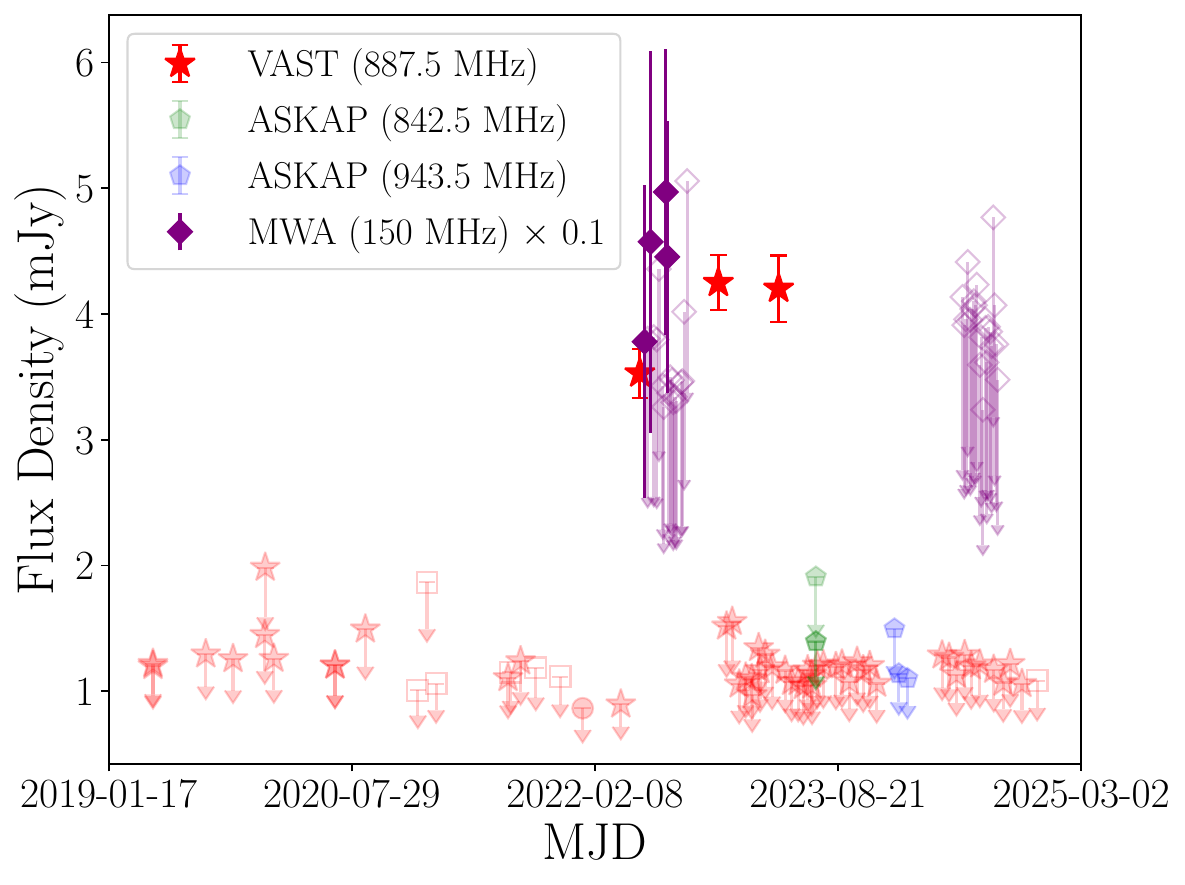}{0.32\textwidth}{(e) \gcrte}
          \fig{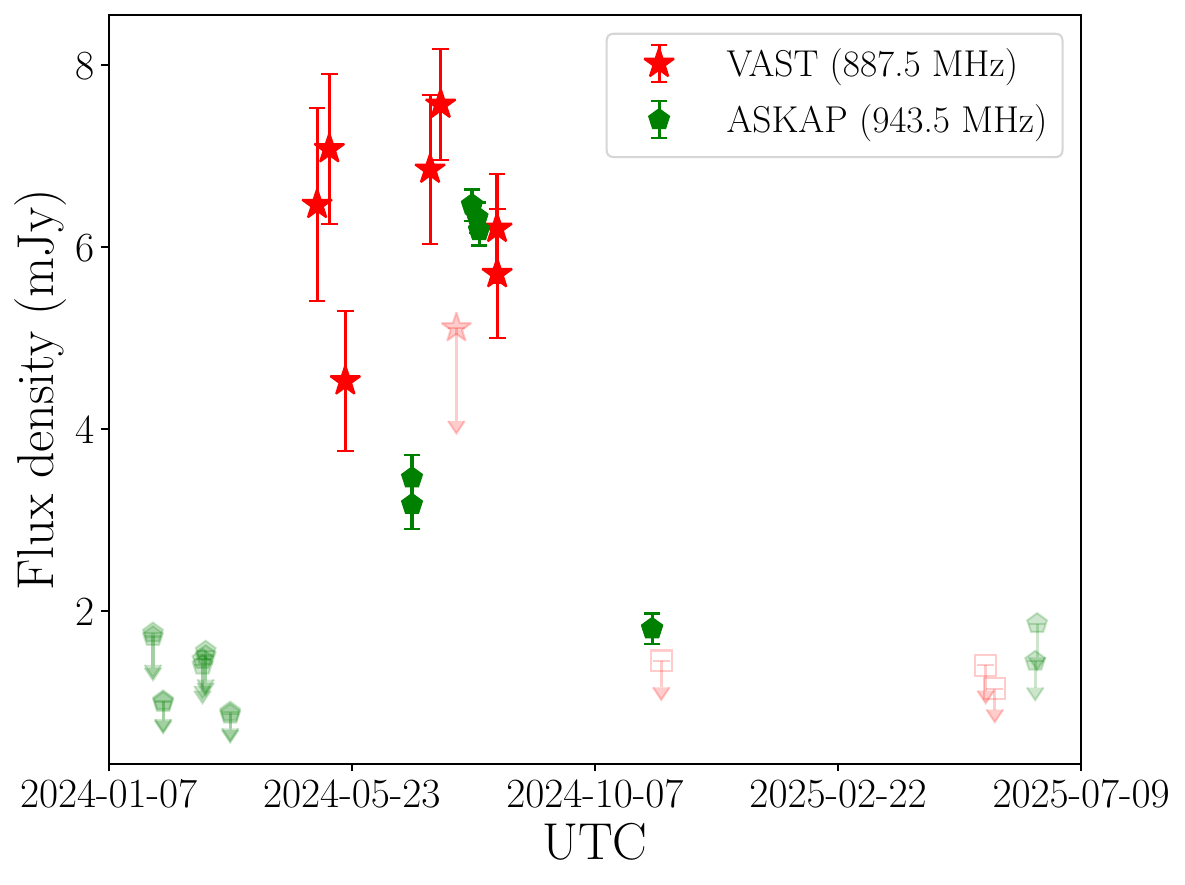}{0.32\textwidth}{(f) \gcrtf}}
\caption{Light curves of the candidates listed in Table~\ref{tab:cands}. Data shown as red stars are the measurements from the VAST Galactic survey, those shown as red downward pointing arrows are 5-$\sigma$ upper limits from non-detections, and those shown in green and blue diamonds are the measurements from 8\,hr target of opportunity ASKAP observations of different sources where our target source happened to be in the field. Shown in purple diamonds are the data from the GPM from the MWA (open symbols are upper limits).}
\label{fig:lcs}
\end{figure*}

\section{Multi-wavelength observations}\label{sec:methods}

Although sources with known multi-wavelength counterparts were excluded, the final sample of radio sources can have unclassified or transient multi-wavelength counterparts, which can provide definitive characteristics of what these sources can be. Hence, we perform a detailed multi-wavelength analysis using available archival data. We provide a summary of various methods (presented by wavelength) that were used to investigate the properties of our sample of sources. 
\subsection{Radio}
\begin{enumerate}
    \item To generate a detailed light curve, we inspected all available public radio data, including ASKAP data that is not taken as part of the VAST survey. We also examined contemporary data from the Very Large Array Sky Survey \citep[VLASS; ][]{vlass}, and the Galactic Plane Monitor from MWA. 
    \item  To study the detailed properties of radio emission, we constructed time and frequency-resolved 2D intensities, or \textit{dynamic spectra}, using the \textsc{dstools} software package \citep{dstoools}. We used the dynamic spectra to look for both short-timescale and narrow-band emission. In particular, frequency-resolved spectra can be used to examine the Faraday rotation of the linear polarization vector and estimate the rotation measure (RM) of the source. Identification of strong polarization (linear and/or circular) can point to the presence of strong magnetic fields. In this article, we follow the standard convention for polarization measurements: the polarization angle is defined as $\chi (\nu)=\frac{1}{2}\rm tan^{-1} \left(\frac{U(\nu)}{Q(\nu)}\right)$ and RM is measured using $\chi (\lambda=c/\nu) = \chi_0 + \rm RM \times \lambda^2$, with a positive value of RM corresponding to net line of sight magnetic field pointing towards the observer and vice-versa. RM is estimating using \textsc{rmsynthesis}\footnote{\url{https://github.com/gheald/RMtoolkit}}.
\end{enumerate}

\subsection{Optical/Infrared}
\begin{enumerate}
    \item  For the optical emission, we inspected  Dark Energy Camera Plane survey \citep[DECam/DECaPS;][]{decaps} images at optical wavelengths, where available. In the absence of these, we utilized the Panoramic Survey Telescope and Rapid Response System \citep[\textit{Pan-STARRS};][]{panstarrs} images. We also extracted the high cadence light curves using the public data from the Zwicky transient facility \citep[ZTF; ][]{ztf}.
    \item For near-infrared (NIR) emission, we inspected Visible and Infrared Survey Telescope for Astronomy (VISTA) Variables in the Via Lactea \citep[VVV;][]{vvv} images and the United Kingdom Infrared Deep Sky Survey \citep[UKIRT/UKIDSS][]{UKIDSS} images. 
    \item At mid-IR wavelengths, we looked at the Wide-field Infrared Survey Explorer \citep[WISE;][]{wise} data, particularly after the \textit{NEOWISE} reactivation \citep{neowise}.
    \item We visually inspected individual images to look for point sources at the target position. In the case of non-detection, where required, we estimated the local noise to get the 5-$\sigma$ upper limit on the instrumental magnitude. Instrumental magnitude is converted to apparent magnitude using official zeropoints if they are provided by the surveys. In their absence, zeropoints are estimated using \textsc{sextractor} \citep{sextractor} and DECaPS/UKIRT/2MASS catalogs as references. All magnitudes in this article are provided in the AB magnitude system, unless explicitly stated otherwise.
\end{enumerate}

\subsection{X-ray}
\begin{enumerate}
    \item  We examined archival \textit{Swift}, 
    \textit{XMM-Newton}, 
    and \textit{Chandra} observations 
    to look for X-ray point sources. We also cross-matched our sample with the \textit{Fermi} Large Area Telescope (LAT) 14-year source catalog \citep{4fgl} to look for a counterpart to within half a degree. In the absence of a detection, we estimated the background count rate and converted it to flux upper limits. None of the reported GRTs have detectable X-ray counterparts, and hence, the X-ray spectral information is unclear. However, given the likely compact object nature of these sources, under the assumption that the magnetospheric processes power the X-ray emission (similar to pulsar/magnetars; \citealp{kaspi_magnetra_review}), the source spectrum can be expressed as a power law. Given the lack of information, we assumed a power law index of $-1$. Since all these sources are in the Galactic plane, we adopt a neutral hydrogen column density $N_{\rm H}$ of $10^{22}$ cm$^{-2}$ to convert count rate limit to the flux limit. 
\end{enumerate}

\section{Results}\label{sec:res}


We present our sample of six sources and their observational properties, along with the archival GRTs in Table~\ref{tab:cands_radio_prop}. In Table~\ref{tab:cands}, we provide their flux measurements/upper limits at radio, optical, NIR, and X-ray wavelengths. In all our searches for a counterpart at optical/NIR/X-ray wavelengths, we did not detect a persistent counterpart; hence, all these flux measurements correspond to upper limits on the persistent emission. Figure~\ref{fig:lcs} shows the radio light curves of the individual candidates. Provided in the Appendix~\ref{app:images} are the Stokes I images of the brightest detection and the stacked image made from all the non-detections to show the transient nature of the sources. The stacked image is generated by combining the radio images rather than a deconvolution of the full visibility data. In Appendix~\ref{app:oir_images}, we also provide optical/NIR composite images of the fields centered on the radio candidates for reference. In what follows, we detail properties of individual sources. 

\begin{deluxetable*}{llrrrcrcccrrl}[!hb]\label{tab:cands_radio_prop}
\rotate
\caption{List of GCRT-like candidates from our search on VAST data and their basic detection statistics. Multi-wavelength flux measurements for these sources are provided in Table~\ref{tab:cands}.}
\tablehead{
\colhead{Name} & \multicolumn{4}{c}{Coordinates} & \multicolumn{2}{c}{Flux} & \colhead{Active} & \colhead{Multiple\tablenotemark{a}} & \colhead{$\chi^2$} & \multicolumn{2}{r}{Polarization\tablenotemark{b}} & \colhead{$\alpha$\tablenotemark{c}} \\
\colhead{} & \colhead{} & \colhead{} & \colhead{} & \colhead{} & \colhead{} & \colhead{} & \colhead{time} & \colhead{events} & \colhead{} & \colhead{} & \colhead{} & \colhead{} \\
\hline
\colhead{} & \colhead{RA} & \colhead{Dec} & \colhead{GL} & \colhead{GB} & \colhead{$\nu$} & \colhead{$\rm S_{\rm peak}$} & \colhead{} & \colhead{} & \colhead{(reduced)} & \colhead{Linear} & \colhead{Circ.} & \colhead{} \\
\colhead{} & \colhead{(hour)} & \colhead{(degree)} & \colhead{($^{\circ}$)} & \colhead{($^{\circ}$)} & \colhead{(GHz)} & \colhead{(mJy)} & \colhead{(days)} & \colhead{} & \colhead{} & \colhead{(\%)} & \colhead{(\%)} & \colhead{}
}
\startdata
J074913$-$155457\tablenotemark{d} & 07h49m12.7s & $-$15d54m58s & 233.7145 & 5.073 & 0.88 & 2.1(2) & 250 & False & 2.3 & False & False & $\lessapprox0$ \\
J075024$-$205945 & 07h50m24.5s & $-$20d59m45s & 238.2543 & 2.755\phm{0} & 0.88 & 12.3(2) & \nodata & True & 244.5 & $\sim 100\%$ & $\sim 50\%$ & $-$1.9(3) \\
J160646$-$513843 & 16h06m46.1s & $-$51d38m43s & 330.7804 & 0.341 & 0.88 & 4.3(4) & \nodata & True & 7.7 & $\sim 100\%$ & False & \nodata \\
J163248$-$420308 & 16h32m48.4s & $-$42d03m08s & 340.6879 & 4.036 & 0.88 & 2.4(2) & 197 & False & 3.3 & False & False & $\lessapprox-3$ \\
J172523$-$303720 & 17h25m23.4s & $-$30d37m20s & 356.2101 & 2.784 & 0.88 & 4.2(2) & \nodata & True & 51.8 & $\sim100\%$ & False & $-$2.5(1) \\
J183417$-$092723 & 18h34m17.3s & $-$09d27m23s & 22.570 & $-$0.532\phm{0} & 0.88 & 8\phm{.0}(1) & 95 & False & 3.4 & False & False & $\approx 0$ \\
\hline
\sidehead{\textbf{Archival sources}}
A1742$-$28 & 17h42m29.3s & $-$28d59m18s & 359.5967 & 0.5562 & 0.96 & 480(70) & $<$120 & False & \nodata & NA\tablenotemark{e} & NA & \nodata\\ 
\text{\citep{davies_transient_1976}} & & & & \phm{0.0000}& & & & & & & & \\
GCT 1742$-$2859 & 17h42m30.6s & $-$28d59m50s & 359.5917 & 0.5475 & 1.4\phm{0} & $\sim$900\phm{(00)} & $\sim$220 &  False & \nodata & False & False & $-1.2$\\
\text{\citep{zhao_transient_1992}} & & & & \phm{0.0000}& & & & & & & & \\
GCRT J1746$-$2757 & 17h46m09.8s & $-$27d57m01s & 0.9040 & 0.4112 & 0.33 & 220(20) & $<150$ & False & \nodata & NA & NA & \nodata\\
\text{\citep{hyman_low-frequency_2002}} & & & & \phm{0.0000}& & & & & & & & \\
GCRT J1745$-$3009 & 17h45m00.5s & $-$30d09m52s & 358.8826 & $-$0.5271 & 0.33 & 1000(110) & 0.054\tablenotemark{f} & True& \nodata  & \nodata & $100$\% & $-$13(3)\\
\text{\citep{hyman_powerful_2005}} & & & & \phm{0.0000}& & & ($<$540) & & & & & \\
GCRT J1742$-$3001 & 17h42m04.7s & $-$30d01m44s & 358.6654 & 0.0835 & 0.23 & 107(6) & 270 & False & \nodata & NA & NA & $<-2$ \\
\text{\citep{hyman_gcrt_2009}} & & & & \phm{0.0000}& & & & & & & & \\
ASKAP  & 17h36m08.2s & $-$32d16m35s & 356.0863 & $-$0.0394 & 0.88 & 15.5(4) & 90 & False & \nodata & $\sim$80\% & $\sim$30\% & $<-2.7$ \\
J173608.2–321635 & & & & \phm{0.0000}& & & & & & & & \\
\text{\citep{wang_discovery_2021}} & & & & \phm{0.0000}& & & & & & & & \\
\enddata
\tablenotetext{a}{Multiple events refer to the scenario where non-detections were interspersed with detections.}
\tablenotetext{b}{A boolean flag represents the non-detection of polarized emission. The reported linear and circular polarizations are not values from the same timestamp within an observation; they just represent the maximal value during the observation.}
\tablenotetext{c}{Lack of data indicates that the signal-to-noise ratio of the detection is too low for constraining the spectral properties.}
\tablenotetext{d}{Likely a scintillating candidate.}
\tablenotetext{e}{NA refers to the case where full polarization data were not available or recorded.}
\tablenotetext{f}{This source showed periodic bursts with a period of $P=77.3$\,min. It was detected on multiple occasions between October 2022 and April 2004 but has been quiet for the past couple of decades.}
\end{deluxetable*}

\begin{deluxetable*}{lrcccccccc}[!htb]\label{tab:cands}
\caption{List of unidentified variable point radio sources from our search and their multi-wavelength properties.}
\tablehead{\colhead{Source} & \colhead{Flux density} & \multicolumn{4}{c}{Optical\tablenotemark{a}} & \multicolumn{3}{c}{Near-Infrared\tablenotemark{a}} & \colhead{X-ray\tablenotemark{a}}\\
\hline
\colhead{} & \colhead{peak} & \colhead{$r$} & \colhead{$i$} & \colhead{$z$} & \colhead{$Y$} & \colhead{$J$} & \colhead{$H$} & \colhead{$Ks$} & \colhead{[0.2-10]\,keV}\\
\colhead{} & \colhead{(mJy)} & \colhead{AB} & \colhead{AB} & \colhead{AB} & \colhead{AB} & \colhead{AB} & \colhead{AB} & \colhead{AB} & \colhead{erg cm$^{-2}$ s$^{-1}$}}
\startdata
J074913$-$155457 & 2.14$\pm$0.22 & 22.5 & 22.4 & 21.9 & 21.0 & 22.8 & \nodata & 22.0 & \nodata\\
J075024$-$205945\tablenotemark{b} & 12.37$\pm$0.21 & 23$\pm$0.2 & 23 & 22.5 & 21.7 & \nodata & \nodata & \nodata & \nodata\\
J160646$-$513843 & 4.26$\pm$0.39 & 22.8 & 22.3 & 21.6 & 20.8 & 21.5 & 20.8 & 20.7 & 2.7$\times 10^{-13}$ \\
J163248$-$420307\tablenotemark{c} & 2.44$\pm$0.24 & 22.9 & 21.6 & 21.4 & 20.5 & 21.1 & 20.5 & 20.2 & 1.2$\times 10^{-12}$ \\
J172523$-$303720 & 4.25$\pm$0.21 & 22.5 & 21.7 & 20.9 & 19.9 & 20.4 & 20.1 & 19.9 & \nodata \\
J183418$-$092723 & 8.37$\pm$1.06 & 22.9 & 22.4 & 21.6 & 20.4 & 19.8 & 18.8 & 17.8 & 1.5$\times 10^{-12}$ \\
\enddata
\tablenotetext{a}{Upper limits.}
\tablenotetext{b}{Source position is saturated by a bright source 10\arcsec\, away.}
\tablenotetext{c}{DECaPS magnitudes for this source correspond to the aperture magnitudes of a faint source towards the edge of the error circle. The limiting magnitudes, however, will be close to these.}
\end{deluxetable*}


\subsection{\gcrta}
\gcrta\, was discovered as a faint transient with the peak flux density of the brightest VAST detection corresponding to a signal-to-noise ratio (SNR) of 9. Forced photometry on the non-detection images revealed multiple weak (3-$\sigma$) detections, showing a gradual rise and decay over several months. We examined the dynamic spectrum of the brightest radio detection and found no strong evidence for bursts or polarized emission. The source is also detected at other frequencies (843\,MHz and 943\,MHz) in ASKAP observations, but due to the narrow frequency range and faint nature of the detections, a clear distinction between thermal ($S_{\nu}\propto\nu^2$) and free-free ($S_{\nu}\propto\nu^0$) emission could not be made. No persistent source at optical/IR wavelengths. 

\begin{figure*}[!htb]
    \includegraphics[width=\linewidth]{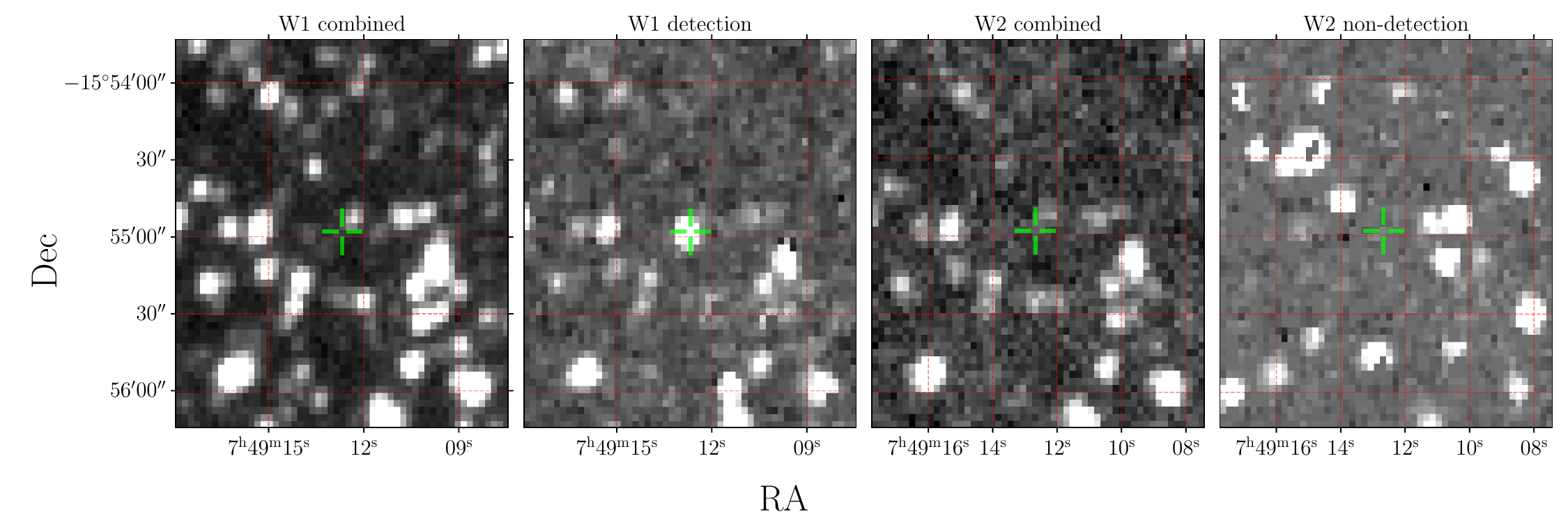}
    \caption{\gcrta: 2.5\arcmin\, cutouts of NEOWISE images. From left to right are the W1 co-add (except for the detection epoch), the detection W1 image, the W2 co-add (excluding the discovery epoch), and the W2 image from the W1 discovery epoch. \gcrta's radio position is marked with the lime crosshairs.}
    \label{fig:j0749_wise}
\end{figure*}

However, examination of \textit{NEOWISE} images revealed the presence of a single flaring episode on 2015-10-26T09:33:34 (see Figure~\ref{fig:j0749_wise}). It is very bright (SNR of 43), fast evolving ($<$3 hours, observations 94\,min before and after this did not detect it), and narrow-band (non-detection in W2). Given the source density in the corresponding \textit{AllWISE} image, the probability of chance coincidence between this IR source and the radio source is $<0.1$\%. 

\subsection{\gcrtb}\label{sec:gcrtb}
\gcrtb\ was discovered as a bright source that switched between periods of detections and non-detections. Using the brightest detection, we found that the emission is almost 100\% polarized, showing both linear and circular polarizations. The estimated rotation measure is $+124.1\pm0.4$\,rad/$\rm m^2$ with the polarization angle being relatively constant. The spectral index is $\alpha=-1.9\pm0.3$, indicating a steep-spectrum source. The light curve shows a decaying behavior towards the end of the observation, suggesting that emission is burst-like, similar to LPTs/\lptgcrt. The sporadic detections also support this with detections corresponding to wide ($>12$\,min) bursts serendipitously coinciding with VAST observations. While a majority of LPTs detected so far have small pulse widths ($\approx$minute; \citealp{radiosky_dawes2026}), there are sources \citep{hurley-walker_long-period_2023,askap1448} where the pulse widths are multiple minutes to even 30\,min, similar to \gcrtb. These temporal, spectral, and polarimetric properties are consistent across all the VAST detections; see Table~\ref{tab:j0750pol} for epoch-wise polarization properties. We caution that a number of these detections are close to the noise floor, and hence, the errors on the corresponding polarization fractions are large. 
\begin{deluxetable}{crcrr}\label{tab:j0750pol}
\caption{Epoch-wise polarisation properties of J075024$-$205945}
\tablehead{\colhead{Date} & \multicolumn{2}{c}{Flux density} & \multicolumn{2}{r}{Polarisation}\\
\hline
\colhead{} & \colhead{I} & \colhead{V\tablenotemark{a}} & \colhead{Circ.\tablenotemark{b}} & \colhead{Lin.\tablenotemark{c}}\\
\hline
\colhead{} & \colhead{(mJy)} & \colhead{(mJy)} & \colhead{(\%)} & \colhead{(\%)}}
\startdata
2023-06-22T04:49:58 & 1.81$\pm$0.18 & $-$0.73$\pm$0.13 & 40 & 33 \\
2023-06-25T04:57:41 & 1.05$\pm$0.18 & $<$0.48 & $<$46 & 45 \\
2023-10-16T20:48:33 & 4.07$\pm$0.22 & $-$1.38$\pm$0.15 & 34 & 43 \\
2023-10-30T20:55:41 & 12.37$\pm$0.21 & $<$0.45 & $<$3 & 100 \\
2023-12-08T18:11:45 & 4.97$\pm$0.20 & $-$0.59$\pm$0.18 & 12 & 90 \\
2024-01-11T16:23:29 & 2.94$\pm$0.21 & $-$1.53$\pm$0.19 & 52 & 71 \\
2024-01-30T16:26:43 & 1.52$\pm$0.20 & $<$0.60 & $<$40 & 73 \\
2024-02-17T15:29:24 & 1.65$\pm$0.24 & $<$0.60 & $<$36 & 81 \\
2024-03-08T13:02:12 & 1.53$\pm$0.23 & $<$0.60 & $<$40 & 21 \\
\enddata
\tablenotetext{a}{Upper limits on Stokes V flux density are quoted as 3-$\sigma$, corresponding to the absolute value.}
\tablenotetext{b}{An estimate of the absolute value is quoted.}
\tablenotetext{c}{Linear polarisation is estimated from the de-rotated light curve.}
\end{deluxetable}


\begin{figure}[!htb]
    \centering
    \plotone{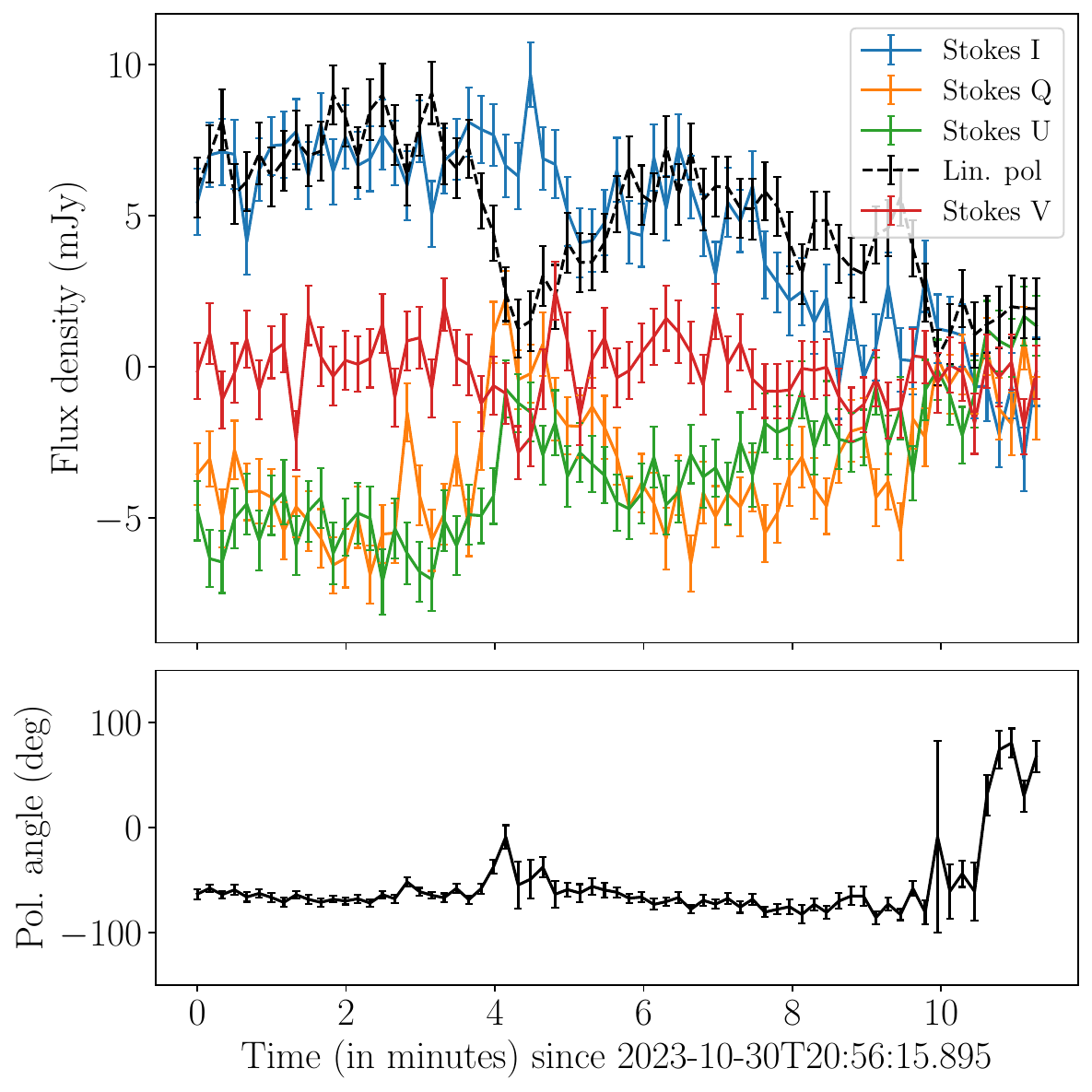}
    \caption{\gcrtb: Light curves of all four Stokes parameters from the brightest detection. The top panel shows the light curves for all polarizations, and the bottom panel shows the polarization position angle. The resolution is 10\,s.}
    \label{fig:j0750_lc}
\end{figure}

In order to search for multiple pulses, we performed two 6\,hr high-frequency observations  at 1.1--3.1\,GHz with the Australia Telescope Compact Array \citep{Wilson2011}. These observations were conducted once in June 2024 and once in January 2025 under the project code P3363.  We also performed one 6-hour low-frequency observation with the MeerKAT telescope \citep{meerkat}. This was performed in July 2025 under the project code SCI-20241101-TM-01 using the Ultra-high frequency (UHF) receiver with a bandwidth covering 588--1088\,MHz. No point source or pulsations were detected in any of the observations, indicating that the period is $>$6\,hrs or the source could have turned off, like \lptgcrt.
However, ASKAP observations separate from VAST, revealed multiple detections over a broader time period (see Figure~\ref{fig:lcs}(b)). Some of these epochs are close to our ATCA/MeerKAT observations, within a month, implying that either the source did not switch off or switched off only for short periods. 

We then tried to estimate the pulse period that could explain the detections, assuming that the pulses are periodic. From the barycentered times (midpoint of the observations), we found support for multiple periods between 15\,min and 7\,hours. However, none of them could simultaneously explain both the detections and non-detections, nor could they explain the lack of pulsations/bursts in our 6\,hour MeerKAT observations.

\begin{figure}
    \centering
    \plotone{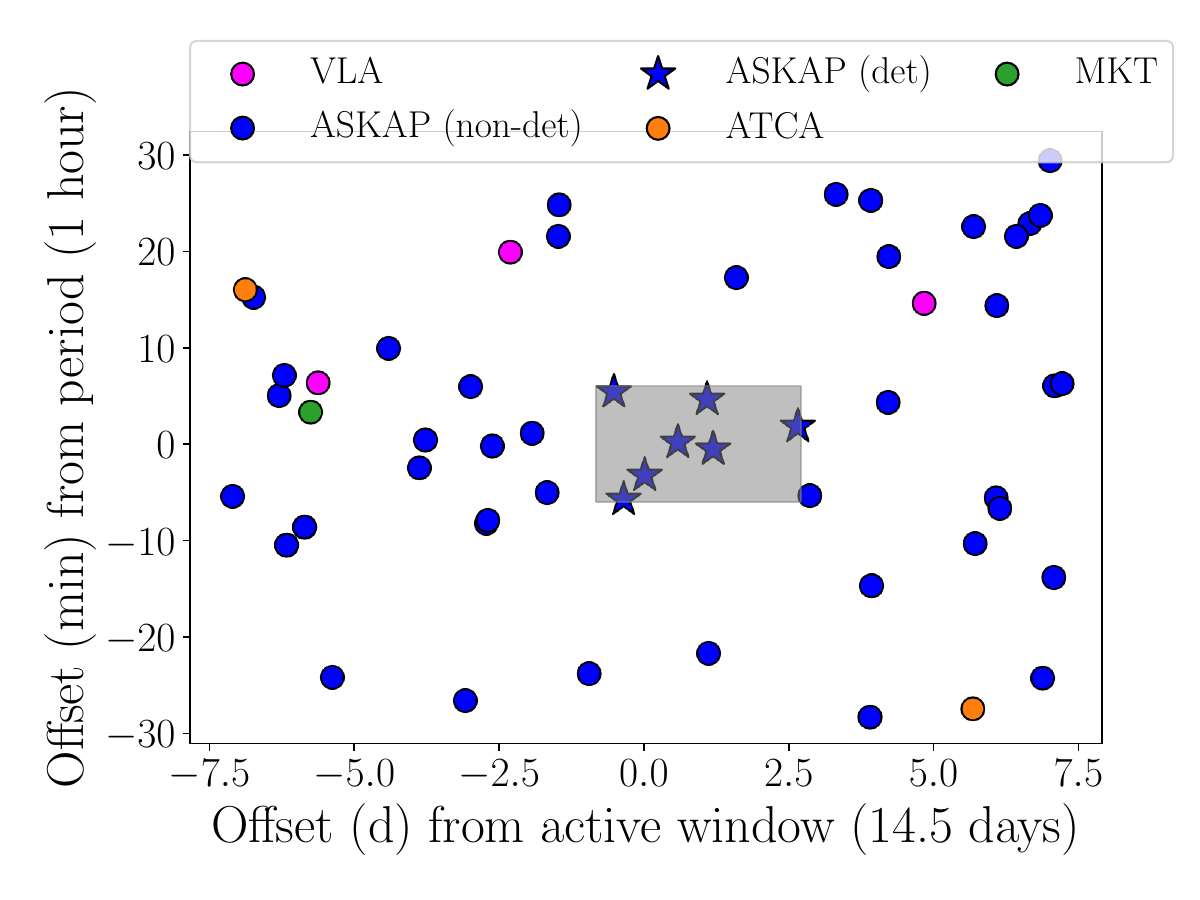}
    \caption{\gcrtb\ and its ``observability window'': For a pulse period of $\sim$60\,min and an ``active window'' of $\sim$3\,days every 14.5\,days, shown are the offsets of mid-point observational times of all observations. In this model, observations that happen to occur $\pm$10\,min from the pulse period are visible in a $\approx \pm$3\,day window every 14.5\,days. The gray shaded regions show this. The star markers represent detections while circles represent non-detections, and observations from different telescopes are marked by different colors.}
    \label{fig:j0750_geometry}
\end{figure}

We posit that this source might only be visible as bursting, during ``active windows'' which periodically recur (similar to \citealp{Horvath2025}). Under this, we find a pulse period of $P\approx1$\,hr and a secondary period of 14.5\,days. We find that the observations that occur during the wide bursts ($\pm$10\,min around pulse period) will result in a detection. However, only those observations that occur in a $\approx$ 3\,day window around the secondary period will result in a successful detection. Figure~\ref{fig:j0750_geometry} shows the offsets of our observational times with respect to this model, and the region on this phase space for when a detection is predicted. As can be seen, all the observations that fell within this region resulted in detections, and those that fell outside this window resulted in non-detections. Our ATCA and MeerKAT observations were performed prior to the identification of the potential ``activity window'' and so were not planned to sample the activity.

\begin{figure}
    \centering
    \plotone{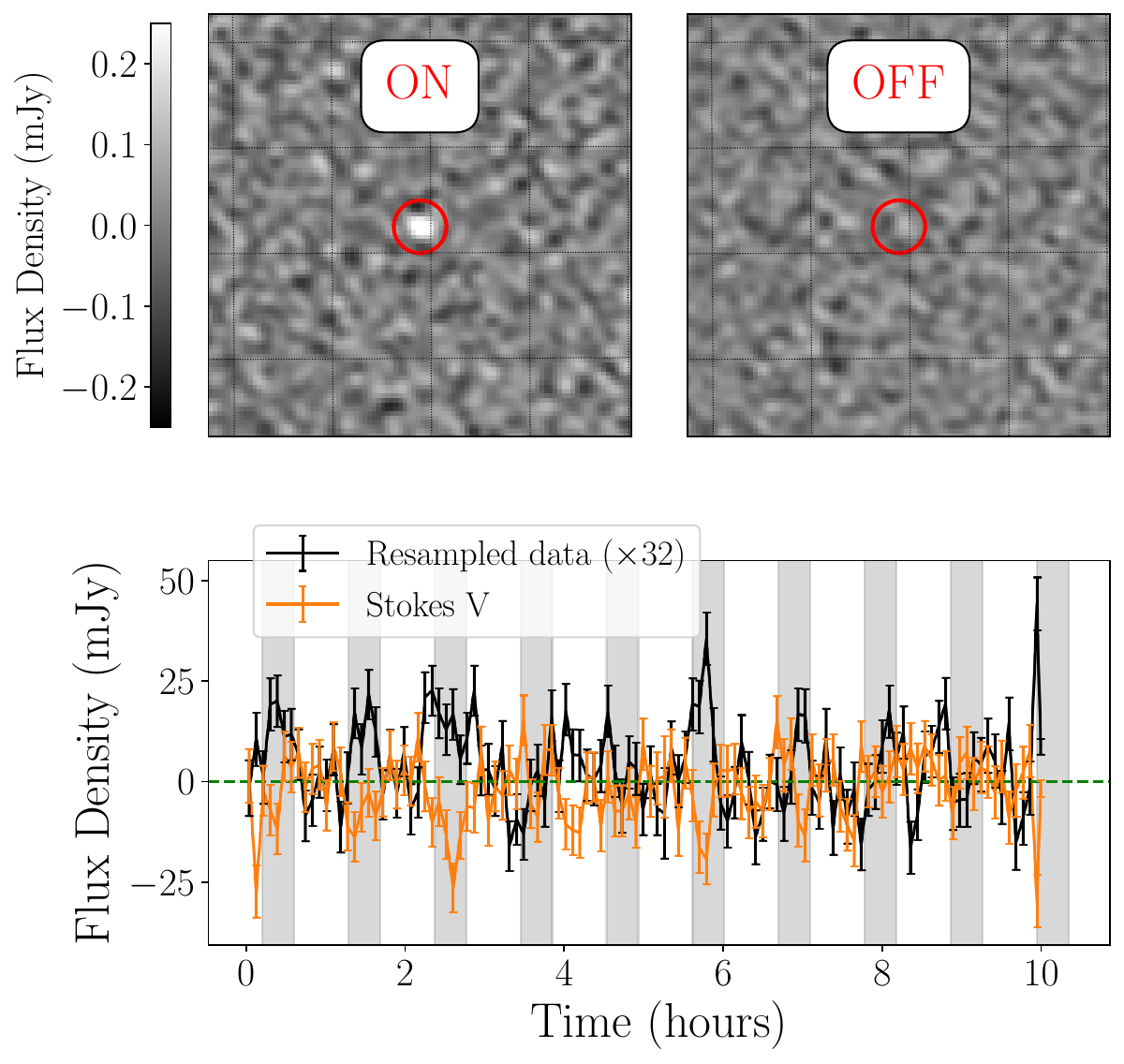}
    \caption{\gcrtb: Image cutouts and light curve from the 10\,hr ASKAP observation. The bottom panel shows the Stokes I and Stokes V light curves, resampled to 300\,s time bins instead of 10\,s native resolution. The gray bands ($\pm$12\,min wide) show the pulse period of 60\,min, which roughly follows the observed burst behavior. Shown in the top panels are the images corresponding to these marked times --- ON and OFF. A bright detection is seen in the image corresponding to the data during the time intervals when the source is bursting, while no flux density excess is seen during the quiescence. The color bar shows the flux density scale of the images.}
    \label{fig:j0750_emu_lc}
\end{figure}

One of the ASKAP detections was from a 10\,hour observation providing a long baseline when the source is active. The dynamic spectrum did not reveal bright short-duration ($<$ few minutes) bursts. Hence, given the low SNR ($\approx8$) in the time and frequency averaged image, we binned the data to finer time resolutions (5\,min) to look for variability. Figure~\ref{fig:j0750_emu_lc} shows the binned light curve indicating variability at the timescale of the pulse period $P_1\approx60$\,min. We imaged the combined data around these pulse periods to make the ``ON'' image showing a bright detection. Imaging the remaining data did not show any excess at the source's location, confirming that the variability is real. In Section~\ref{sec:discussion}, we speculate on the nature of the system that can give rise to such a behavior.

No optical/infrared counterpart was detected in the archival data. However, this source is very close to (and hence contaminated by) a very bright ($K_s\approx$9\,AB) star, 10\arcsec\, away (see Figure~\ref{fig:all_gcrt_cnads_oir}). 



\subsection{\gcrtc}\label{sec:gcrtc}
\begin{figure}[!ht]
    \centering
    \plotone{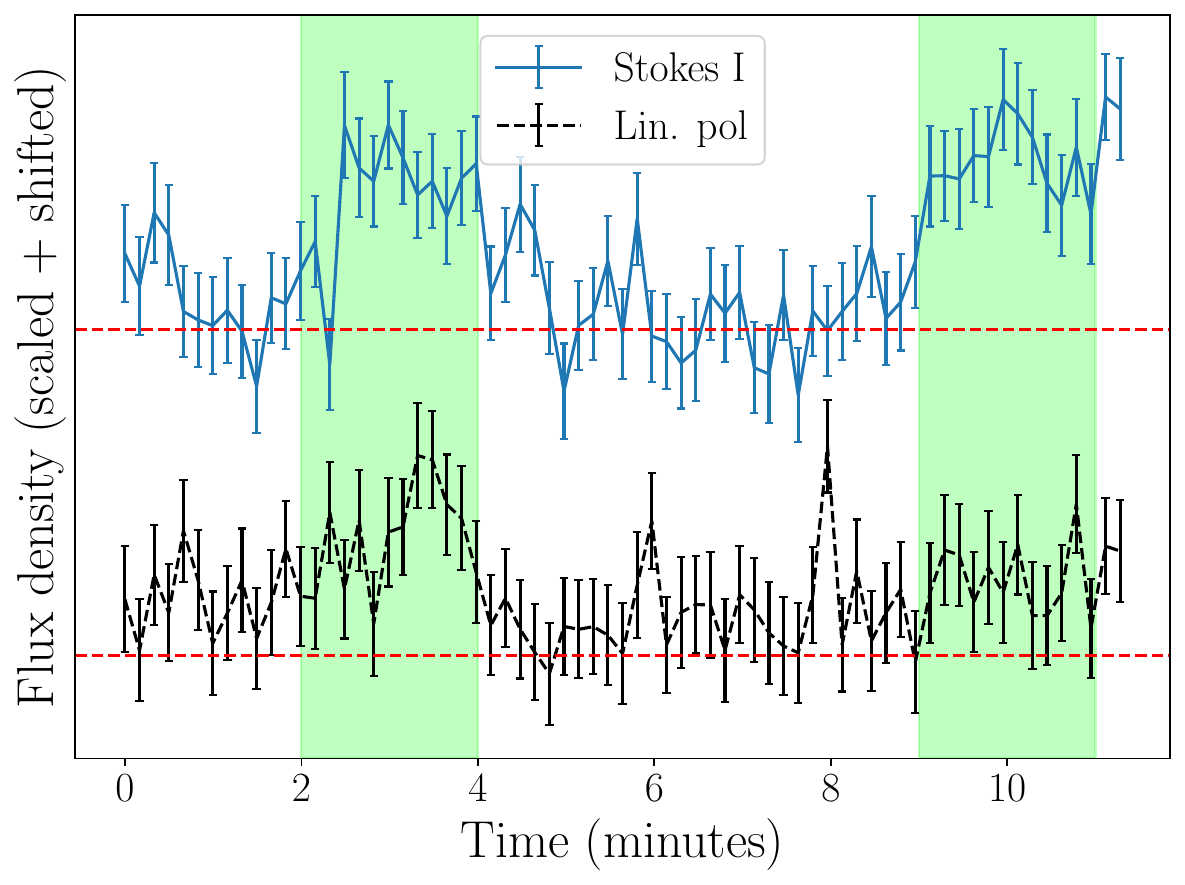}
    \caption{\gcrtc: Light curves of total and linearly polarized light for the brightest detection, showing two weak bursts marked by lime regions. The linear polarization light curve is manually shifted for visual purposes. The red dashed lines mark the zero intensity level.}
    \label{fig:j1606_lc}
\end{figure}

\gcrtc\, was discovered as a relatively faint transient (brightest VAST detection of 4.2\,mJy) and was detected during three different occasions. Inspection of the light curve for the brightest epoch revealed what could be consistent with two two-minute duration bursts, where the second one is probably cut off; see Figure~\ref{fig:j1606_lc}. Combining the emission from the two bursts, we derive an RM$\approx-157$\,rad/m$^2$. This assumes that the value of polarization angle or even the RM does not change during/within the bursts. Given the weak detections and limited SNR, we could not constrain these independently between the bursts. Using this estimate, we find that the first burst is $\approx$100\% linearly polarized and the second burst is $\approx$50\% polarized. We did not detect any significant circular polarization during the bursts. 

In addition to the discovery epoch, \gcrtc\ was detected on two other occasions (see Figure~\ref{fig:lcs}(c)) --- 7 months after the initial burst and a year after the second burst. It was found to be active only for short periods of time. While \gcrtc's behavior is qualitatively similar to that observed in \gcrtb, given the very small number of detections, analysis similar to \gcrtb\ was not possible to estimate any robust period estimates or ``active window'' durations. Combining all the non-detections, we do not find any underlying weak persistent source at 5-$\sigma$ limit of 330\,$\mu$Jy. 

\subsection{\gcrtd}

\gcrtd\, was found to be active for $\sim$6\,months between 2023 January and June in the VAST data (see Figure~\ref{fig:lcs}). 
However, longer (10\,hr) ASKAP observations of this field revealed multiple detections at a constant flux level of 0.8\,mJy, implying that there is an underlying persistent source, albeit at a different frequency. The near-simultaneous 943\,MHz ASKAP and 888\,MHz VAST observations around June 2023 suggest a steep spectral index $\lessapprox -3$. The dynamic spectra did not reveal any signature of bursts or any polarized emission.  

We examined the DECaPS images and VVV images and found a faint source both in DECaPS data and VVV data towards the edge of the error circle of \gcrtd. Figure~\ref{fig:all_gcrt_cnads_oir_2} shows the 30\arcsec\, optical and NIR composite \textit{z}-band and \textit{J}-band cutouts showing this source. Using aperture photometry, we find the following magnitudes for the source near the edge of the error circle: \textit{r}=22.9$\pm$0.2\,mag, \textit{i}=21.6$\pm$0.2\,mag, \textit{z}=21.4$\pm$0.2\,mag, \textit{Y}=20.5$\pm$0.2\,mag, \textit{J}=21.1$\pm$0.1\,mag, \textit{H}=20.5$\pm$0.1\,mag, {$K_s$}=20.2$\pm$0.1\,mag. We also note that given the source density in the field (from VVV images), the chance of random association of a background/foreground source is 96\%, so the object in \gcrtd's error circle can easily be a false association, and hence we do not consider this to be the counterpart to \gcrtd. 



\subsection{\gcrte}
\begin{figure}[!tb]
    \centering
    \plotone{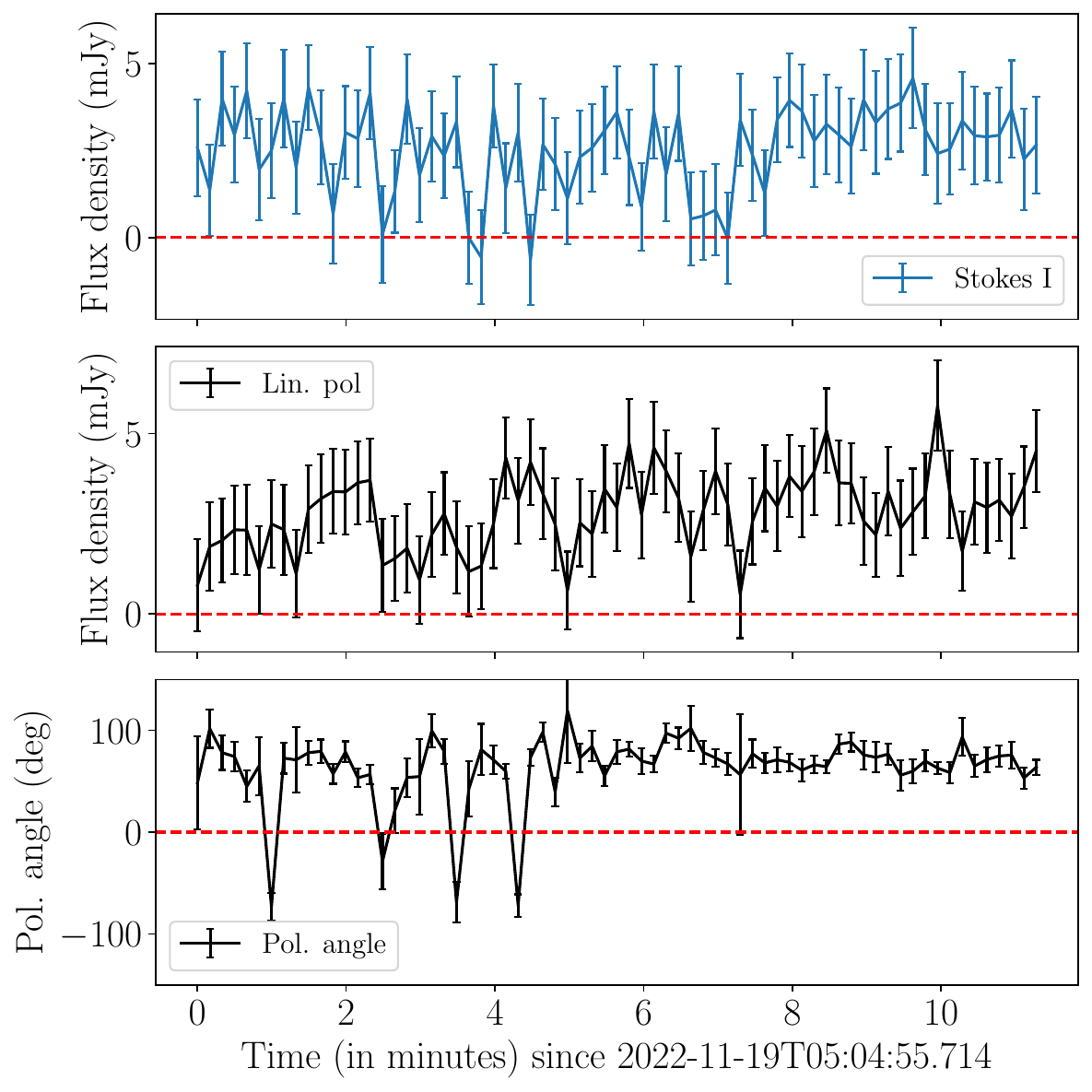}
    \caption{\gcrte: The top panel shows the Stokes I (total intensity) light curve, and the middle panel shows the linearly polarized component. The bottom panel shows the polarization position angle. The dashed red line marks the zero level.}
    \label{fig:j1725_lc}
\end{figure}

\gcrte\ was detected three times (around 4\,mJy) in VAST/RACS observations, interspersed with non-detections, similar to \gcrtb. The light curve did not show any burst-like features during the 12\,min observations (Figure~\ref{fig:j1725_lc}). The time-integrated frequency spectrum revealed strong linearly polarized emission with an RM of $-$438$\pm$1\,rad/m$^2$. The emission was almost 100\% linearly polarized during the entire observation. The polarization angle remained flat for most of the emission (similar to \gcrtb). We did not detect any circular polarization signature (both in V and $|\rm V|$). We examined the other two detections and found similar conclusions (similar RM, 100\% linear polarization), suggesting that the properties of the emission do not vary during the recurrence.

The source's position was serendipitously covered by MWA multiple times over a few weeks on a bi-weekly cadence, once in 2022 when the source was active, and once in 2024. The 2022 observations yielded weak detections (3-$\sigma$) of this source at 150\,MHz in multiple observations. The source appeared to be persistent between observations (see Figure~\ref{fig:lcs}) with a flux density of 40$\pm$10\,mJy. Using the near-simultaneous observations of MWA and ASKAP, we estimate the spectral index of the source to be $-2.5\pm 0.1$. No persistent radio counterpart was found in the stacked VAST data down to a 5-$\sigma$ limiting flux density of 175\,$\mu$Jy. The sporadic detections, the steep spectral and the polarized nature draw qualitative similarities with \gcrtb, but the lack of longer ($>6$\,hr) observation and the limited number of detections mean that this couldn't be established conclusively. 


There is a very faint source right on the edge of the error circle (Figure~\ref{fig:all_gcrt_cnads_oir_3}) with $J=20.3\pm0.2,\ H=19.9\pm0.2,\ K_s=19.73\pm0.2$\,AB mag, but given the source density, the probability of a chance coincidence is $\sim 80$\%, and hence that source is likely be a background source. 


\subsection{\gcrtf}
\gcrtf\, was discovered as a rising transient (peak flux density of 8\,mJy) in the VAST data. It showed a sudden rise in April 2024, which decayed over the next few weeks. Longer (and more sensitive) 8-hr ASKAP observations (observing blocks SB63296, SB64280, SB64328, SB64345, and SB67737) revealed that the source faded away within a few months by November 2024 (see Figure~\ref{fig:lcs}(f)). Extrapolating 888\,MHz VAST observations (using linear temporal evolution) to the near-simultaneous times of these 943\,MHz ASKAP observations (in July 2024) suggested a flat spectral index ($\alpha\approx0$, where $S_{\nu}\propto\nu^{\alpha}$). 
The dynamic spectra did not reveal any variability or polarization, both on 12\,min and 8\,hr timescales. The most constraining upper limit on the quiescent flux density, 450\,$\mu$Jy (5-$\sigma$), is roughly a factor of 20 times fainter than the brightest detection. This makes it unlikely that the variation we see is purely due to scintillation.  

\begin{figure}[!t]
    \centering
    \plotone{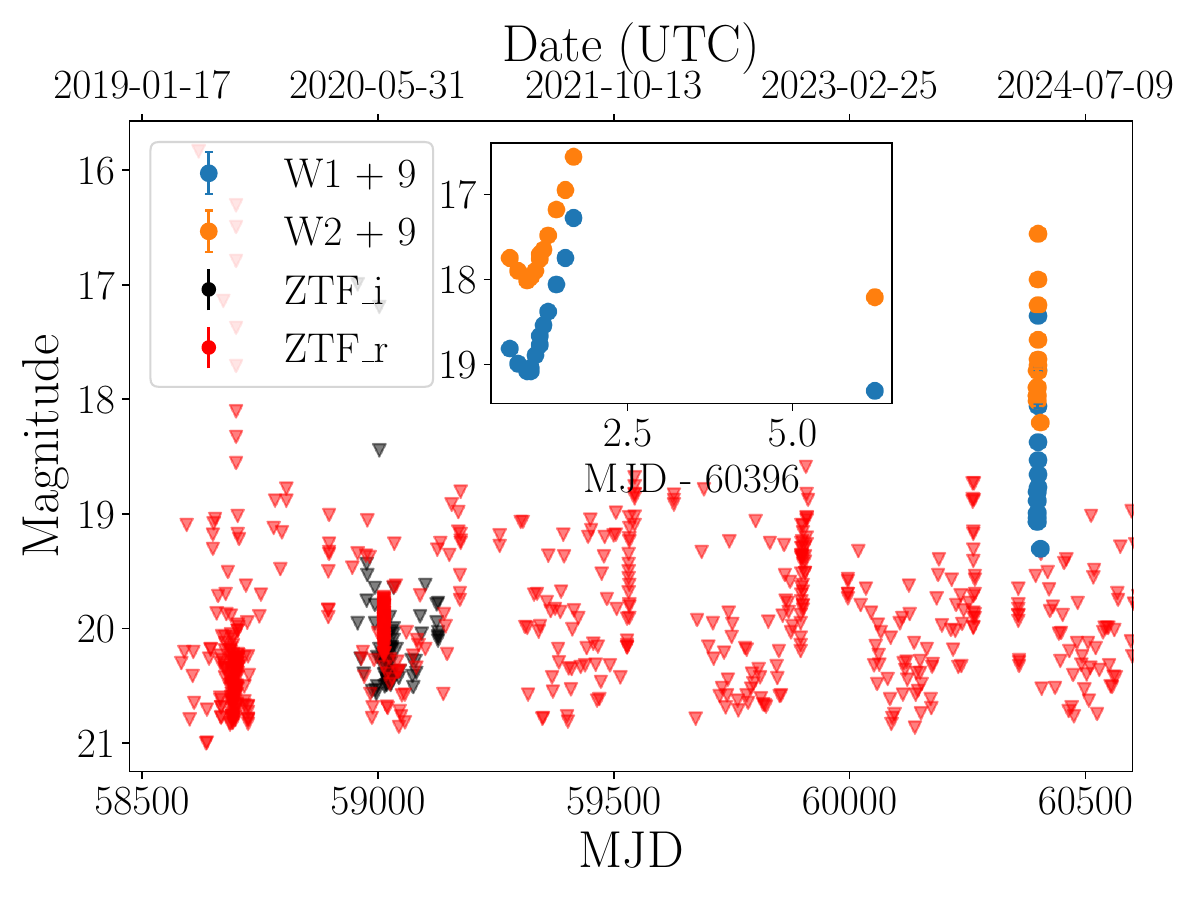}
    \caption{OIR light curve of \gcrtf: Figure shows the ZTF and NEOWISE light curve for \gcrtf\ over a multi-year baseline. The NEOWISE flare occurred a month before the first VAST detection (May 2025). The inset shows the MIR flare that evolved over a few days. Over the same period, ZTF doesn't show any simultaneous optical flare.}
    \label{fig:j1834_oir}
\end{figure}

Inspection of the NEOWISE data indicated a MIR flare a month before the radio outburst that rose and decayed over a few days (see Figure~\ref{fig:j1834_oir}). Although optical data from over a multi-year span did not reveal any flares, there is a 30-day gap between observations, during which the NEOWISE flare occurred. Hence, we can only constrain the evolution timescale at optical wavelengths to be $<$30 days. Despite the MIR flare seen in NEOWISE data, no persistent source in the AllWISE data. 

\section{Discussion}\label{sec:discussion}

\subsection{What is the nature of sources in this radio sample?}\label{sec:radio-sources}
We start by summarizing the observed properties of our sample of six sources. The sources presented here can be broadly classified into two classes. One of them shows highly polarized, burst-like (variable in 12\,min VAST integrations) emission and is sporadically detected. The other shows non-polarized emission that is persistent on shorter timescales, and variable on longer timescales (days--months). Based on this grouping, \gcrtb, \gcrtc, and \gcrte\ fall into the former category (referred to as ``fast flaring type'') while \gcrta, \gcrtd, and \gcrtf\ fall into the latter one (referred to as ``slow flaring type''). One interesting note is that sources from both these classes are observed to be steep-spectral, except for \gcrtf. This steep spectrum suggests that the radio emission is not thermal in nature. This could mean that in polarized sources, the emission might be coherent and coming from magnetospheric processes in strong magnetic fields (at least $\sim$kG) sources, and for those that are non-polarized, it might be synchrotron emission from shock-accelerated processes. A search for counterparts at other wavelengths did not yield any confident persistent counterparts (X-ray/optical/IR), but with the caveat that these observations are archival and not time-simultaneous with the radio activity. We discuss in detail in Appendix~\ref{app:nature} how we disfavor various Galactic sources like stars, magnetars, pulsars, and X-ray binaries, all but in one source (although unlikely, pulsar intermittency can not be currently ruled out in \gcrte).

Below, we discuss the similarities between our sample of sources and Galactic WD binaries.

\subsection{WD binaries}
The observed sample can belong to a population of Galactic WD binaries, some that are magnetic WD binaries (MWD), which can show periodic radio emission, and some that are cataclysmic variables (CVs), which can flare on $\sim$month timescales. 

\subsubsection{MWD binaries -- WD pulsars, LPTs}

\begin{figure*}[!htb]
    \centering
    \plottwo{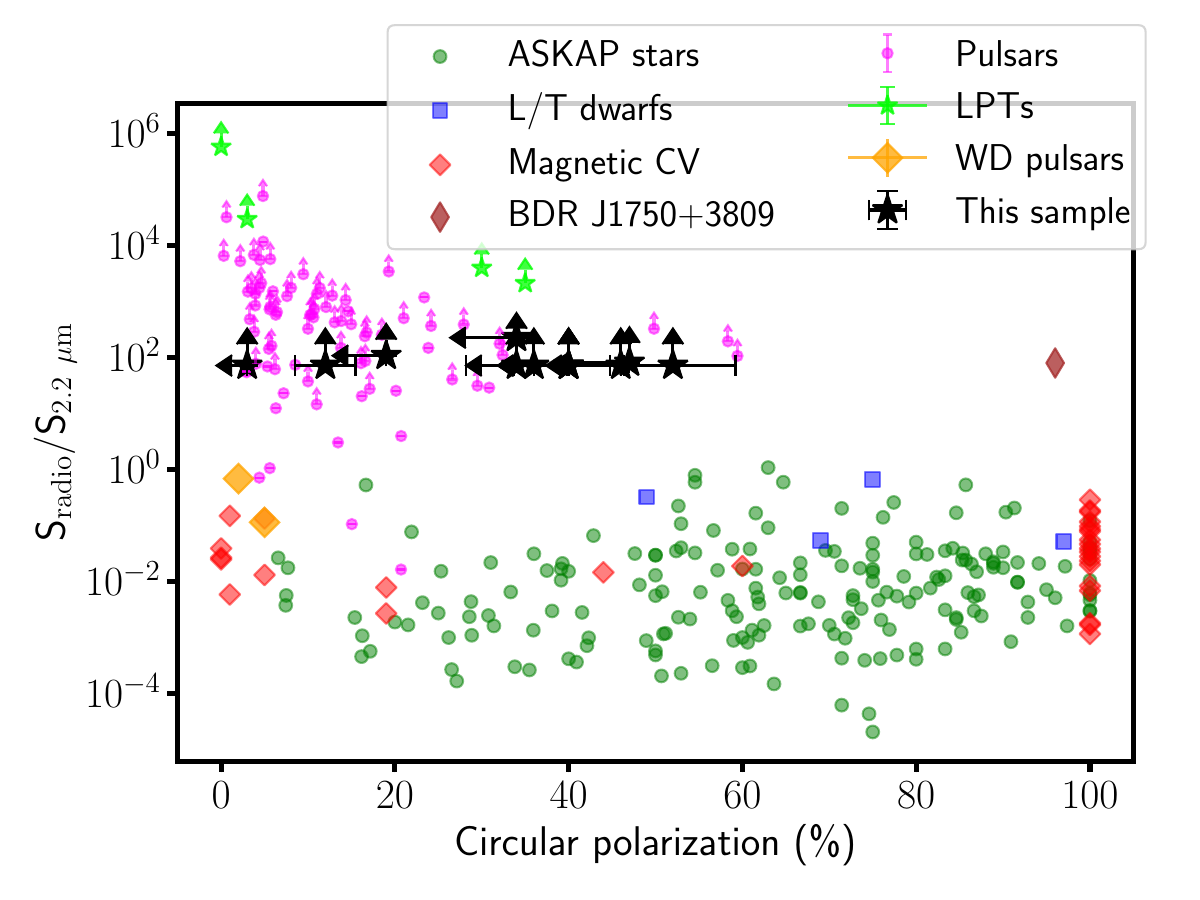}{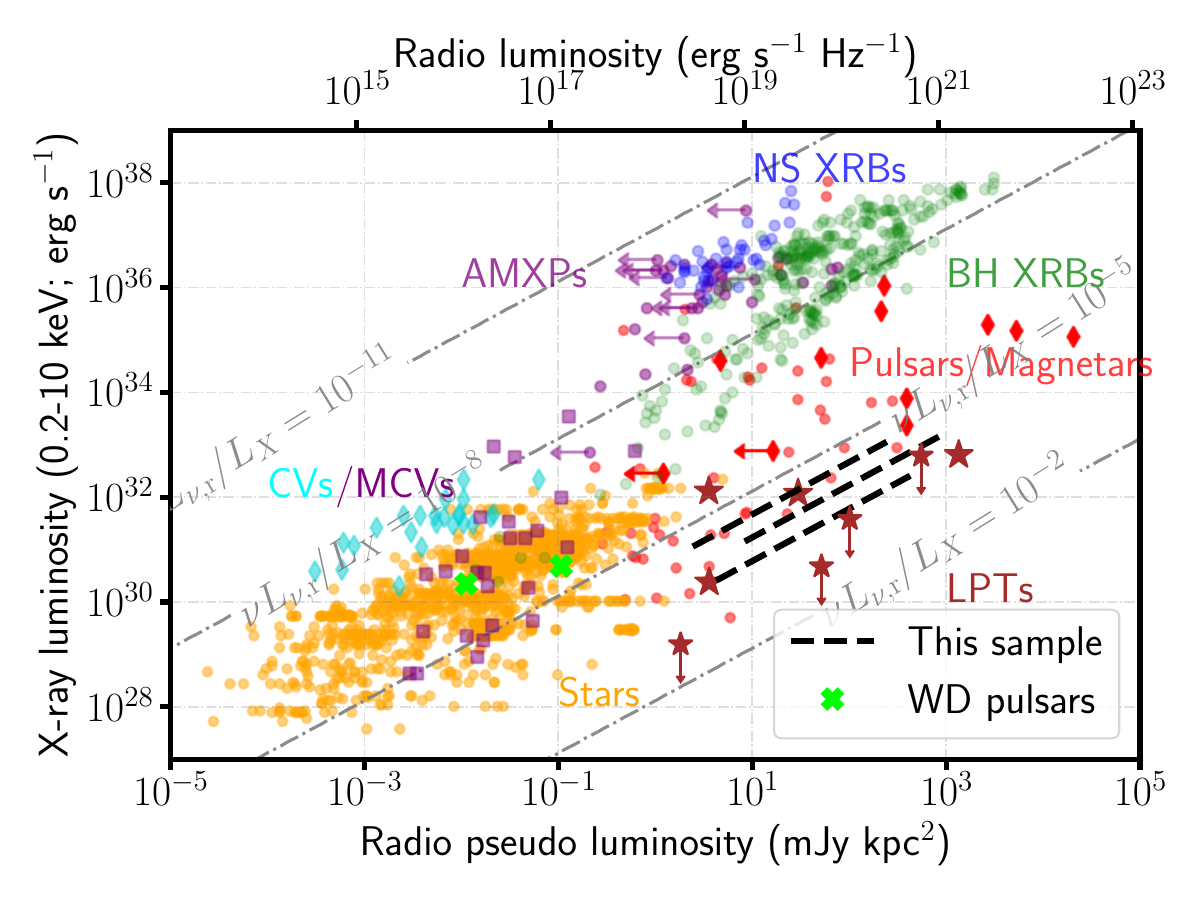}
    \caption{Population properties of our sample of VAST sources: \textbf{\textit{Left}}: Phase space of circular polarization vs the radio to \textit{Ks}-band flux ratio (relative strength of radio to IR emission) showing different source classes. The black stars show the IR upper limits obtained from the VVV data and the circular polarization from the VAST data. Multiple measurements from the same source are shown as different points on the plot. \textbf{\textit{Right}}: Phase space of X-ray vs radio luminosity showing different classes of sources. The current sample of sources is highlighted in black dashed lines. Since we do not have distance estimates for our sample, we show them as a function of distance as they traverse the black dashed lines or lie below them as the distance varies from 100\,pc to 10\,kpc. Adapted from \citep{askap1448}.}
    \label{fig:gcrt_pop}
\end{figure*}

Magnetic WD binaries, like polars and intermediate polars (IPs), are known to emit circularly polarized radio emission \citep{Barrett2017,Barrett2020}, which is hypothesized to be due to a coherent emission mechanism like electron cyclotron maser emission \citep[ECME;][]{Treumann2006,melrose2017}. However, all such known radio-emitting systems are very close by (within a few hundred pc) and have optical counterparts. On the contrary, the lack of a persistent counterpart, similar to those systems for our sample, places them at $>$kpc. This implies that the radio emission has to be orders of magnitude (at least 2--4) brighter than that observed in the current polar sample. This is also reflected in the relative radio to OIR/X-ray emission that is not consistent with the current sample of radio-emitting polars (see Figure~\ref{fig:gcrt_pop}), either indicating a larger energy reservoir for the radio emission or a different emission mechanism in play. 

Two other classes of sources have been recently discovered to exhibit highly polarized radio emission -- WD pulsars and LPTs. In the case of WD pulsars, the MWD spins asynchronously with the orbit ($\sim$1000 times faster) and the pulsed radio emission is thought to arise from the interaction between the pulsar-like WD beam and the companion. While the origin of LPTs, as a whole, is still uncertain and possibly diverse, the optically bright sub-sample of LPTs (which we refer to as WD-LPTs) at least are WD binaries in which the WD is hypothesized to be synchronized (or close to) with the orbit \citep{hurley-walker_29-hour_2024,de_ruiter_white_2024,bloot2025,Horvath2025}, with the origin of radio emission still debated. 

Our fast-flaring sample, especially \gcrtb, bears a lot of resemblance to WD pulsars/WD-LPTs, in terms of its periodic radio emission and highly polarized radio pulses. While a conclusive OIR counterpart is lacking in the case of \gcrtb, the observed radio emission can be understood in the paradigm of a MWD binary. The two observed periods, of $P_1\approx$1\,hr and of $P_2\approx$14.5\,days, can be understood as the spin period and the orbital period respectively, meaning that the WD spins asynchronously with the orbit (similar to WD pulsars but with longer spin and orbital periods)\footnote{Such $\sim$day timescale orbital periods are observed commonly in millisecond pulsar binaries and hence \gcrtb\ can be thought to have a similar evolutionary path but with a less massive primary star resulting in the formation of a WD instead of a NS.}. The active window of $\approx$3\,days once every orbital period, then reflects the range of orbital phases in which the interaction between the components occurs. 

More recently, using the long-term radio data for an optically faint LPT, GPM J1839-10, \cite{Horvath2025} demonstrated the existence of a secondary 8.5\,hrs period (to the 18\,min primary period), which they proposed to be the orbital period in that system. \cite{Horvath2025} proposed that the observed period and the radio pulse intermittency can be explained by a model in which GPM J1839$-$10 is a wide-orbit WD binary, with the binary interaction producing radio pulses around a certain orbital phase. Such a model can naturally explain the observed radio emission in \gcrtb, meaning that \gcrtb, could also be a wide orbit WD-binary. If true, this points to a population of radio-bright WD binaries that can manifest themselves in the current all-sky radio surveys as intermittent or transient LPTs. 

While the two other sources in our fast-flaring sample \gcrtc\ and \gcrte\ lack evidence for the presence of a secondary period, the multiple emission episodes hint at a behavior that is similar to \gcrtb. In the possible parameter space of secondary orbital period and the span of active window, it is possible that a combination different from \gcrtb\ can lead to the limited number of serendipitous VAST detections. For example, \gcrtb\ was discovered serendipitously in $<$15\% of our observations, and an uninformed targeted search on all three different occasions resulted in non-detections. Such a scenario is possible for these two sources. If this is the case, then our fast-flaring sample of sources could represent a population of wide-orbit MWD binaries, with the periodic WD+companion interaction powering the radio pulses.

\begin{figure}
    \centering
    \plotone{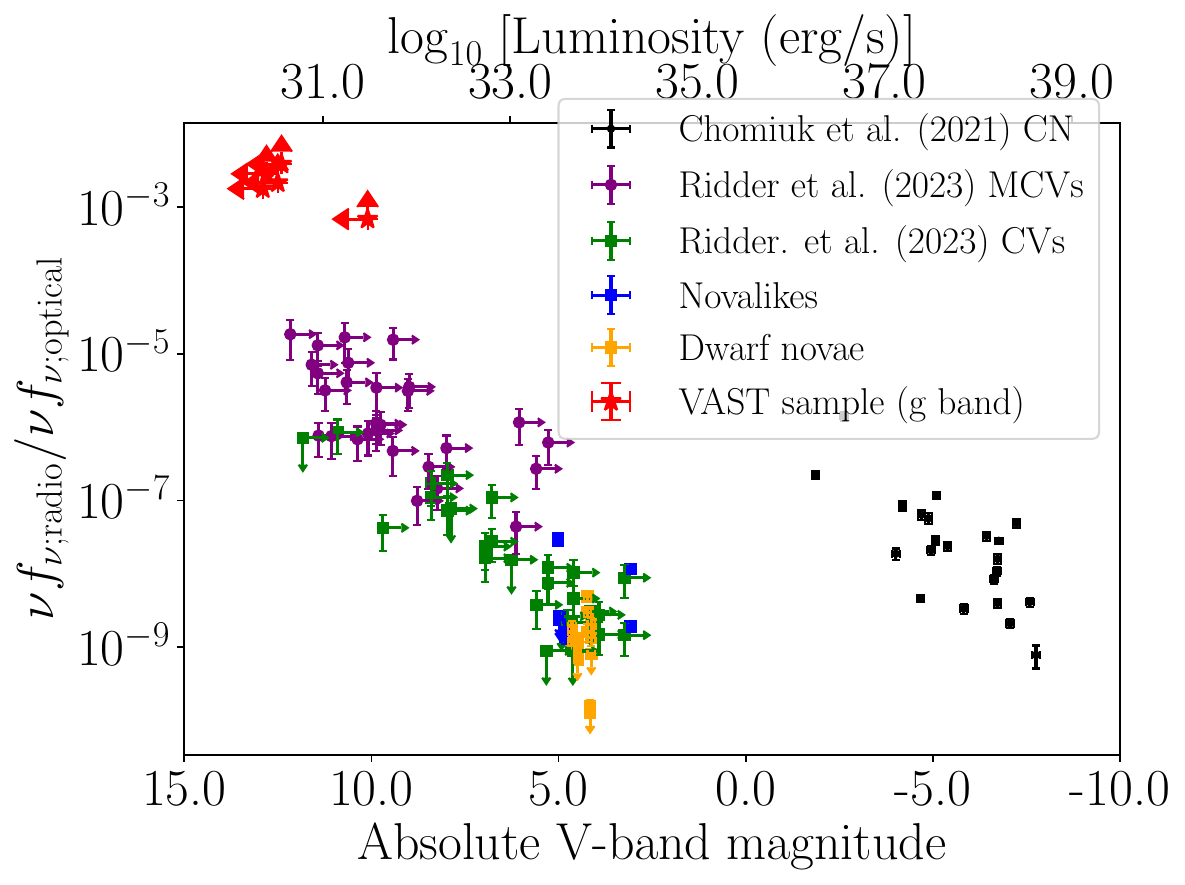}
    \caption{Population properties of radio bright CVs: For a sample of quiescent  MCVs/CVs \citep{Ridder2023}, novalikes \citep{Coppejans2015}, DNe \citep{Coppejans2016}, and classical novae \citep{chomiuk2021}, shown are the radio vs optical flux ratio as a function of peak V-band absolute magnitude. The red stars show our sample of VAST GRTs (assuming a distance of 1\,kpc). The optical luminosity corresponding to the V-band magnitude is shown at the top.}
    \label{fig:opt_radio_cvs}
\end{figure}



\subsubsection{WD binaries -- CVs}
While a population of wide-orbit MWD binaries can potentially explain the fast-flaring sub-sample of our sources, there are also sources in the sample that do not show short-time variable radio emission, but show $\sim$week-long flares. Such unpolarized or weakly polarized radio flares are attributed to CVs and XRBs. The lack of X-ray outbursts at the location of these sources makes it unlikely that these are radio counterparts to XRBs (for more details, see Appendix \ref{app:nature}). Figure~\ref{fig:gcrt_pop} shows the radio to IR/X-ray behavior in various Galactic sources, including XRBs, and it can be seen that the sample of sources presented here is orders of magnitude X-ray faint compared to known XRBs, unless an X-ray flare is missed. However, outburst activity in XRBs lasts for months, and hence it is unlikely that a bright flare is missed by the X-ray instruments.

Weakly or unpolarized radio emission from CVs can result from multiple scenarios --- dwarf novae \citep[DNe; ][]{Coppejans2016}, novalikes \citep{Coppejans2015}, and classical novae \citep[CN; ][]{chomiuk2021}. In the case of dwarf novae and novalikes (hosting a non-magnetic WD), the radio emission is short-lived (a few days), simultaneous with the optical outburst, but in most cases is very faint ($\lesssim$0.1\,mJy). There are individual cases where bright radio flares are observed at late times during/post the decay of optical flares \citep{Mooley2017,Fender2019}. Such flares are usually attributed to synchrotron emission from a transient jet \citep{Kording2008} or discrete plasma ejection \citep{Mooley2017}. In the case of classical novae, \cite{chomiuk2021} found that the timescale of radio evolution can vary from $\approx$100\,days to a few years. However, all of the above cases, although can qualitatively explain the radio behavior (DNe/novalikes/CN), are inconsistent with our sample when comparing the optical properties. Figure~\ref{fig:opt_radio_cvs} shows how the radio and optical properties of known CVs (both in quiescence and during outbursts) and our sample. It can be seen that the VAST sample is at least two orders of magnitude radio-bright compared to known CV populations.

Most quiescent CV/MCV systems are close by (within a few hundred pcs), and hence a distant ($\gtrsim$\,kpc) quiescent CV/MCV may explain the lack of a persistent counterpart. For sources at low Galactic latitudes, dust extinction can also contribute to the non-detection of the persistent emission in these systems. However, the lack of an optical transient flare in this case, from high cadence surveys like ZTF preceding the radio flares, is surprising. Such non-detection can result from a fast optical flare ($\sim$few days) being missed by current surveys; however, most DNe recur on a few day-month timescales, and hence if these are distant DNe, it is unlikely that every such flare is missed in the $>$5\, year-long ZTF data. More importantly, such transient flares can be missed in optical if these sources sources that are deep in the Galactic plane and hence heavily dust-obscured. 

Recently, NEOWISE data (and other contemporary IR surveys) have been used to uncover a sub-population of sources resembling classical novae that are optically faint but very bright at IR wavelengths \citep{Lucas2020,Zuckerman2023}. These have been interpreted as analogs of novae but occurring in highly dust-obscured environments (due to their location in the plane). But, the radio emission in all three slow-flaring sources in our sample evolves faster than most of the radio-detected novae ($\gtrsim$100\,d to years). In addition, the MIR flare seen in \gcrtf\ is extremely fast --- $t_{w,2}$, the time taken for the magnitude to drop by 2 units from peak in WISE bands is $<$5\,days, faster than the entire population discovered by \citep[$\sim$weeks to months;][]{Zuckerman2023}. This makes it an outlier even for IR-discovered CVs in terms of its evolution. However, similar dust-obscured instances in DNe might explain our sample of slow-flaring sources. The timescale of a bright MIR flare is consistent with the typical flare durations in typical DNe. Radio emission from DNe is also short-lived \citep{Kording2008,Coppejans2016}, a few days, and is usually attributed to synchrotron emission from transient jets/discrete outflows, which is consistent with the spectrally flat and unpolarized radio emission observed in \gcrtf. While these IR and radio properties are broadly consistent with DNe, the non-recurrence in NEOWISE data over 10\,yrs is also slightly perplexing. The observing cadence of NEOWISE is roughly 6 months, and hence it is possible that one of the DNe episodes coincided with NEOWISE observations while others were missed. As such, detailed modeling of both IR and radio data and the exact nature of \gcrtf, are deferred to a subsequent paper.

\gcrta\ shows a single IR flaring episode that is even faster than the one observed in \gcrtf, and more perplexing. The single band W1 detection faded by $>2$\,mags (i.e., factor of $>10$) in $<90$\,min. A stellar flare can be ruled out due to the lack of a persistent OIR counterpart. But, even in a CV scenario, such a short-duration and narrowband flare is unexpected. A coherent narrowband flare (e.g., from the cyclotron frequency or ``hump'' in magnetic WD systems) can potentially explain this, but the lack of multiple detections makes it difficult to establish this. In addition, the presence of quiescent emission, both in \gcrta, and \gcrtd, at lower flux density levels, is also perplexing, especially given the lack of one at other wavelengths. This implies that while the transient radio emission is qualitatively consistent with flaring from dust-observed DNe/WD binaries, the persistent radio emission (seen in \gcrta\ and \gcrtd) is atypical of them. Multi-wavelength (OIR/X-ray) discoveries, either during the radio flaring episodes or the underlying persistent source, will be needed in order to characterize these sources further.


Putting this sample in the broader context of radio transients, such a sample can potentially explain the so-called ``GRT'' (or GCRT) phenomenon --- short-lived radio transients that are absent at other wavelengths \citep{hyman_low-frequency_2002,hyman_powerful_2005,hyman_gcrt_2009,wang_discovery_2021}. Similarities can be drawn between our slow-flaring subsample of sources and GRTs, both in terms of radio behavior and multi-wavelength behavior. Given this, we speculate that such sources might be Galactic WD binaries undergoing outbursts/long flaring episodes. However, due to the location of these sources in the plane, optical/high-energy emission can suffer from severe extinction, while the IR emission can be bright. If this is true, then the ongoing (and upcoming) high cadence IR + radio surveys are poised to unveil a (sub-)population of such sources.



\subsection{Event rate}
Since the number of unidentified Galactic radio transients discovered so far is small, the rate of such events is not well constrained. This is where surveys like VAST, with its regular coverage, can help. But to estimate the true rate of such events, one needs to account for different biases at each step, leading to a final sample of candidates. Such biases can stem from two places --- the observational setup of VAST and the manual filtering steps imposed to recover these transients (see \S\ref{sec:sample}). Here, we estimate the unbiased rate of events similar to our slow-flaring sample. Similar estimates for fast-flaring sources require information about the underlying period and active duration distribution, which are currently unknown. Hence, we defer this to future studies. 


To estimate the observational and selection effects, we performed simulations using the actual observing setup to recover the detection rate. We injected 1000 sources per year into the Galactic plane following the spatial distribution of the CV population --- \cite{Rowell2011,canbay2023} --- thin disk model in radial, and exponential distribution in the vertical directions. 

\begin{equation}
\begin{split}
    p(r) \propto \exp{\left(-\frac{r}{r_{L}}\right)},\ \ p(\theta) &= \frac{1}{2\pi}\,,\nonumber\\
    p(z) \propto \exp{\left(-\frac{|z|}{z_L}\right)}\nonumber
\end{split}
\end{equation}
where $r$ is the Galactocentric distance to the source, $\theta$ is the Galactic polar angle, $z$ is the height from the plane, $r_L$=2.6\,kpc, $z_L$=250\,pc, and the overall normalization corresponding to the local rate of CVs \citep{Rowell2011}.

\begin{figure*}[!htb]
    \plottwo{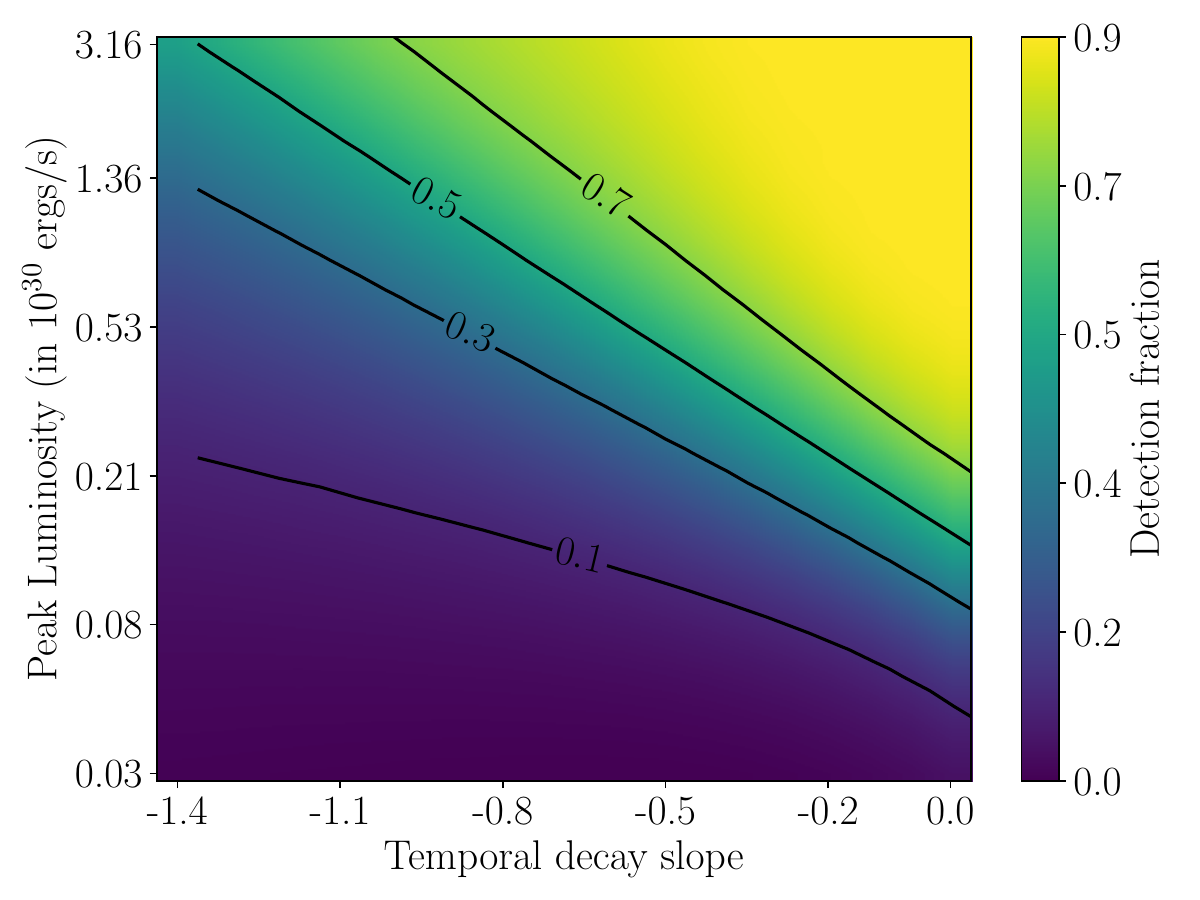}{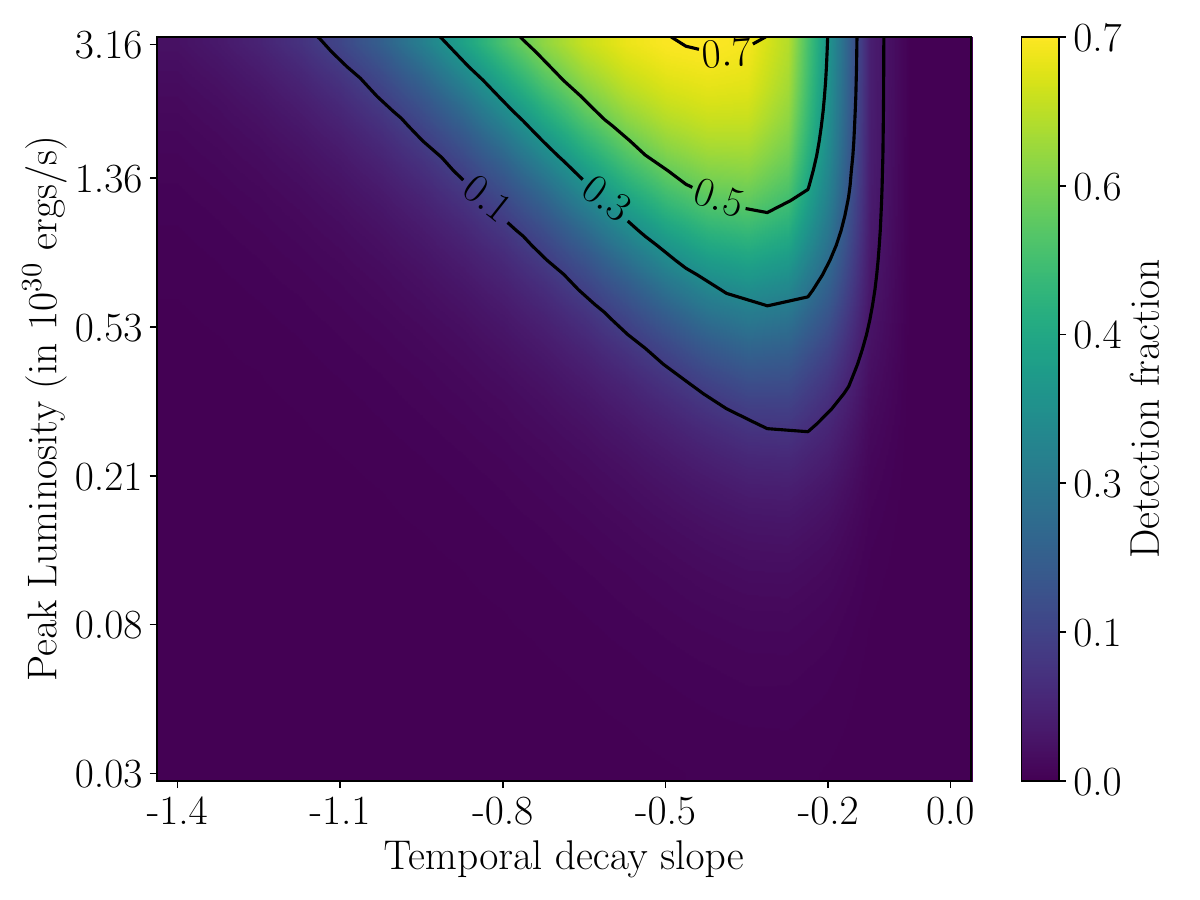}
    \caption{\textit{\textbf{Left:}} Detection fraction of GRTs over the phase space of peak luminosity vs the temporal decay slope accounting for the VAST discrete sampling. \textit{\textbf{Right:}} Detection fraction after imposing the various selection criteria (see \S\ref{sec:sample} for details).}
    \label{fig:det_frac}
\end{figure*}

We randomly select outburst times for these sources to be uniform within 1.5 years of VAST's observation time, to match the time frame of our sample. We assume that all these sources at the start of the outburst have the same peak luminosity and temporally evolve following a power-law decay 
\begin{equation}
    L(t) = L_0 \left(\frac{t-t_0}{1\, \rm day}\right)^{-\tau}\nonumber
\end{equation}
where $L_0$ is the peak luminosity at the onset of outburst $t_0$, $t$ is the time of observation and $\tau$ is the temporal power-law index\footnote{The normalization factor 1\,day is purely by choice and is covariant with $L_0$.}. We then impose the VAST observing setup (using the history of VAST that dictates when a particular sky location is observed) to sample the flux density of these sources and obtain the detection fraction as the fraction of total sources with flux density above the detection threshold (0.25\,mJy). We iterate this process for 1000 trials for a given luminosity and decay index to account for Poisson errors. 

Figure~\ref{fig:det_frac} (left panel) shows this detection fraction as a function of the peak luminosity and the temporal decay slope. Following the expectations, the detection fraction increases along the diagonal on which both the peak luminosity and the decay rate increase as sources are brighter and decay more slowly, and hence the probability of detection is higher.

We then apply the selection criteria that were used to filter candidates (\S\ref{sec:sample}). The only cuts that are relevant here are the demands of the light-curve variability, through $\eta$ and $V$ metrics, and a minimum of three detections. For each source in the simulation, we track the flux decay and measure these statistics. We then filter sources based on these, and the net detection fraction is shown in the right-side panel in Figure~\ref{fig:det_frac}. Two effects stand out here: i) Sources with shallower declines ($\tau \sim$0) are down-weighted since the light curve evolves much more slowly and hence variability is not strong, and ii) Sources with steeper declines $\tau\leq-1$ are also down-weighted since they decay faster and go below VAST's sensitivity within three epochs. The net effect is that sources with moderate decay slope $\tau\approx-0.5$ and higher peak luminosities are preferentially selected.

\begin{figure}[!htb]
    \centering
    \includegraphics[scale=0.4]{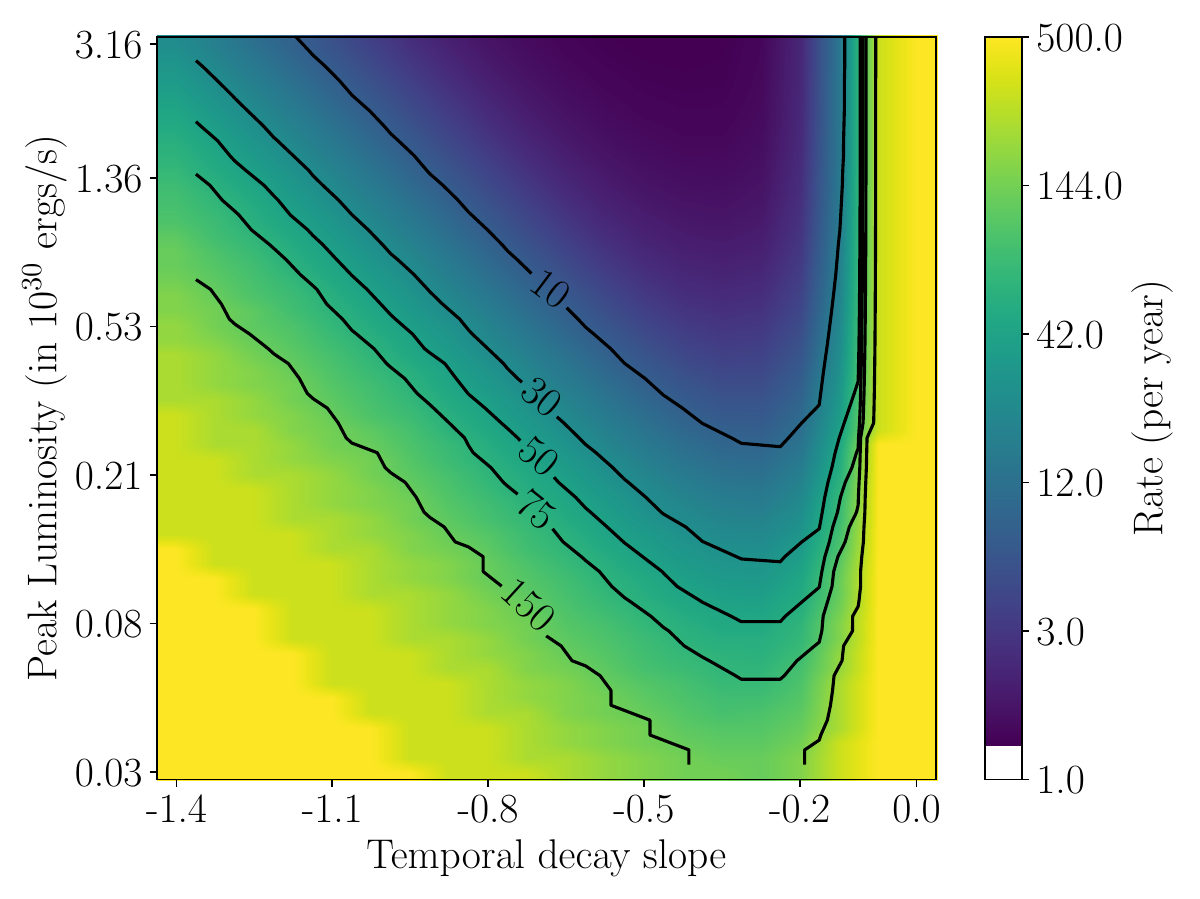}
    \caption{Estimated minimum rate (95\% confidence limits) of GRTs as a function of peak luminosity and the temporal decay spectrum using the detection probability derived in Figure~\ref{fig:det_frac}.}
    \label{fig:rate}
\end{figure}

However, the absence of distance estimates and hence the luminosities for our sample of sources makes it difficult to convert this net detection fraction to an overall rate or derive the luminosity function for GRTs. For example, we do not know if all the sources in our sample stem from the same peak luminosity or follow a certain distribution. Hence, we provide the 95\% confidence limit on the minimum rate of such events as a function of peak luminosity and the decay slope, which can produce one GRT candidate seen by VAST. If all three flaring type GRTs in our sample stem from the same peak luminosity, then this overall rate will be three times the minimum rate predicted above. Figure~\ref{fig:rate} shows this minimum rate as a function of peak luminosity and the temporal power-law index. The overall rate of GRTs can then be estimated with the knowledge of luminosity and decay index. For example, if the sources observed here are drawn from a population with a peak luminosity of 10$^{30}$ erg/s and a decay index of $\tau=-0.7$, the rate of these events can be constrained to a minimum of 4 per year that produces 1 VAST GRT (12 if all three VAST GRTs are alike)

Next generation of radio telescopes like the Deep Synoptic Array \citep[DSA;][]{DSA}, the Square Kilometer Array \citep[SKA;][]{SKA}, the Canadian Hydrogen Observatory and Radio Transient Detector \citep[CHORD;][]{CHORD}, are poised to push the sensitivity by at least 2 orders of magnitude beyond current VAST limits. Hence, the detection rate of GRTs could potentially increase by orders of magnitude as they discover fainter analogs. However, much depends on the survey strategy, as good cadence is required to discover these rapidly-evolving Galactic transients. For example, monthly cadence observations of the plane with DSA/CHORD can result in $\mathcal{O}(100)$ discovery of VAST GRT-like sources (brighter than 1\,mJy) per year and $\mathcal{O}(1000)$ fainter events (if such sources exist). For a lower cadence of one year, this will result in the discovery of $\mathcal{O}(10)$ VAST GRT-like sources and $\mathcal{O}(100)$ fainter sources per year, although most of them will be single epoch detections. While $\mathcal{O}(10)$ VAST GRT-like sources will be at least 100-$\sigma$ detections in DSA, and thus rule out effects like scintillation (for poorer cadences), allowing for robust filtering, such as a similar SNR requirement in $\mathcal{O}(100)$ fainter sources will result in $\mathcal{O}(10)$ discoveries per year.

\section{Conclusion}\label{sec:conclusion}
We used the Galactic plane survey data from the first 1.5\,yrs of the VAST survey and discovered a sample of six unusual radio transients. While we selected candidates purely based on temporal variability, our sample seemingly revealed what may be two different types of radio sources --- sources that are sporadic, steep-spectral, highly polarized, and show radio pulses, and those that are smoothly evolving, and unpolarized with slow-flaring behavior. We show that our fast-flaring subsample can be understood in the current LPT paradigm, but originating from wide orbit MWD binaries. If so, this could mean that transient LPTs might actually be wide-orbit binaries, with the intermittency coming from the orbital phases preferred for radio emission. If such a conclusion is established in a wider sample of LPTs, this could hint at a unified model for LPTs as WD binaries. Meanwhile, for our slow-flaring subsample of sources, we draw similarities between the outbursts/flares from CVs, particularly DNe-like, but those that are dust obscured (due to the position in the plane). Combining these two samples, we speculate whether the ongoing radio surveys are uncovering radio emission in sub-populations of WD binaries that are previously unexplored. A larger sample of such radio-bright sources from ongoing surveys, but those with simultaneous discoveries at other wavelengths, will help in conclusively answering this question. %


\begin{acknowledgments}
We thank the anonymous referee for providing useful and constructive feedback. AA and DLK were supported by NSF grant AST-1816492. DLK is further supported by NSF grant AST-2511757.  Parts of this research were conducted by the Australian Research Council Centre of Excellence for Gravitational Wave Discovery (OzGrav), project number CE230100016. 
This scientific work uses data obtained from Inyarrimanha Ilgari Bundara / the Murchison Radio-astronomy Observatory. The Australian SKA Pathfinder is part of the Australia Telescope National Facility (\url{https://ror.org/05qajvd42}) which is managed by CSIRO. Operation of ASKAP is funded by the Australian Government with support from the National Collaborative Research Infrastructure Strategy. ASKAP uses the resources of the Pawsey Supercomputing Centre. The establishment of ASKAP, the Murchison Radio-astronomy Observatory, and the Pawsey Supercomputing Centre are initiatives of the Australian Government, with support from the Government of Western Australia and the Science and Industry Endowment Fund.
\end{acknowledgments}

\vspace{5mm}

          

\appendix 
\restartappendixnumbering
\section{Stokes I images}\label{app:images}
Shown below are the Stokes I images of the brightest detections and the stacked image corresponding to all the non-detections for the candidates in this sample.

\begin{figure*}
    \gridline{\fig{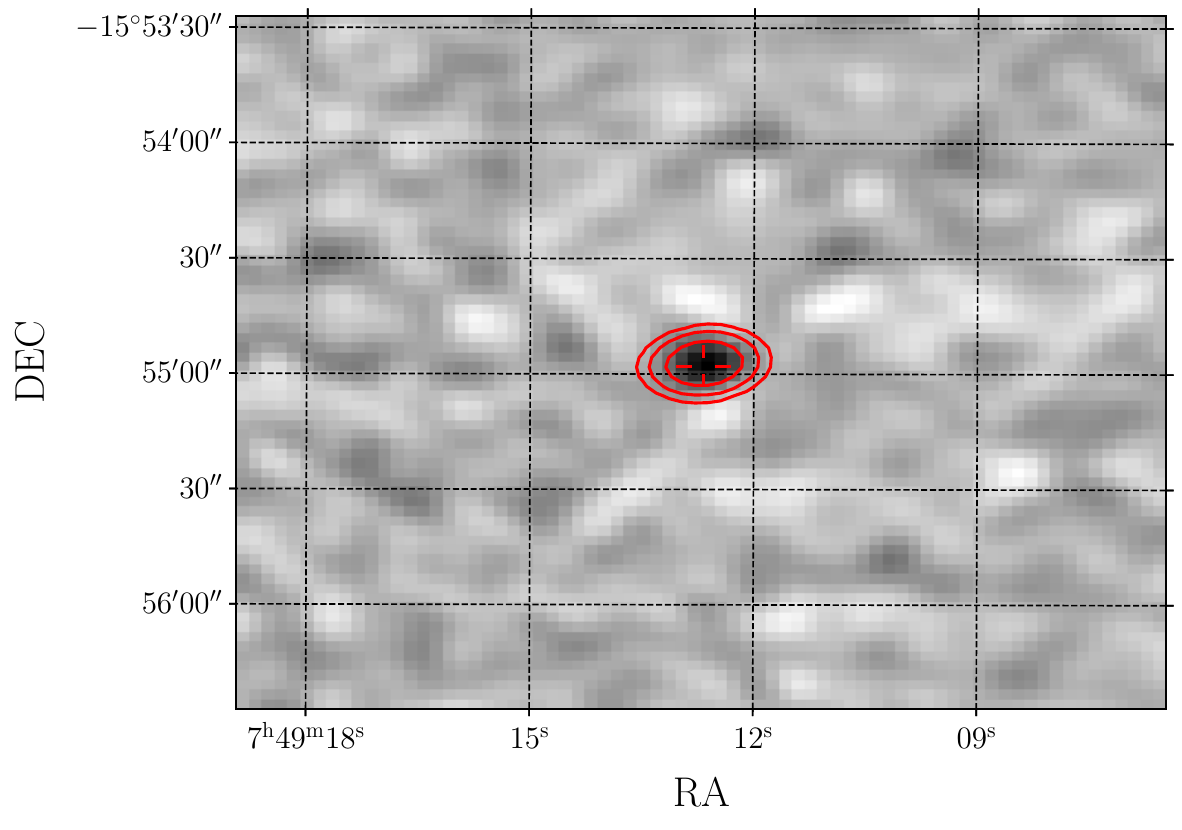}{0.5\textwidth}{(a) J074913$-$155457 brightest detection}
          \fig{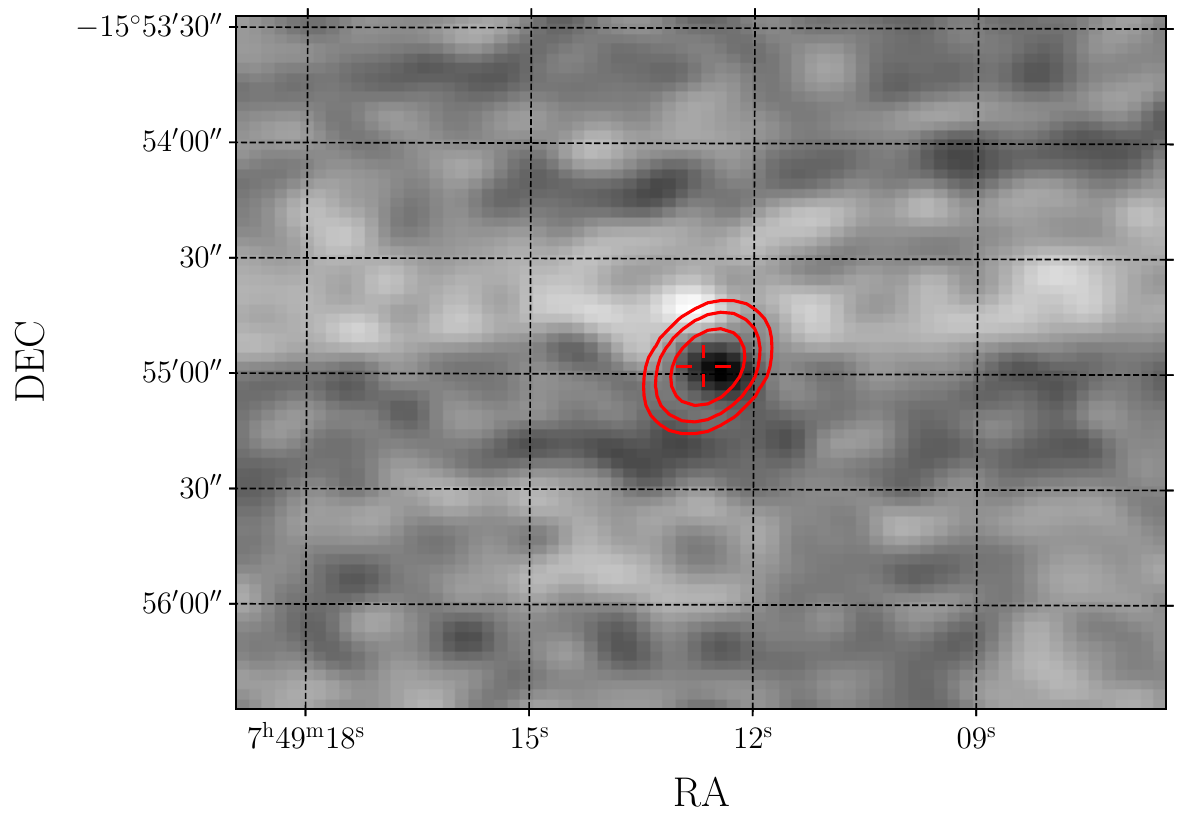}{0.5\textwidth}{(b) J074913$-$155457 stacked image}}
    \gridline{\fig{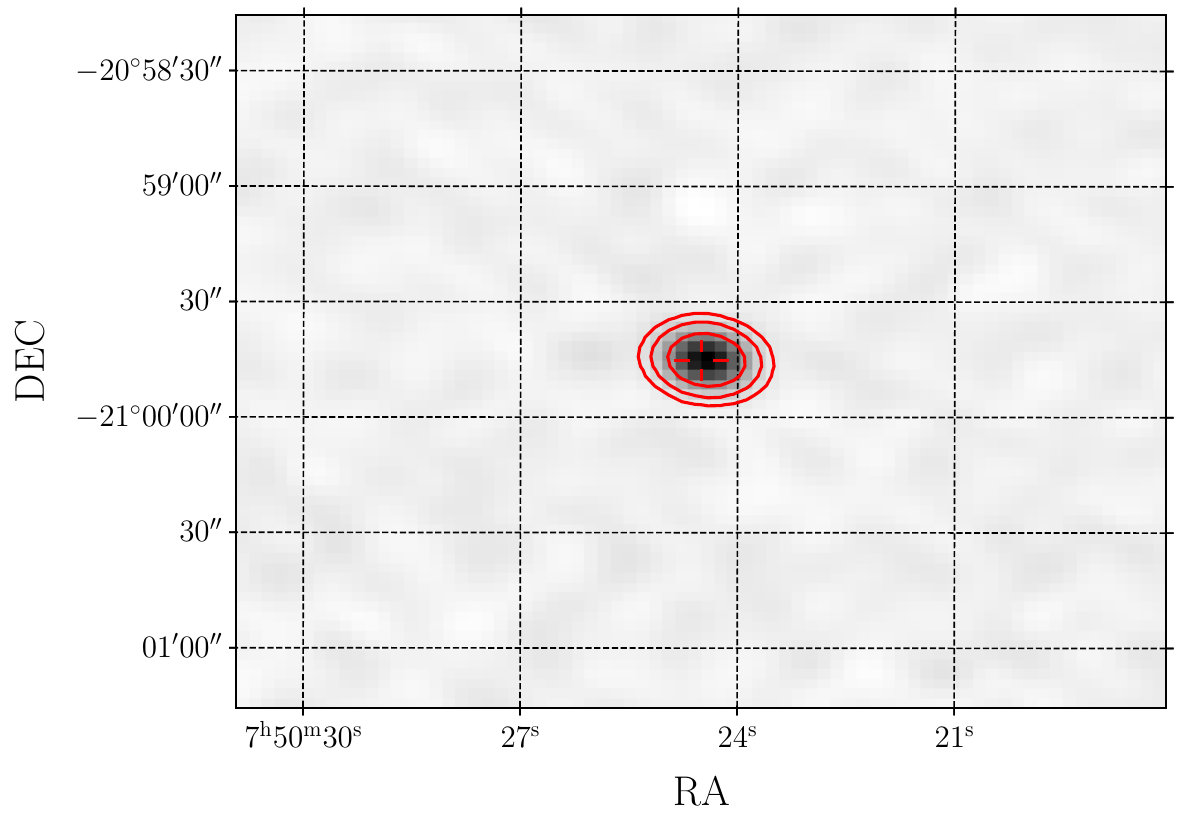}{0.5\textwidth}{(c) J075024$-$205945 brightest detection}
          \fig{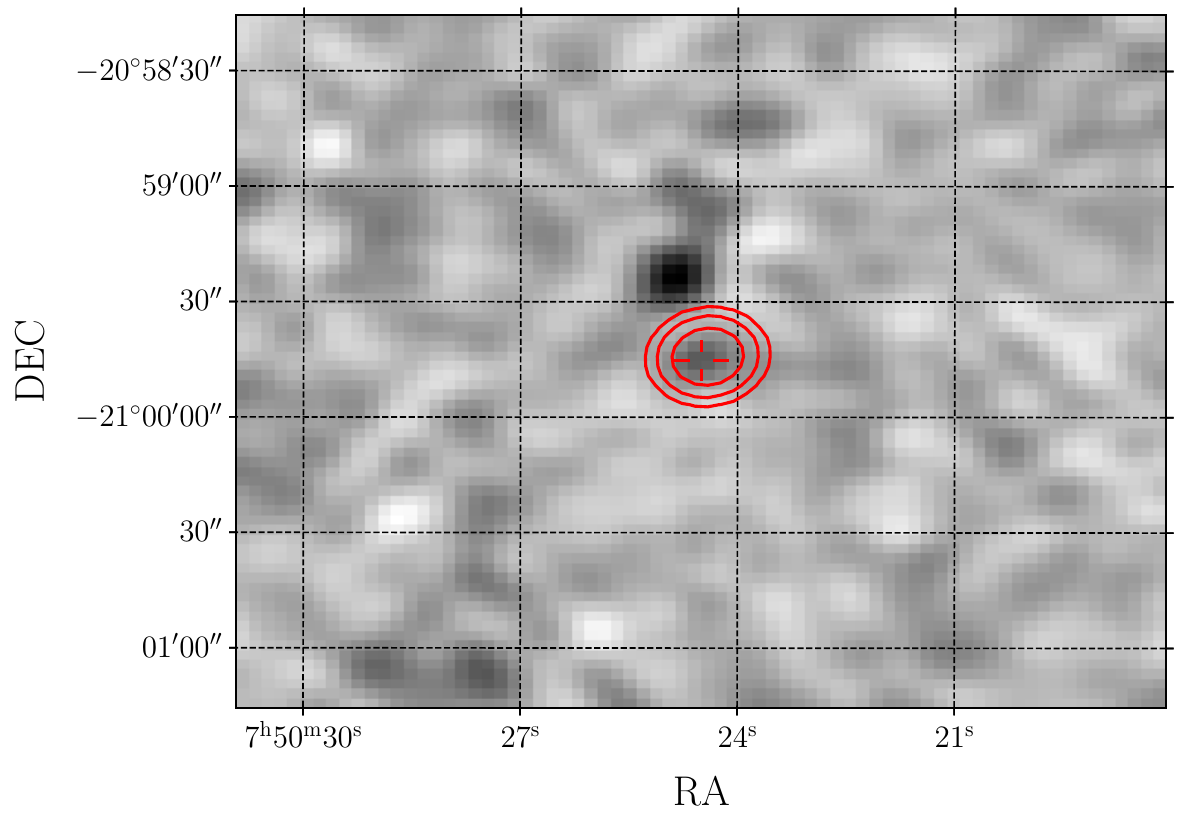}{0.5\textwidth}{(d) J075024$-$205945 stacked image}}
    \gridline{\fig{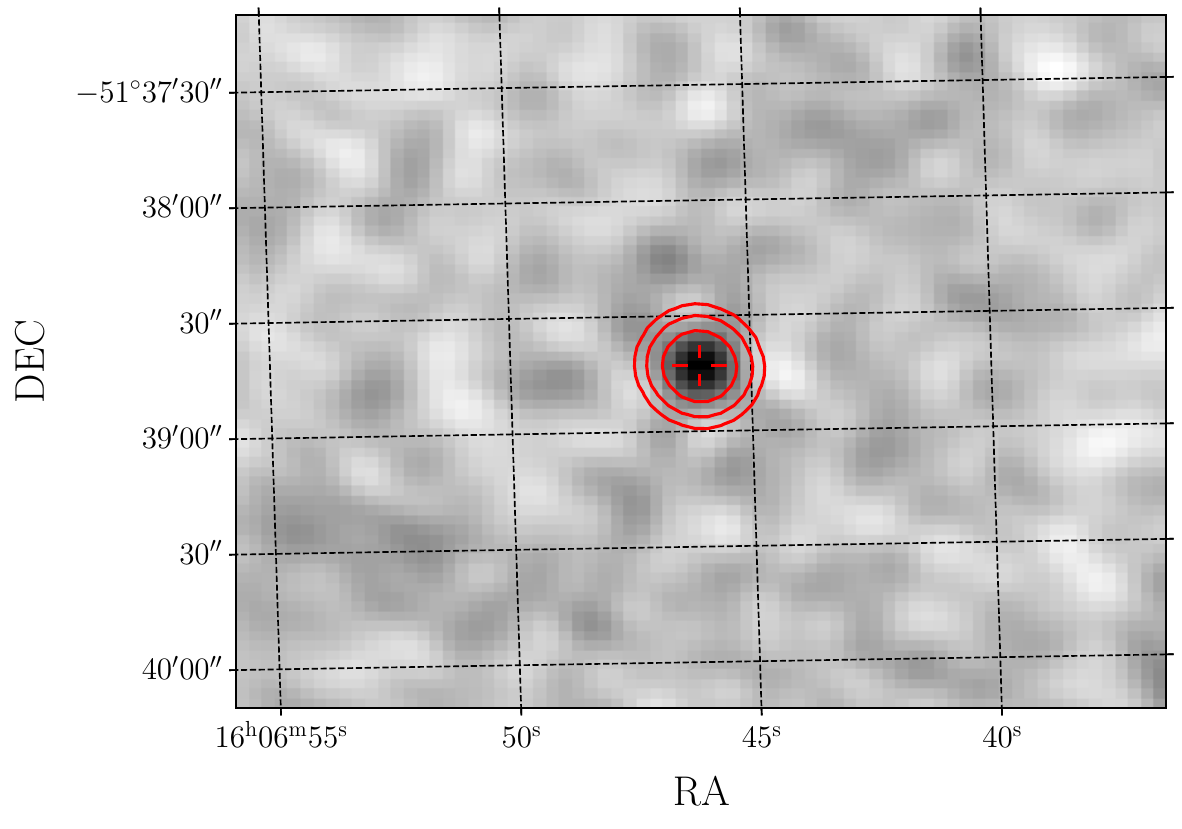}{0.5\textwidth}{(e) J160646$-$513843 brightest detection}
          \fig{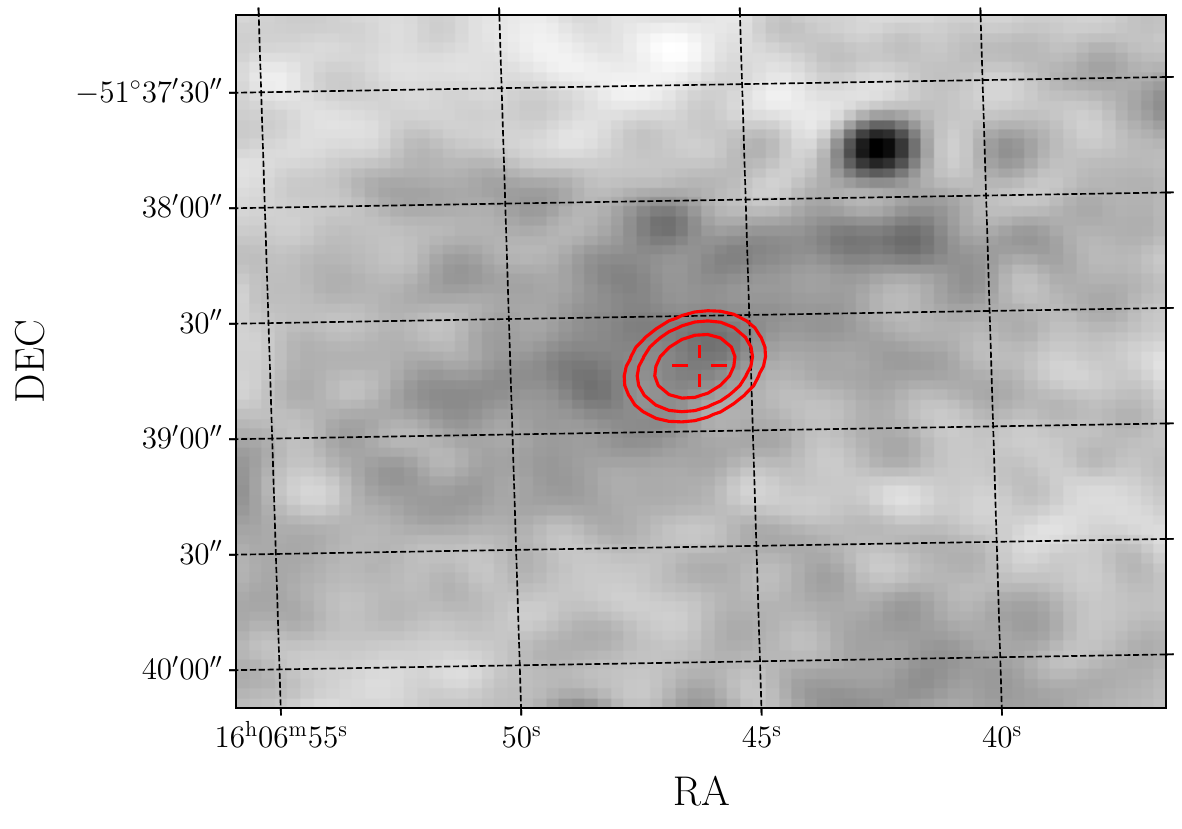}{0.5\textwidth}{(f) J160646$-$513843 stacked image}}
    \caption{Brightest detection image and the stacked image from all the non-detections}
    \label{fig:all_pulsar_res}
\end{figure*}
\begin{figure*}
    \gridline{\fig{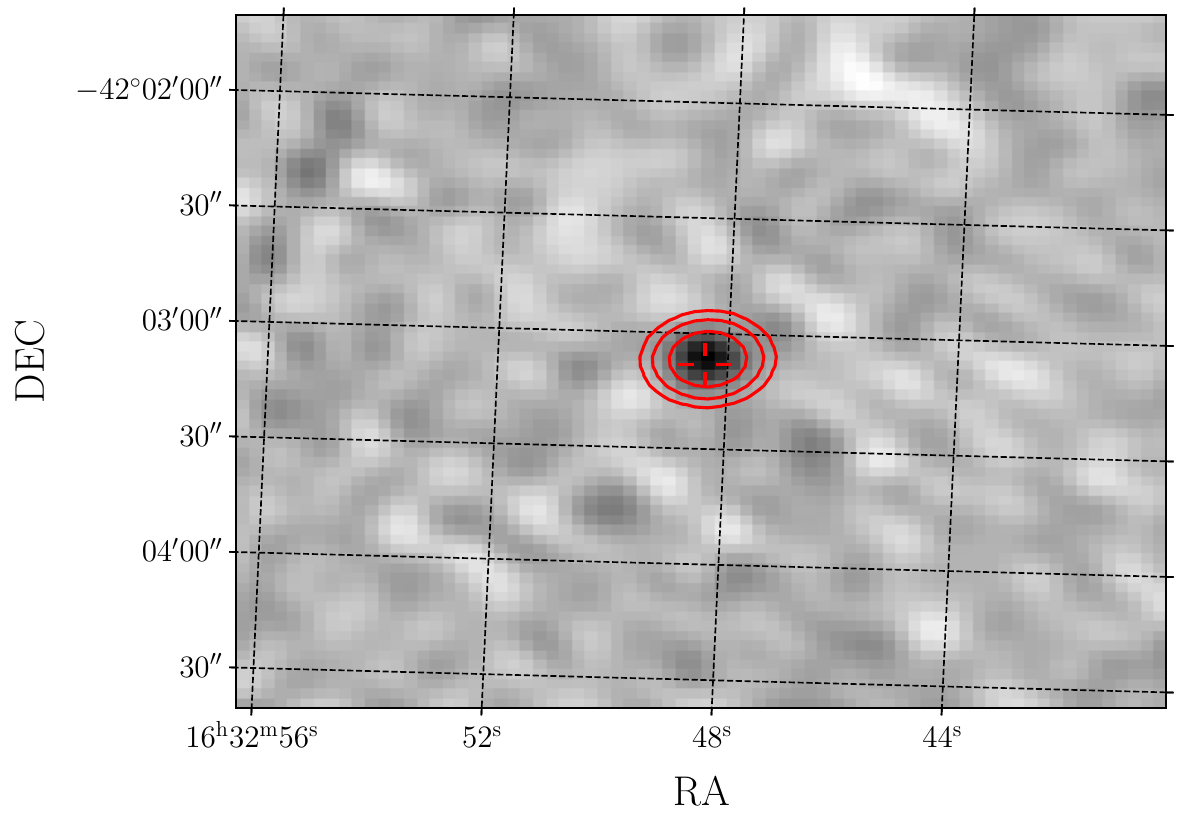}{0.5\textwidth}{(a) J163248$-$420307 brightest detection}
          \fig{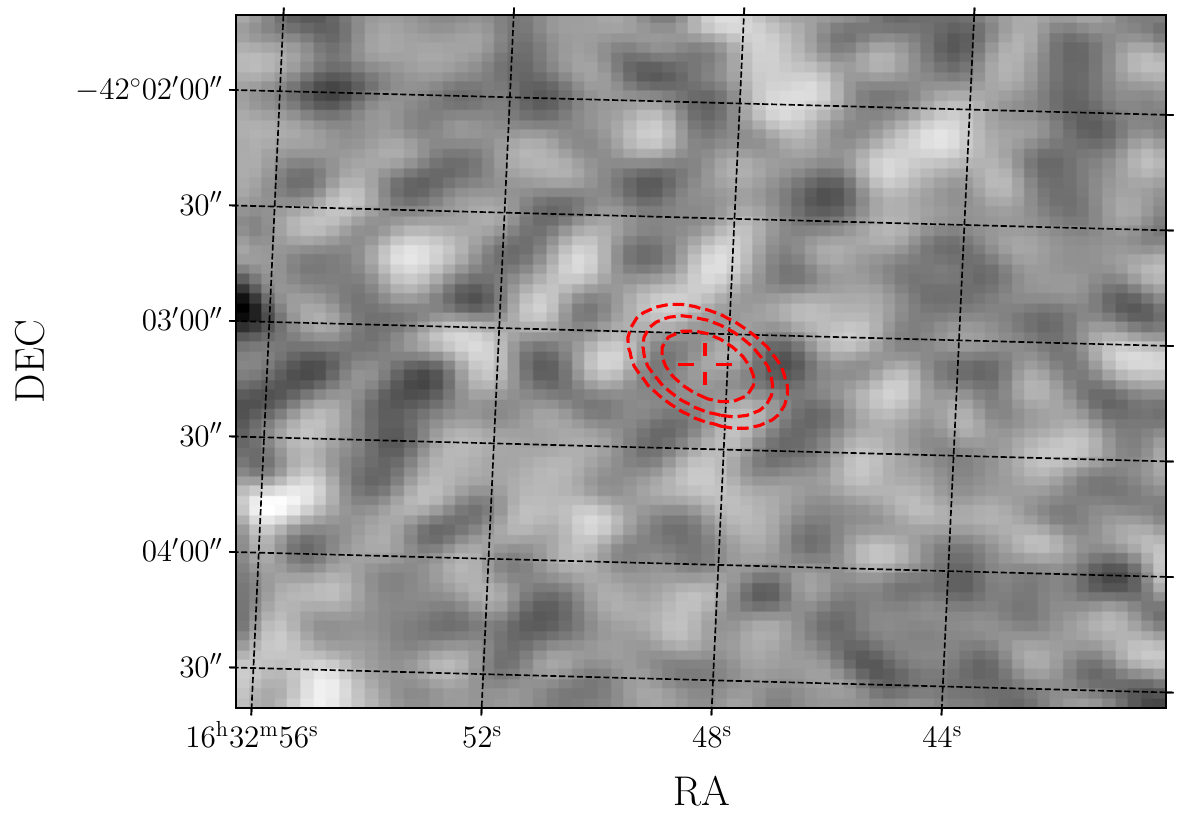}{0.5\textwidth}{(b) J163248$-$420307 stacked image}}
    \gridline{\fig{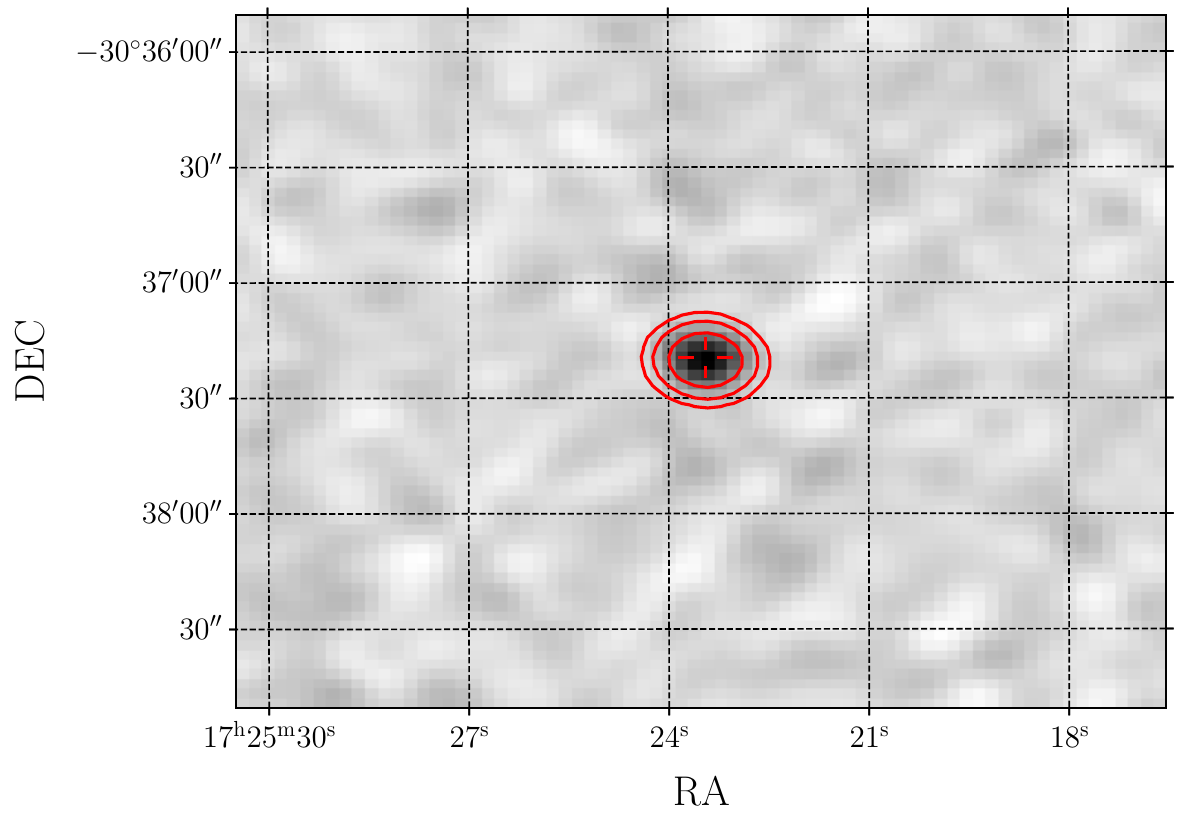}{0.5\textwidth}{(c) J172523$-$303720 brightest detection}
          \fig{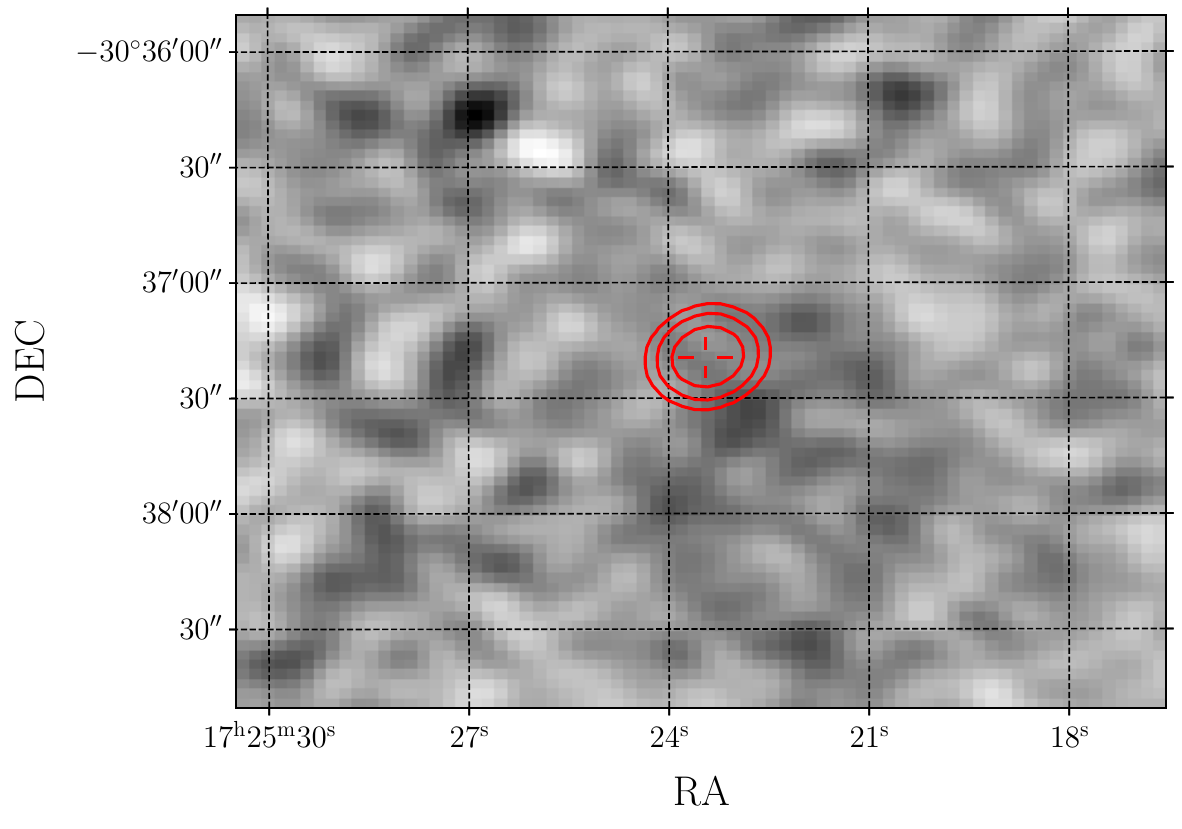}{0.5\textwidth}{(d) J172523$-$303720 stacked image}}
    \gridline{\fig{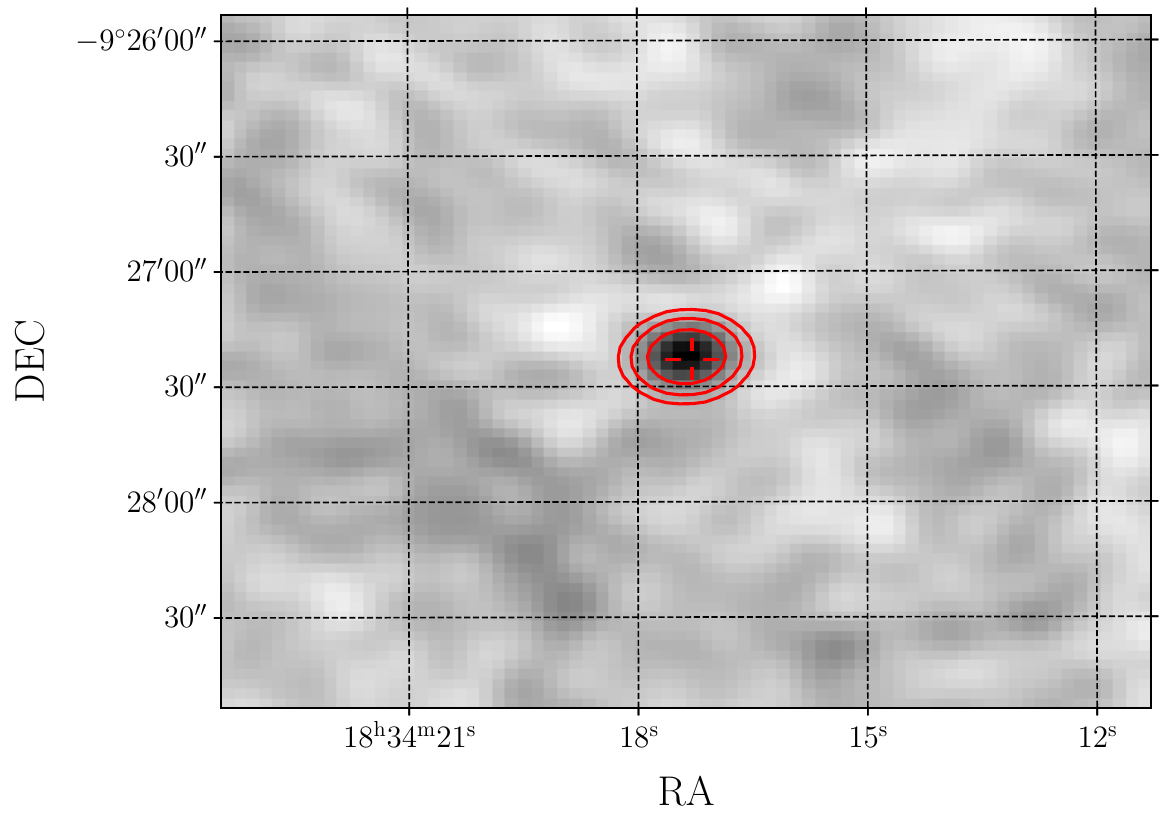}{0.5\textwidth}{(e) J183418$-$092720 brightest detection}
          \fig{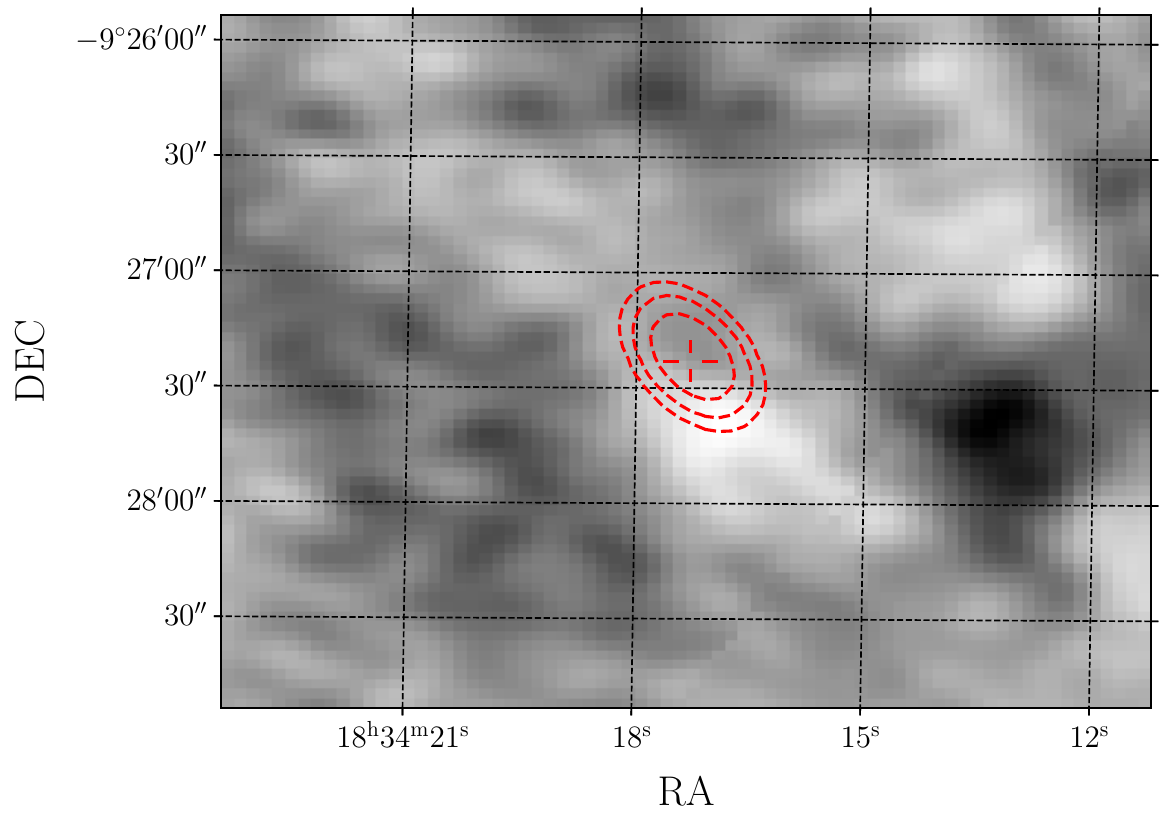}{0.5\textwidth}{(f) J183418$-$092720 stacked image}}
    \caption{Brightest detection image and the stacked image from all the non-detections}
    \label{fig:all_pulsar_res_2}
\end{figure*}

\section{Optical/Infrared composite images}\label{app:oir_images}
Shown below are the optical ({DECaPS/PANSTARRS} --- where available {DECaPS} was preferred since these observations were deeper, and in the absence of these observations we chose {PANSTARRS} observations), and Infrared (similarly we chose both \textit{VVV} and {UKIRT} observations but VVV observations were preferred because of the depth) composite images. At Optical wavelengths, the composite images was made using \textit{r, i, z} band observations and at infrared wavelengths, the composite image was made using $J$, $H$, and $K_s$ observations.
\begin{figure*}
    \gridline{\fig{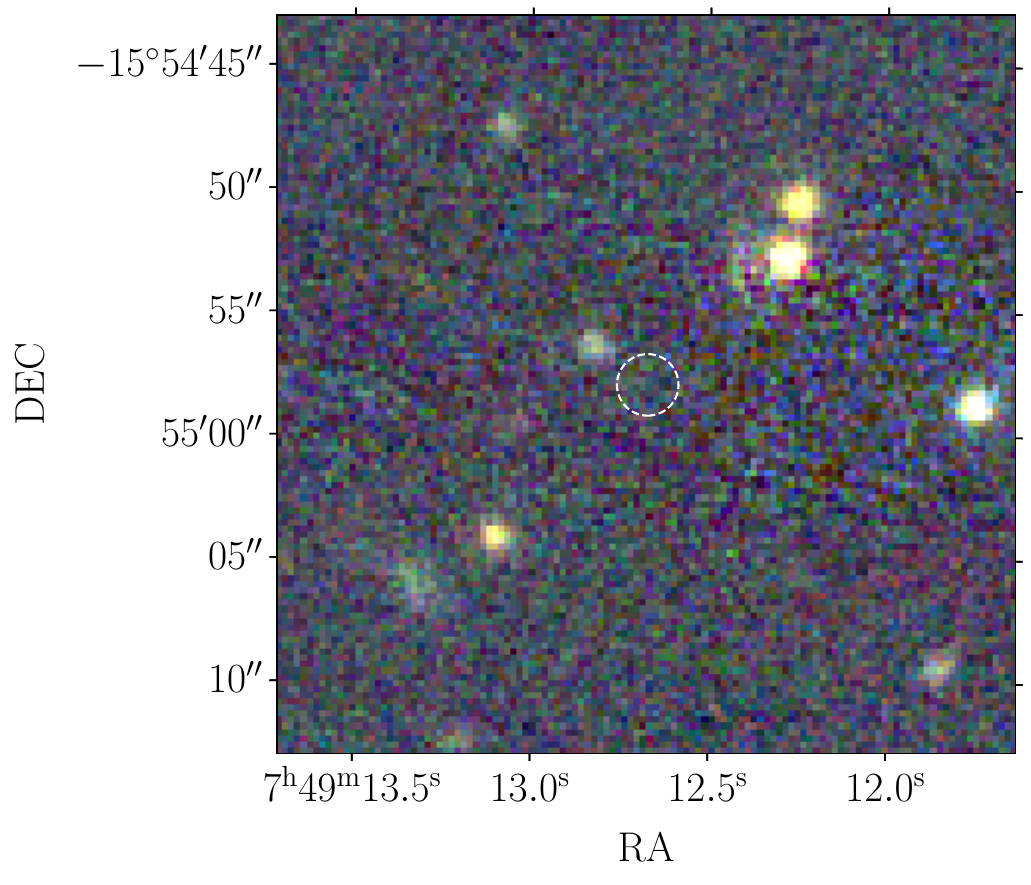}{0.5\textwidth}{(a) J074913$-$155457 {PANSTARRS} (\textit{r, i, z}) composite image}
          \fig{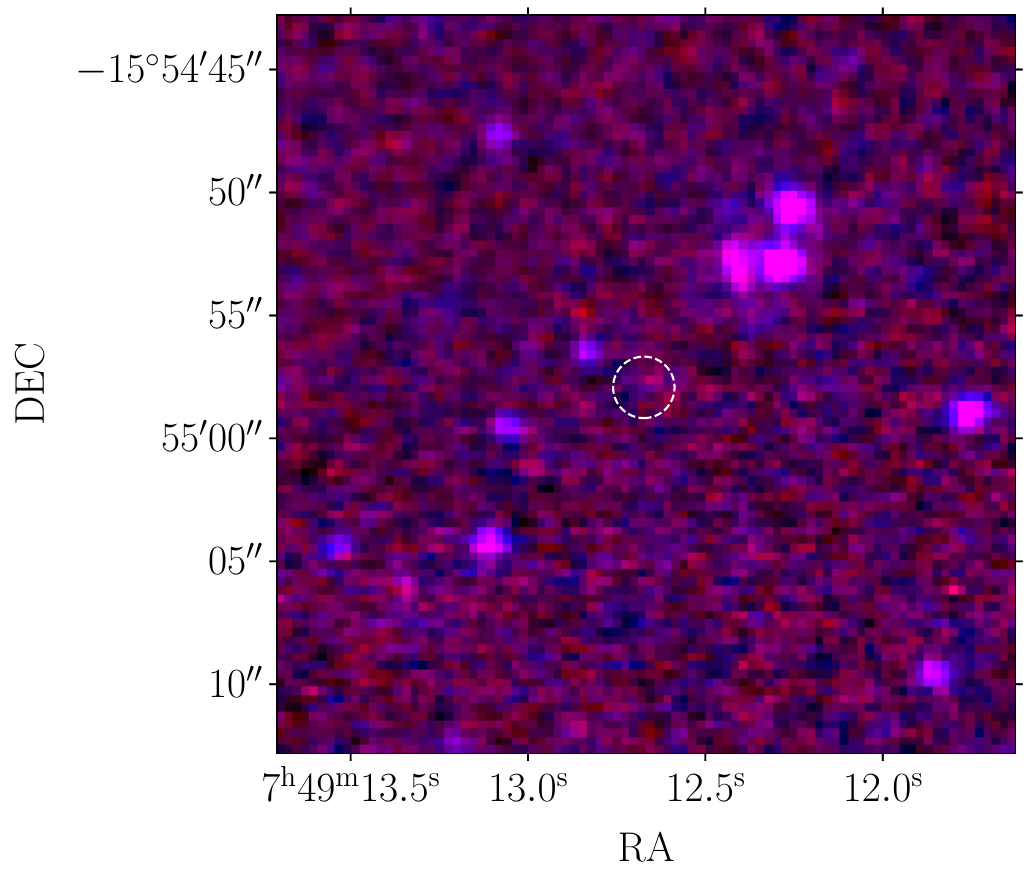}{0.5\textwidth}{(b) J074913$-$155457 {VVV} ($J$, $K_s$) composite image}}
    \gridline{\fig{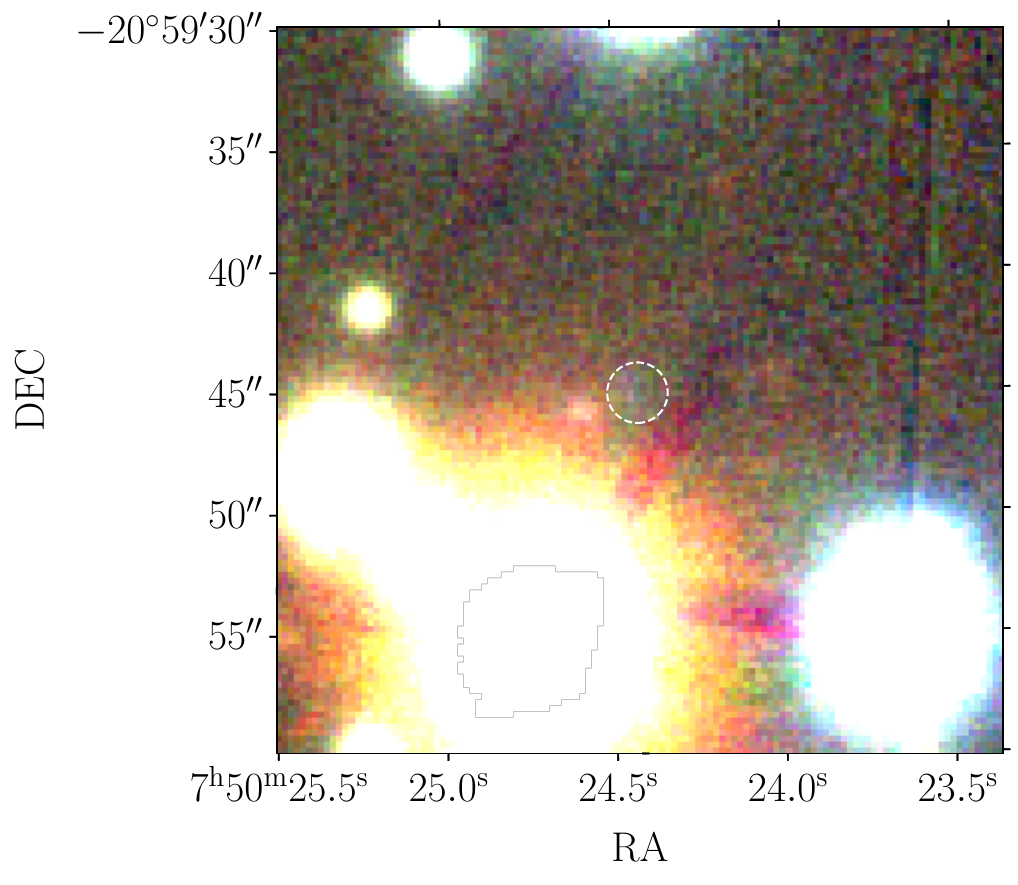}{0.5\textwidth}{(c) J075024$-$205945 {PANSTARRS} (\textit{g, r, i}) composite image}
          \fig{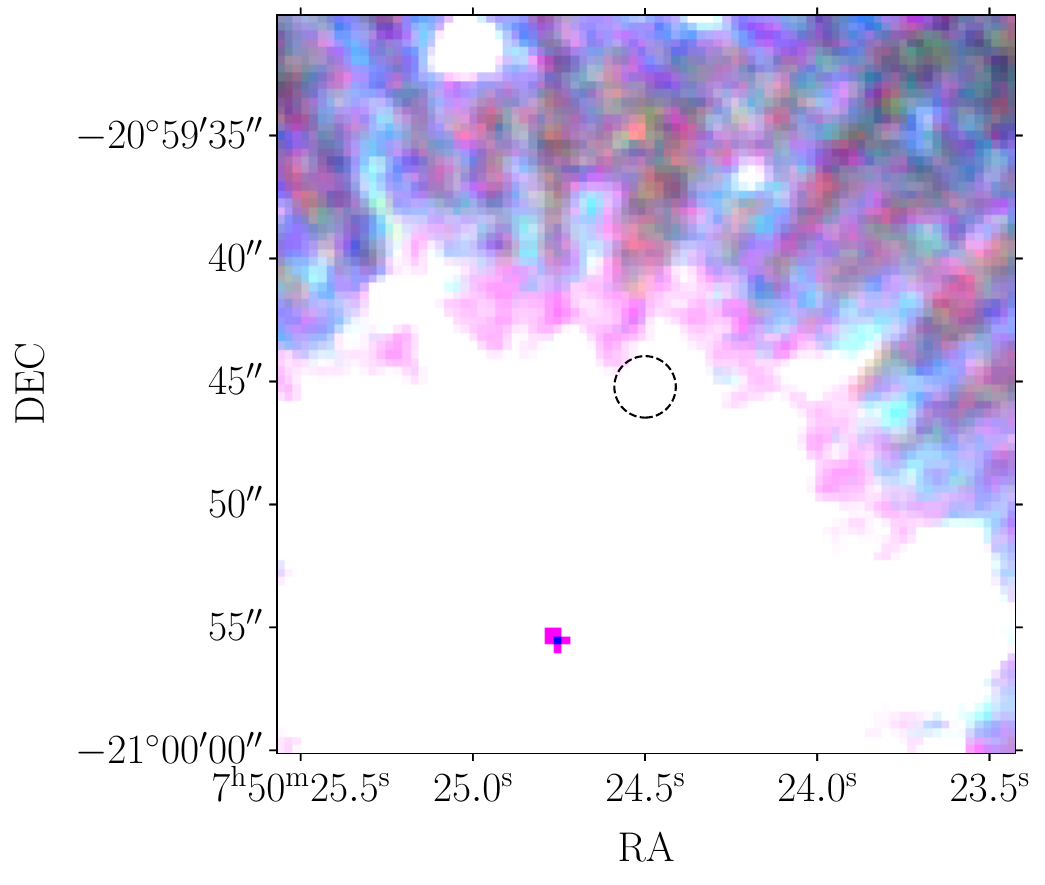}{0.5\textwidth}{(d) J075024$-$205945 {VVV} ($J$, $H$, $K_s$) composite image}}
    \caption{Optical/infrared composite images. The white/black dotted circle shows the 2.5\arcsec\, error circle surrounding the source.  In both cases, the images are $30\arcsec$ on a side, with north up and east to the left.}
    \label{fig:all_gcrt_cnads_oir}
\end{figure*}
\begin{figure*}
    \gridline{\fig{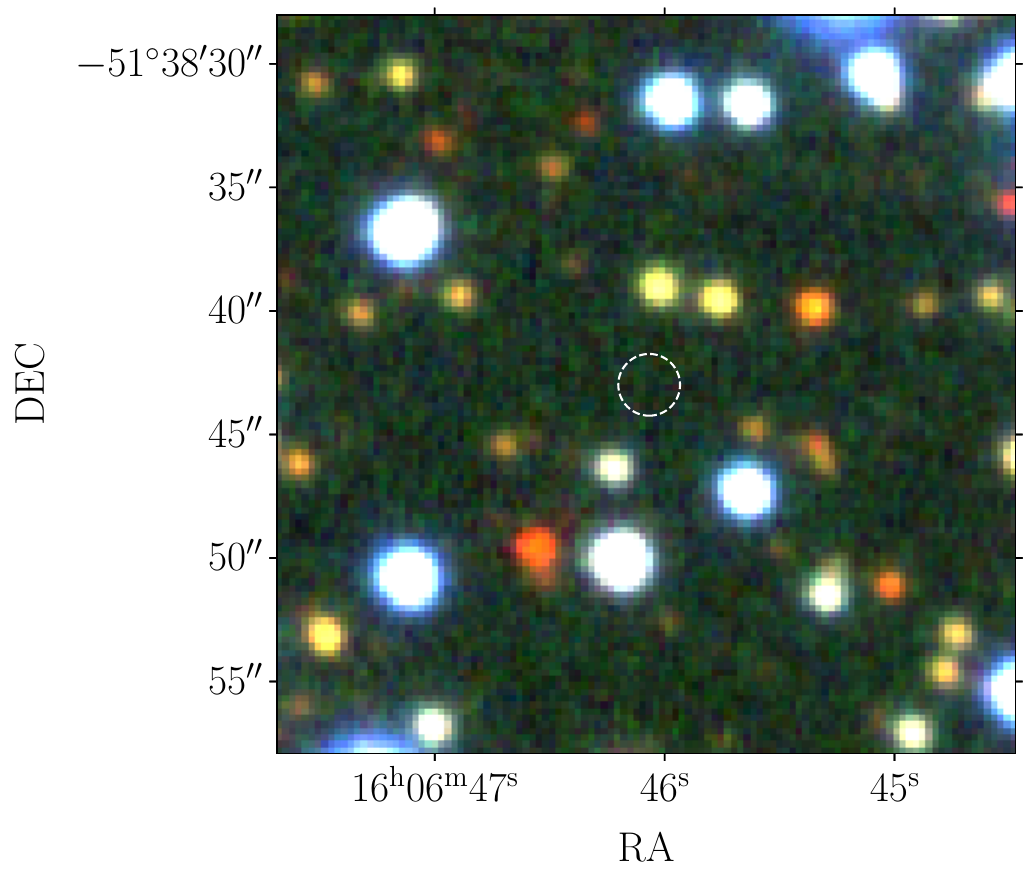}{0.5\textwidth}{(a) J160646$-$513843 {DECaPS} (\textit{r, i, z}) composite image}
          \fig{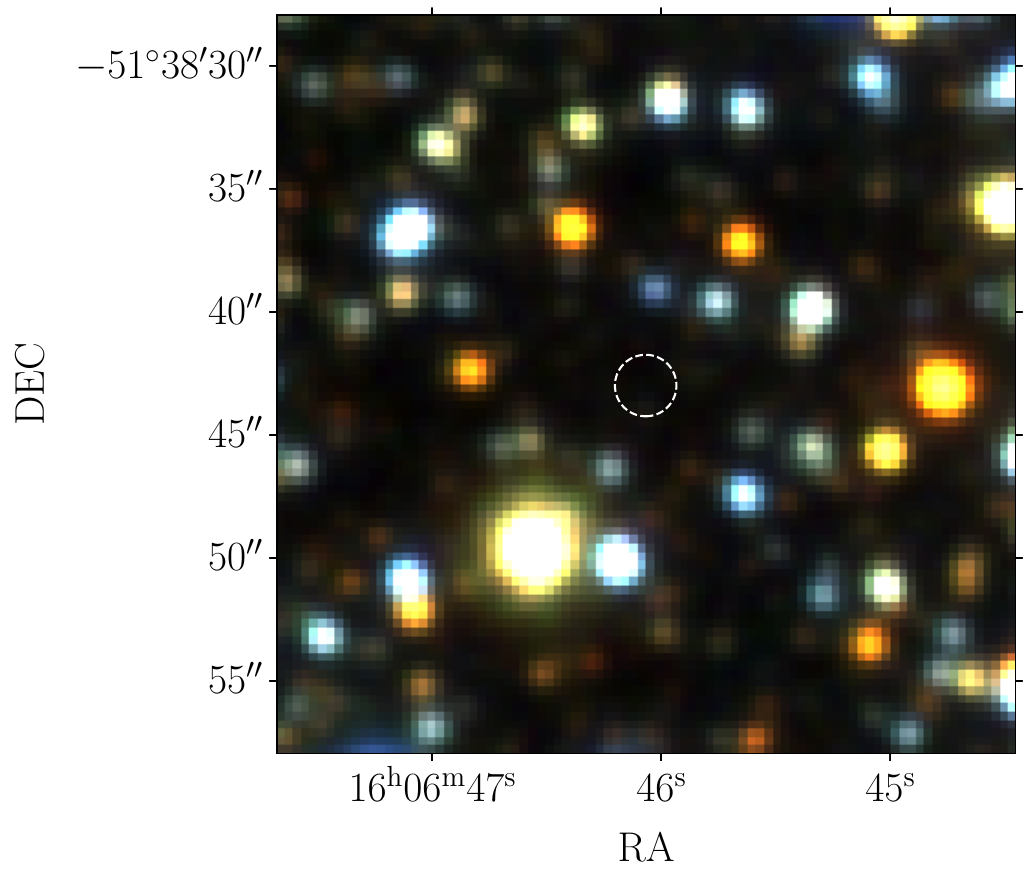}{0.5\textwidth}{(b) J160646$-$513843 {VVV} ($J$, $H$, $K_s$) composite image}}
    \gridline{\fig{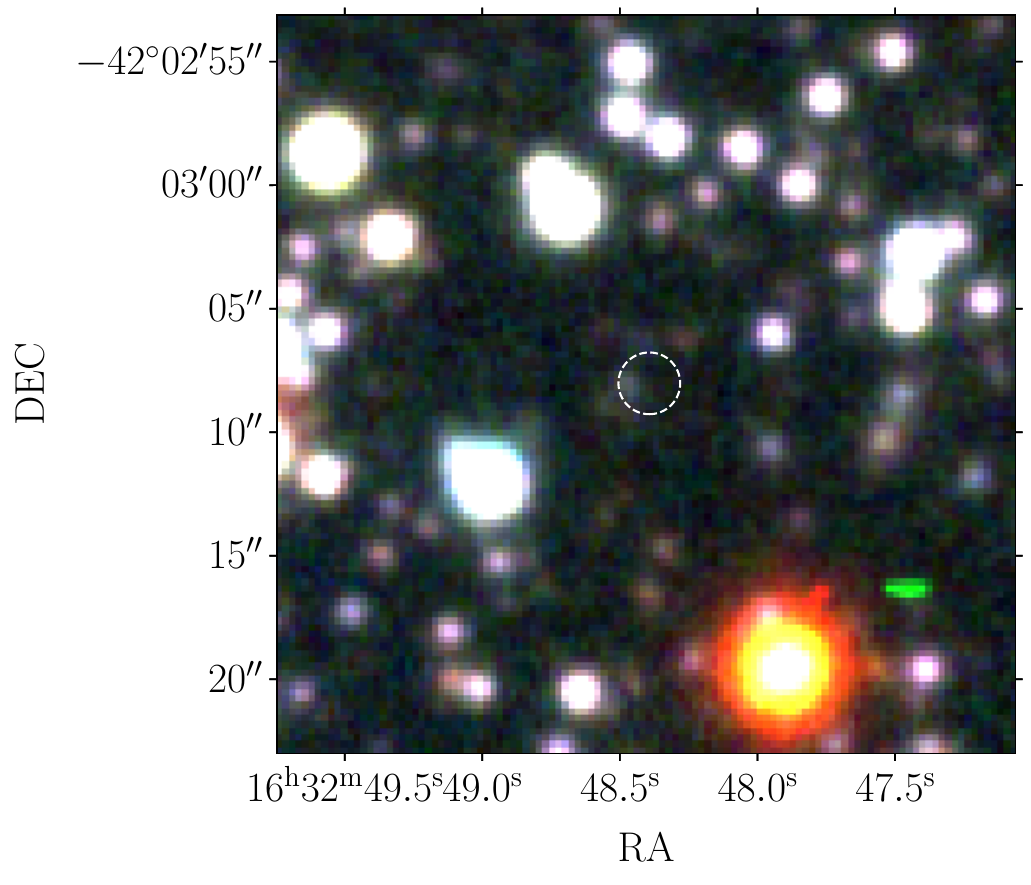}{0.5\textwidth}{(c) J163248$-$420307 {DECaPS} (\textit{r, i, z}) composite image}
          \fig{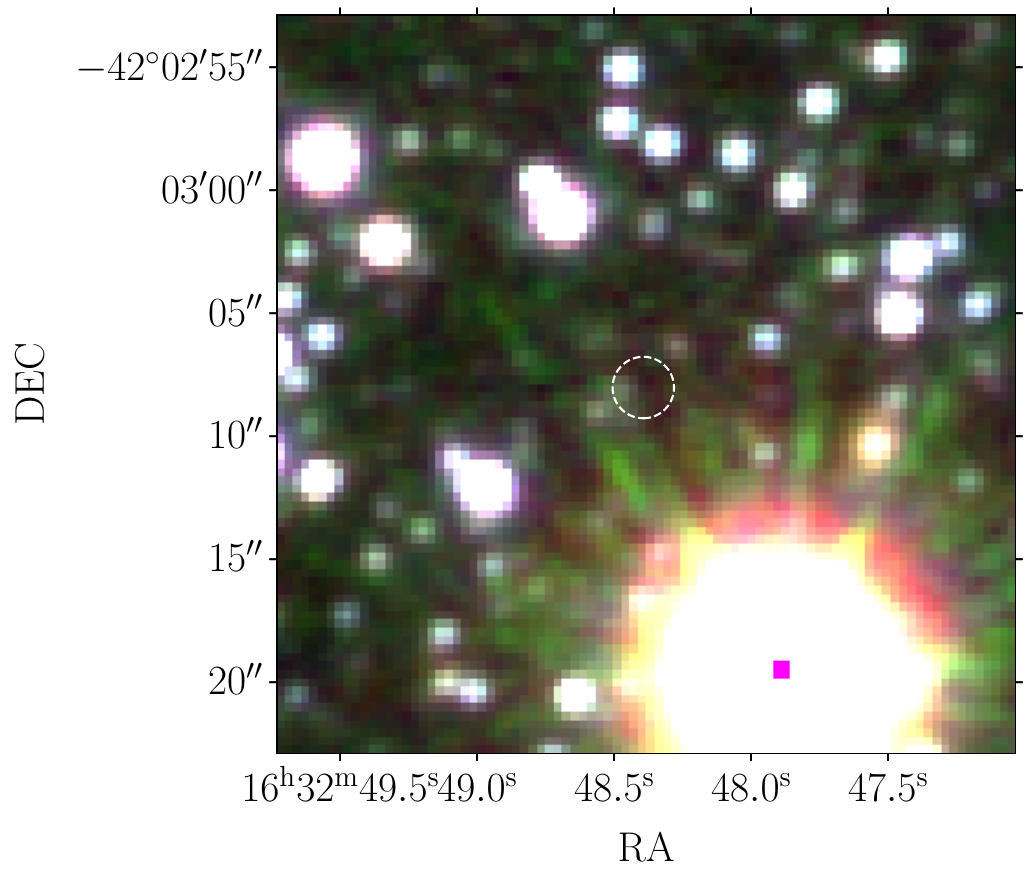}{0.5\textwidth}{(d) J163248$-$420307 {VVV} ($J$, $H$, $K_s$) composite image}}
    \caption{Optical/Infrared composite images. The white/black dotted circle shows the 2.5\arcsec\, error circle surrounding the source.  In both cases, the images are $30\arcsec$ on a side, with north up and east to the left.}
    \label{fig:all_gcrt_cnads_oir_2}
\end{figure*}
\begin{figure*}
    \gridline{\fig{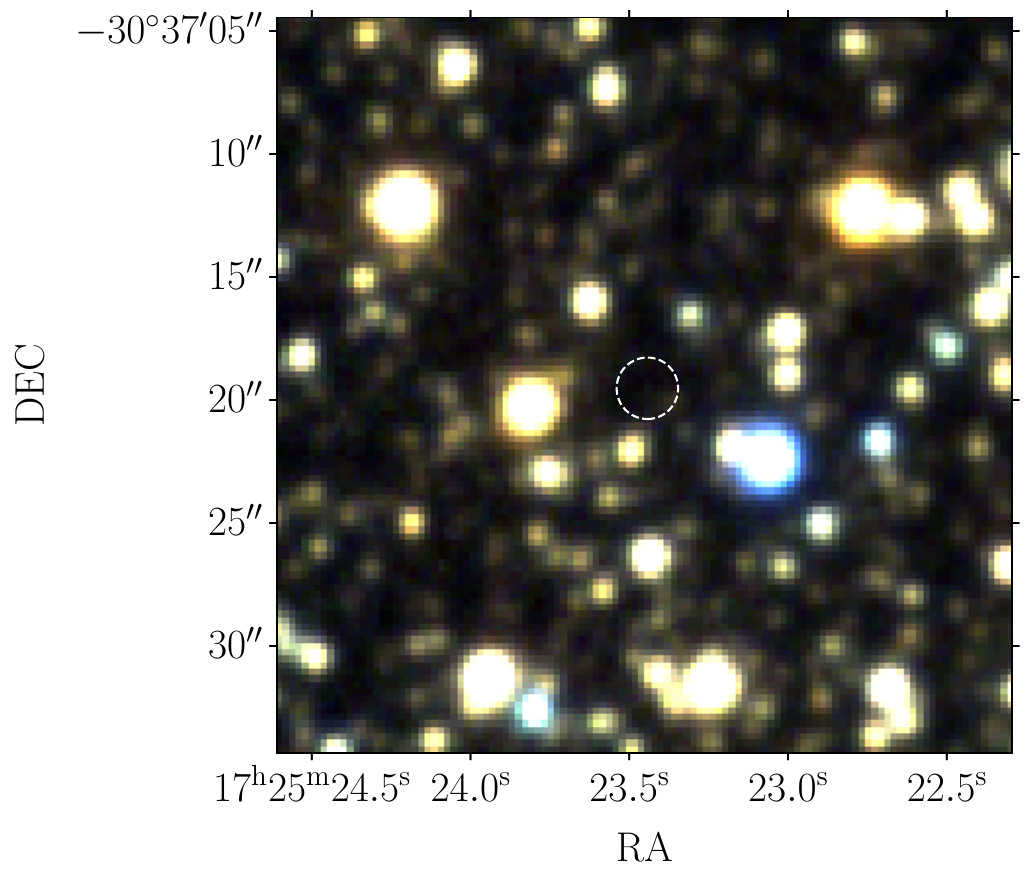}{0.5\textwidth}{(a) J172523$-$303720 \textit{DECaPS} (\textit{r, i, z}) composite image}
          \fig{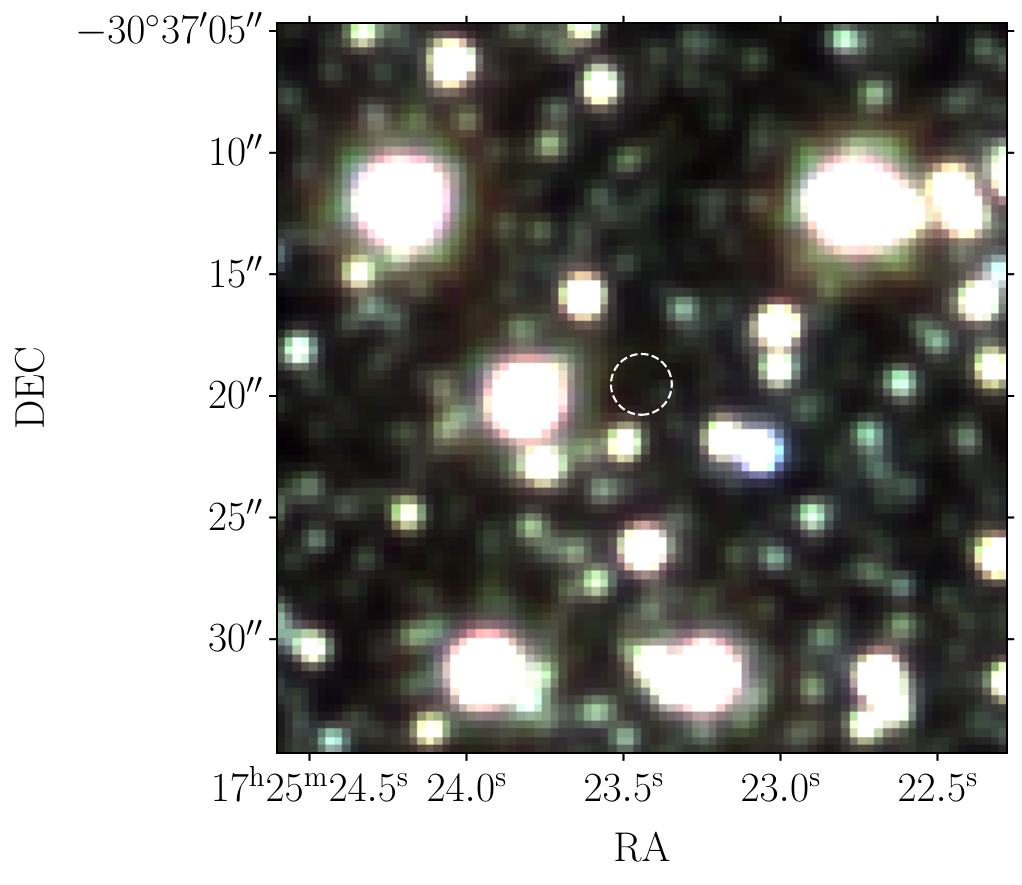}{0.5\textwidth}{(b) J172523$-$303720 {VVV} ($J$, $H$, $K_s$) composite image}}
    \gridline{\fig{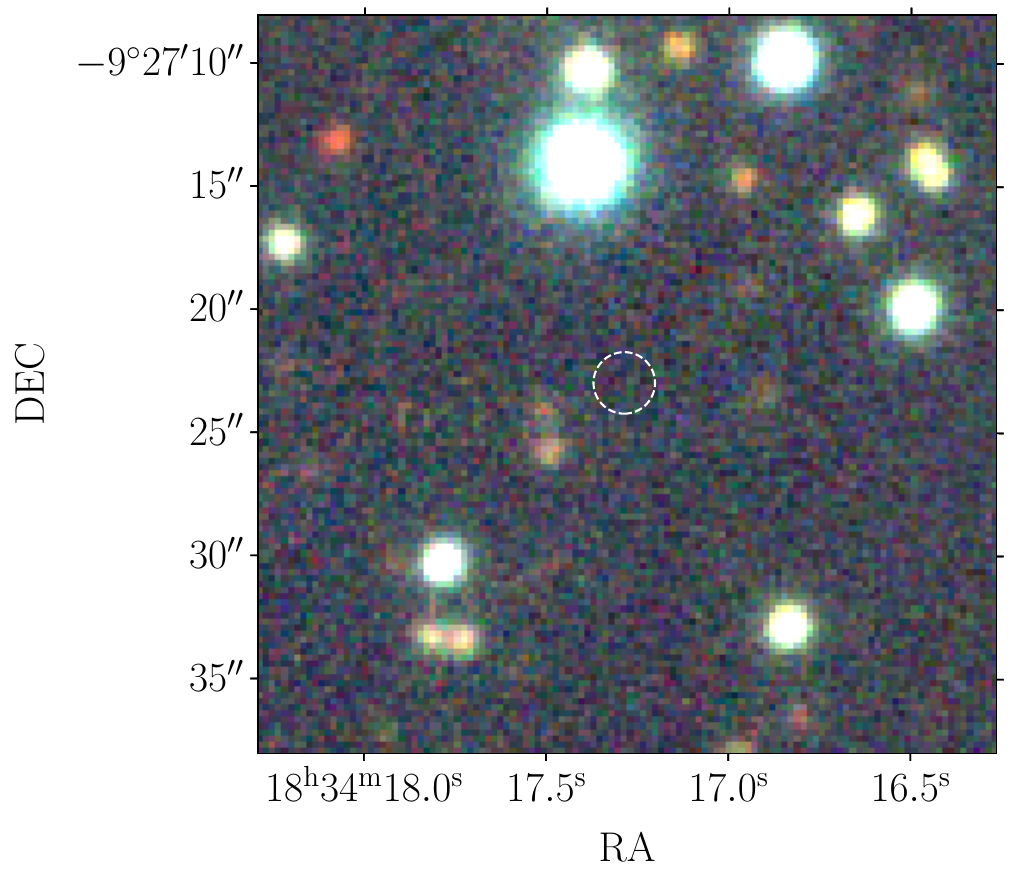}{0.5\textwidth}{(c) J183418$-$092720 {PANSTARRS} (\textit{r, i, z}) composite image}
          \fig{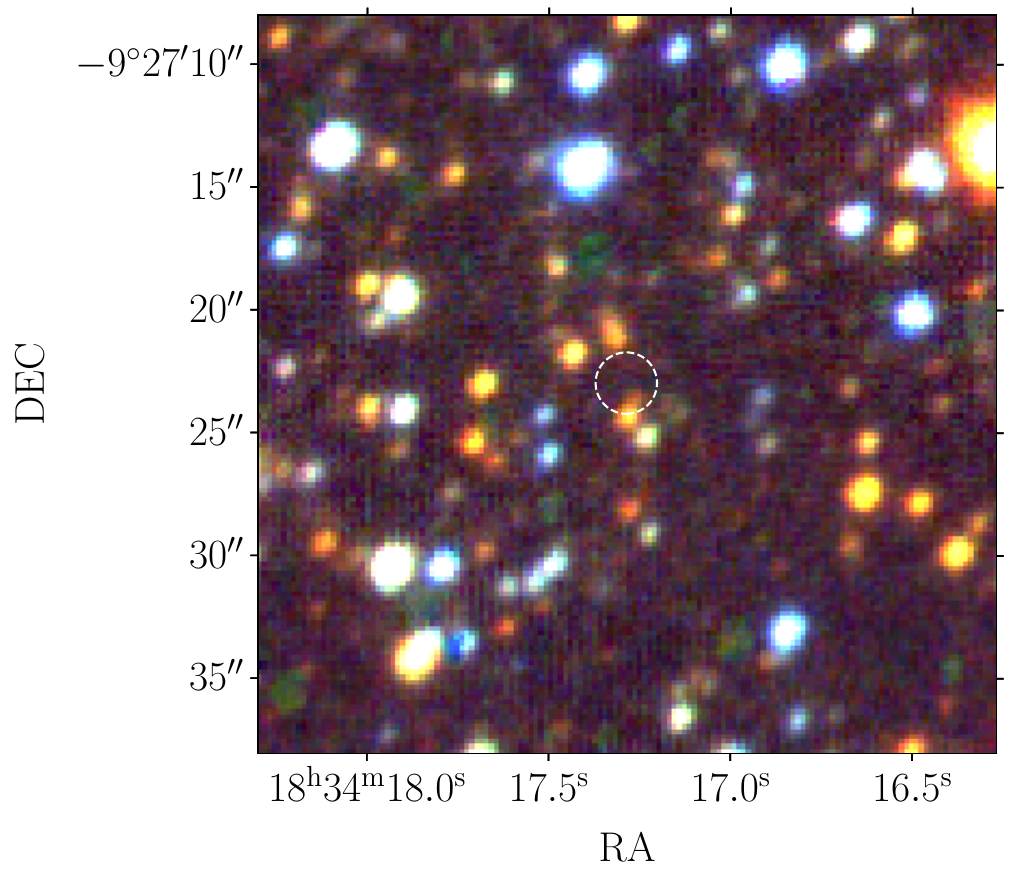}{0.5\textwidth}{(d) J183418$-$092720 {VVV} ($J$, $H$, $K_s$) composite image}}
    \caption{Optical/Infrared composite images. The white/black dotted circle shows the 2.5\arcsec\, error circle surrounding the source.  In both cases, the images are $30\arcsec$ on a side, with north up and east to the left.}
    \label{fig:all_gcrt_cnads_oir_3}
\end{figure*}

\section{Nature of sources}\label{app:nature}
Here, we discuss in more detail how we disfavor various possibilities regarding what our sample of sources can be. 

\subsection{Background extragalactic sources}
Variability at radio wavelengths that VAST is sensitive to can stem from a wide variety of radio transients \citep{vast}. The most common sources of confusion are the active galactic nuclei \citep[AGNs;][]{Padovani2017}, where both intrinsic or extrinsic (propagation) effects result in stochastic/structured temporal variability. Variations (a few tens of percent) intrinsic or arising from refractive interstellar scintillation (RISS) are common in AGNs \citep{Marscher85,Ulrich97,Hovatta2007,Rickett2002,Lovell2003}. Large variations (from a factor of a few to a couple of orders of magnitude) are uncommon at shorter timescales but are seen over decadal timescales involving transitions from radio-quiet to radio-loud \citep{Nyland2020}. Although radio emission from AGN jets can show both linear \citep{agn_lin_pol} and circular \citep{agn_circ_pol} polarizations, the polarization fraction is typically $<10\%$ \citep{Mesa2002,Sullivan2015}. In addition, none of the sources presented here has an identifiable optical or IR counterpart, indicative of AGNs. Hence, a combination of significant polarization, fast temporal variation, and lack of an OIR counterpart makes the current sample unlikely to be an unidentified background AGN. 

Similar arguments can be made for extragalactic transients, such as afterglows to explosive events like supernovae, gamma-ray bursts, and tidal disruption events. Combining polarization, fast evolution, and lack of an optical/X-ray counterpart (host galaxy) can be used to disfavor extragalactic transients.

\subsection{Galactic sources -- Stars}

\begin{figure}
    \centering
    \includegraphics[width=\linewidth]{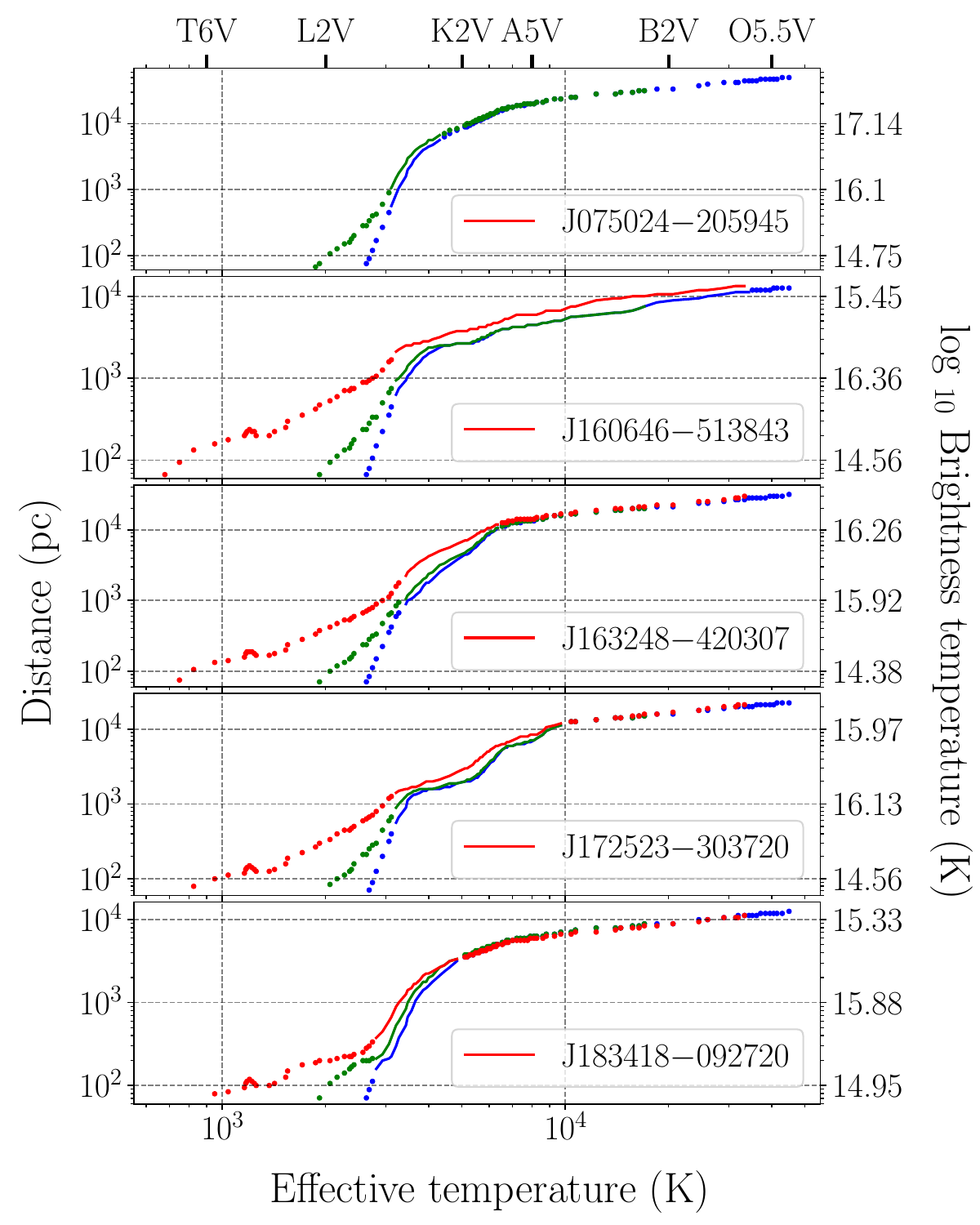}
    \caption{Distance upper limits on stars of various spectral classes using the optical upper limits for sources in our sample. Each panel shows the limits for a single source. Absolute magnitudes in $V$, $g$, $J$-bands for standard stars in each spectral class are adopted from \citep{Pecaut2013}, and the distance upper limits are estimated separately for each band. The blue, green, and red curves show the estimates for $V, g, J$ bands, respectively. For $V$-band, we used the $g$-band upper limits to calculate the distance, since we do not have archival data in this band. We used the 3D extinction model from \texttt{DUSTMAPS} \citep{green2018,Green2019,Zucker2025} to account for extinction along the line of sight. The solid part of the curves shows the distances for which these maps give reliable extinction estimates, and the dotted parts of the curves are extrapolated by extending the extinction using a linear model (both for distant and nearby regions). The right-hand side of the figure shows corresponding radio luminosities for these distance upper limits.}
    \label{fig:dis_upp}
\end{figure}

Radio emission from stars can be due to multiple emission phenomena, including both incoherent emission, like synchrotron emission \citep{dulk1985}, and coherent emission like plasma and maser emission \citep{melrose2017}. Incoherent (gyro--)synchrotron radio emission, which usually dominates in main-sequence stars can be disfavored based on the high levels of polarization and expected correlation between its X-ray and radio emission \citep{Guedel1993} --- $L_X=\nu\,L_{\nu, R}\times 10^{15.5}$, where $L_X$ is the X-ray luminosity (in erg/s) and $L_{\nu,R}$  is the specific radio luminosity (in erg/s/Hz) --- where the X-ray upper limits are at least two orders of magnitude deeper than this predicted value. Coherent emission is observed in magnetic stars --- predominantly in low mass stars (M-dwarfs and ultra-cool dwarfs), but also some hot magnetic stars \citep{Das2022} ---  and in most cases attributed to ECME. However, ECME is observed as narrow pulsations, every rotational period, as opposed to slow ($\sim$weeks) outbursts. While this can explain a subset of our sample that is fast-flaring, other factors, like the lack of a counterpart (and its implications on the brightness temperature), can be used to disfavor this. 

Exploiting the OIR non-detections, we can estimate the distance lower limit for stars of different spectral types. Figure~\ref{fig:dis_upp} shows these distance limits (and the corresponding radio brightness temperatures) for different spectral type stars. Some of the hot magnetic stars (O/B type) can be ruled out based on their distance limits ($>$20\,kpc). In the remaining sample, primarily dwarf stars (M/L/T/Y-type), the lower limit on the estimated brightness temperature is almost comparable to the lower end of the pulsar distribution, which makes it difficult to explain the underlying energy budget for this radio emission. Hence, although not impossible, it is unlikely that the sources in our fast-flaring sample are dwarf stars/ultracool dwarfs. 

\subsection{Galactic sources -- Isolated compact objects}
Other Galactic sources that show variable radio emission are pulsars and magnetars, although through different mechanisms. The optical/NIR non-detections are in line with these objects. Figure~\ref{fig:gcrt_pop} shows the NIR/X-ray and radio properties of known sources, and it can be seen that our sample of sources is consistent with the population of pulsars/magnetars. However, four of the five sources in our sample are covered by the Parkes Multi-Beam Pulsar Survey (PMPS), but without any pulsar/magnetar detection. The lack of pulsations means that if they are indeed pulsars, then they might be highly scattered or very broad pulse profile pulsars. In addition, continuum emission from pulsars can be variable due to scintillation. But variations are modest ($<$ a factor of 2), and so our fast-flaring sample is unlikely to be a result of scintillation of fainter pulsars. In addition, the sporadic nature of scintillation is also inconsistent with month-long live outbursts in our slow-flaring sample. This makes the pulsar interpretation difficult to reconcile with our sample (both fast-flaring and slow-flaring). Magnetars, on the other hand, are observed to undergo flaring episodes during which they show substantial linear polarization. 
In the case of magnetars, the lack of X-ray detections is difficult to reconcile since none of the radio bright magnetars are X-ray quiet. This means that our sample of sources is at least 1--4 orders of magnitude brighter in terms of the radio to X-ray luminosity compared to known magnetars (albeit the limits are not simultaneous in time and hence weak; see Figure~\ref{fig:gcrt_pop}). 

While usual cases of pulsars can be disfavored, there are more extreme cases, which include intermittent pulsars \citep{Kramer2006,intermittent_pulsars}, and rotating radio transients \citep[RRATs;][]{McLaughlin2009}. Both interpretations can be disfavored in our slow-flaring type, which resembles slow-evolving synchrotron flares. For the fast-flaring sample, if our sources are similar to RRATs, the dynamic spectra should show unresolved emission in a single time bin (10\,s) or a few isolated time bins, as opposed to the observed behavior, making it unlikely. In intermittent pulsars, emission in the ``on'' state should resemble standard pulsar emission and hence strong variability on min--hour timescales, like observed in \gcrtb, and \gcrtc, is not expected, since continuum variations are often attributed to scintillation and scattering and are modest. However, pulsar intermittency can not be completely ruled out in \gcrte.

\subsection{Galactic sources -- NS/BH binaries}
The slow-flaring sources in our sample can draw similarities with X-ray binaries (XRBs), in which radio emission is usually observed during X-ray outbursts in the form of synchrotron radiation, and is weakly polarized, and which shows a smooth rise and decline over a few weeks. But no X-ray outburst was detected at the position of these sources, which makes it difficult for these sources to be analogs of XRBs. In addition, Figure~\ref{fig:gcrt_pop} shows that our sample of sources is at least three orders of magnitude brighter in the relative radio to X-ray luminosity, compared to known XRBs, but we caution again that our X-ray limits are not time simultaneous with our radio observations. 


\bibliography{references}{}
\bibliographystyle{aasjournal}

\end{document}